\documentclass[trackchanges,twocolumn]{aastex63}
\usepackage{graphicx} 
\usepackage{txfonts} 
\usepackage{natbib}
\usepackage{booktabs}

\submitjournal{\apj}
\shorttitle{Kinematics of multiple populations in globular clusters} 
\shortauthors{G.\,Cordoni et al.} 
\begin{document} 

\title{Gaia and Hubble unveil the kinematics of stellar populations in the Type II globular clusters $\omega$\,Centauri and M\,22.}  
\author{G.\,Cordoni} 
\affiliation{Dipartimento di Fisica e Astronomia ``Galileo Galilei'' -
  Univ. di Padova, Vicolo dell'Osservatorio 3, Padova, IT-35122}
\author{A.\ P.\,Milone}
\affiliation{Dipartimento di Fisica e Astronomia ``Galileo Galilei'' -  Univ. di Padova, Vicolo dell'Osservatorio 3, Padova, IT-35122}
\affiliation{Istituto Nazionale di Astrofisica - Osservatorio Astronomico di Padova, Vicolo dell'Osservatorio 5, Padova, IT-35122}
\author{A.\ F.\,Marino} 
\affiliation{Dipartimento di Fisica e Astronomia ``Galileo Galilei'' - Univ. di Padova, Vicolo dell'Osservatorio 3, Padova, IT-35122}
\affiliation{Istituto Nazionale di Astrofisica - Osservatorio Astronomico di Padova, Vicolo dell'Osservatorio 5, Padova, IT-35122}
\affiliation{Centro di Ateneo di Studi e Attivit\`a Spaziali ``Giuseppe Colombo'' - CISAS, Via Venezia 15, Padova, IT-35131}
\author{G.\ S.\,Da Costa}
\affiliation{Research School of Astronomy and Astrophysics, Australian National University, Canberra, ACT 0200, Australia}
\author{E.\,Dondoglio}
\affiliation{Dipartimento di Fisica e Astronomia ``Galileo Galilei'' -  Univ. di Padova, Vicolo dell'Osservatorio 3, Padova, IT-35122}
\author{H.\,Jerjen}
\affiliation{Research School of Astronomy and Astrophysics, Australian National University, Canberra, ACT 0200, Australia}
\author{E.\ P.\,Lagioia }
\affiliation{Dipartimento di Fisica e Astronomia ``Galileo Galilei'' -
  Univ. di Padova, Vicolo dell'Osservatorio 3, Padova, IT-35122}
\author{A.\, Mastrobuono-Battisti}
\affiliation{Department of Astronomy and Theoretical Physics, Lund Observatory, Box 43, SE--221 00, Lund, Sweden}
\author{J.\ E.\,Norris}
\affiliation{Research School of Astronomy and Astrophysics, Australian National University, Canberra, ACT 0200, Australia}
\author{M.\,Tailo}
\affiliation{Dipartimento di Fisica e Astronomia ``Galileo Galilei'' -
  Univ. di Padova, Vicolo dell'Osservatorio 3, Padova, IT-35122}%
\author{D.\,Yong}
\affiliation{Research School of Astronomy and Astrophysics, Australian National University, Canberra, ACT 0200, Australia}
\affiliation{ARC Centre of Excellence for Astrophysics in Three Dimensions (ASTRO-3D), Canberra, ACT 0200, Australia}

\correspondingauthor{G.\ Cordoni}
\email{giacomo.cordoni@phd.unipd.it} 
\begin{abstract}
The origin of multiple stellar populations in Globular Clusters (GCs) is one of the greatest mysteries of modern stellar astrophysics. $N$-body simulations suggest that the present-day dynamics of GC stars can constrain the events that occurred at high redshift and led to the formation of multiple populations.  Here, we combine multi-band photometry from the \textit{Hubble Space Telescope} (\textit{HST}) and ground-based facilities with \textit{HST} and Gaia Data Release 2 proper motions to investigate the spatial distributions and the motions in the plane of the sky of multiple populations in the type II GCs NGC\,5139 ($\omega\,$Centauri) and NGC\,6656 (M\,22).  We first analyzed stellar populations with different metallicities. Fe-poor and Fe-rich stars in M\,22 share similar spatial distributions and rotation patterns and exhibit similar isotropic motions. Similarly, the two main populations with different iron abundance in $\omega\,$Centauri share similar ellipticities and rotation patterns. When analyzing different radial regions, we find that the rotation amplitude decreases from the center towards the external regions. Fe-poor and Fe-rich stars of $\omega\,$Centauri  are radially anisotropic in the central region and show similar degrees of anisotropy.   We also investigate the stellar populations with different light-element abundances and find that their N-rich stars exhibit higher ellipticity than N-poor stars. In $\omega\,$Centauri both stellar groups are radially anisotropic. Interestingly, N-rich, Fe-rich stars exhibit different rotation patterns than N-poor stars with similar metallicities.  The stellar populations with different nitrogen of M\,22 exhibit similar rotation patterns and isotropic motions. We discuss these findings in the context of the formation of multiple populations.
\end{abstract} 
 

\section{Introduction}\label{sec:intro}
 
An increasing amount of research is providing evidence for the presence of two main classes of GCs \citep[e.g.\,][]{marino2009, marino2019, milone2017}. 
While stars in the majority of Type I Galactic GCs exhibit homogeneous abundances of heavy elements \citep[e.g.\,][]{carretta2009a}, a small but significant number of `anomalous' clusters (Type II GCs) show internal variations in metallicity and in those elements that are associated to $s$ processes \citep[e.g.\,][]{yong2008, dacosta2009, yong2014, marino2015, johnson2015, marino2019}. 

Type II GCs exhibit distinctive photometric features, including multimodal sub-giant branches (SGBs) in Color-Magnitude Diagrams (CMDs) made with optical filters \citep[e.g.\,][]{milone2008, marino2009, piotto2012}, and multimodal red-giant branches (RGBs) and SGBs in the $I$ vs.\,$U-I$ or $V$ vs.\,$U-V$ CMDs, with metal-rich stars populating red RGBs and faint SGBs \citep[e.g.\,][]{marino2011, lee2015, lee2020}. 

Based on multi-band photometry of 58 GCs, \citet[][]{milone2017} find that Type II GCs make up 17\% of the studied clusters.
The fact that Type II clusters exhibit star-to-star metallicity variation, suggests that they have been able to retain a small amount of the material ejected by supernovae. In this respect, they differ from Type I GCs, where supernova yields seem to have no effect on the chemical composition of second-population stars. 

Due to their large total masses and the complexity of their stellar populations, it has been suggested that Type II GCs formed in the environment of dwarf galaxies, e.g. in their nucleus. 

These galaxies are then tidally destroyed by the interaction with the Milky Way. This possibility is supported by the observation that the Type II GC M\,54 lies in the nucleus of the Sagittarius dwarf galaxy \citep{bellazzini2008} and by the fact that the class of Type II GC includes NGC\,5139 ($\omega\,$Centauri), which is often considered the remnant of a dwarf \citep[e.g.\,][]{bekki2003}. Moreover, based on the integrals of motion of their orbits, at least half of the known Type II GCs (seven out thirteen clusters) are associated with the Enceladus Galaxy thus demonstrating their extragalactic origin \citep{milone2020}. 

Remarkably, the evidence that both metal-rich and metal-poor stars of most Type II GCs host stellar populations with different light-element abundances \citep[e.g.\,][]{marino2009, marino2011}, indicates that independent processes are responsible for the heavy-element enrichment and for the variation of light elements. Insights on the formation processes can be gained via the study of the kinematics of stellar populations with different chemical compositions.

Indeed, the various scenarios on the formation of multiple populations in GCs, suggest that second-generation stars are born in the cluster center, in a high density subsystem embedded in a more-extended first generation \citep[][and references therein]{dercole2008, calura2019}.
$N$-body simulations \citep[e.g.][]{mastrobuono2013, mastrobuono2016, vesperini2013, henault2015, tiongco2019} demonstrate that the dynamical evolution of second-generation stars should be significantly different from that of the first generation and the signature of the different initial conditions could be detected in present-day GC kinematics of GCs where the stars are not fully mixed. Hence, the present-day dynamics of stellar populations with different metallicities and light-element abundances provide a unique window into the origin of multiple populations in Type II GCs. \\

In recent papers, we exploited Gaia Data Release 2 \citep[DR2,][]{gaia2018a} proper motions to investigate the kinematics of stellar populations with different light-element abundances of Type\,I GCs    
 \citep{milone2018, cordoni2019}. 
We find that multiple stellar populations of various GCs, NGC\,0288, NGC\,6121, NGC\,6752 and NGC\,6838, share similar internal kinematics and morphology, in contrast with what is observed in NGC\,104, NGC\,5904 and NGC\,6254.   
Indeed, when we select the main groups of N-poor and N-rich stars (called first and second population, respectively)  we find that both populations of NGC\,104 share similar rotation patterns in the central region and hints of different rotation in the cluster outskirts \citep[Figure~10 in][]{cordoni2019}. Moreover, N-rich stars of NGC\,104 exhibit  show stronger radial anisotropies than the first population \citep[][see their Figure~5 and 10, respectively]{milone2018, cordoni2019}.
The rotation curves of N-poor and N-rich stars of NGC\,5904 seem to exhibit different phases with a statistical significance of $\sim$2.5$\sigma$ \citep[][see their Figure~8]{cordoni2019} and N-rich stars of NGC\,5904 and  exhibit higher ellipiticy than N-poor ones, in close analogy with what is observed in NGC\,6254 \citep[][Figure~5 and 6]{cordoni2019}.

Here, we extend the analysis to the Type II GCs $\omega\,$Centauri, and NGC\,6656 (M\,22), to study the internal kinematics of stellar populations with different metallicities and light-element abundances. The main physical parameters of these two clusters, which share similar nucleosynthetic enrichment processes despite their different masses \citep[e.g.\,][]{dacosta2011}, are listed in Table~\ref{tab:parameters}. In particular, we note that the long half-mass relaxation time of $\omega\,$Centauri, which exceeds the Hubble time \citep{baumgardt2018}, makes this cluster an ideal target to infer the initial configuration of multiple stellar populations. On the contrary,  The half-mass relaxation time is shorter in M\,22   \citep[$t_{\rm h}\sim 3$\,Gyr, e.g.\,][]{baumgardt2018}.

The paper is organized as follows. In Section~\ref{sec:data} we introduce the dataset and describe the method to select stars with high-precision proper motions and in Section~\ref{sec:cmds} we identify multiple stellar populations along the CMDs. We discuss the properties of multiple populations with different iron content in Section~\ref{sec:fe subpop}, such as their spatial distributions, rotation and velocity profiles.  
In Section~\ref{sec:1P2P} we extend the analysis to the stellar populations with different light-element abundances. Finally, the summary and the discussion of the results are provided in Section~\ref{sec:summary}.

\section{Data and data analysis}\label{sec:data}
To investigate the kinematics and the spatial distributions of stellar populations in M\,22 and $\omega\,$Centauri, we combined the exquisite catalogues of proper motions and stellar positions provided by Gaia DR2, with multi-band wide-field photometry from \citet{stetson2019}. Photometry and proper motions are available for stars of M\,22 and $\omega\,$Centauri with radial distances smaller than $\sim$8.4 and $\sim$28.5 arcmin from the center, respectively.
 Most stars within $\sim$1.7 arcmin from the center of M\,22 and within $\sim$2.5 arcmin from the center of $\omega\,$Centauri  have poor-quality Gaia DR2 proper motions because of crowding. Hence, for the stars in these central regions we used multi-band photometry and relative proper motions from {\it HST} images.

Gaia DR2 proper motions are affected by systematic errors that depend on the positions and the colors of the stars \citep[e.g.][]{lindegren2018}.
We followed the method by \citet{vasiliev2019}, which accounts for systematic errors by enlarging the uncertainties associated with proper-motion determinations.
 As a consequence, as discussed by \citet{cordoni2019}, the error bars provided in this work overestimate the true errors. Indeed, our main focus  is the relative motion of the multiple stellar populations in $\omega\,$Centauri and M\,22, which share similar colors in the Gaia passbands and have, in first approximation, similar spatial distributions. Hence, the effect systematic errors on the relative motions of the distinct population may be partially cancelling out. 
 In the following, we provide details for the data from ground-based facilities and {\it HST}.

\subsection{Ground-based dataset}
We used the catalogues obtained by \citet{stetson2019}, which provide high-precision photometry of stars in the $U$, $B$, $V$ and $I$ bands over a wide field of view. Details on the dataset and on the data reduction are provided by \citet[][]{stetson2005, monelli2013, stetson2019}.
 $U$-band photometry of M\,22 is taken from \citet{marino2015} and was derived from images collected with the Wide Field Imager of the ESO/MPI telescope at Cerro Tololo Inter-American Observatory (WFI@2.2m). We refer to the paper by Marino and collaborators for details on their photometric catalogue. 
 The photometry is calibrated on the photometric system by \citet{landolt1992}. \\
 
\citet{bellini2009} used multi-epoch data acquired by WFI@2.2m to derive proper motions of stars in the field of view of $\omega\,$Centauri, which are suitable to separate field stars from cluster members.   
Due to crowding, stellar proper motions from GAIA DR2 are not available for most of the stars in the central region of $\omega\,$Centauri. To increase the sample size, we identified the stars without Gaia DR2 proper motions that according to \citet{bellini2009} have membership probabilities larger than $90\,\%$ and included these stars in the analysis of the spatial distribution of multiple stellar populations of $\omega\,$Centauri. 
  
We emphasize that proper motions from Bellini and collaborators are not included in our study on the kinematics of $\omega\,$Centauri.
Instead, as we will widely discuss in the next sections, the internal kinematics of multiple stellar populations in $\omega\,$Centauri and M\,22 are investigated by using high-precision proper motions from {\it HST} images and from Gaia DR2 alone.
 
\subsection{{\it HST} dataset}
{\it HST} photometry and relative proper motions are used to investigate stellar populations of M\,22 and $\omega\,$Centauri with radial distances smaller than $\sim$1.7 and $\sim$2.5 arcmin, respectively.

To identify the stellar populations along the RGB of M\,22 and $\omega\,$Centauri, we used the catalogues by \citet{milone2017} and \citet{milone2018}, which include photometry collected through the F275W, F336W, F438W and F814W bands of the Ultraviolet and Visual Channel of the Wide Field Camera 3 (UVIS/WFC3). 

The main properties of the images that we used to derive relative stellar proper motions are summarized in Table~\ref{tab:data}. 
To derive the photometry and the astrometry of all the stars we used the FORTRAN software package KS2 developed by Jay Anderson, \citep[see, e.g.][for details]{anderson2008, sabbi2016}. 
Since we are interested in proper motion determination, we reduced the images collected in different epochs independently, and measured the position of stars at each epoch.  
Stellar positions have been corrected for geometrical distortion by using of the solutions provided by \citet{bellini2009b, bellini2011}.  
We measured proper motions as in \citet[][see their Section 4]{piotto2012} by comparing the distortion-corrected stellar positions at different epochs. To derive the proper motion of each star and minimize the effect of any residual distortion, we used the sample of 45 nearest cluster members as reference stars to fix the zero point of the motion. 
Hence, our measurements from {\it HST} data provide proper motions relative to the average local cluster motion.  
  
\begin{table*}
  \caption{Description of the {\it HST} images used in the paper to derive stellar proper motions.}
\centering
\begin{tabular}{c c c c l l}
\hline \hline
 CAMERA & FILTER  & DATE &  N$\times$EXPTIME & PROGRAM & PI \\
\hline
\hline
  &   &      &       M\,22           &          &    \\
\hline
 ACS/WFC   & F606W  & Jul 01 2006     & 3s$+$4$\times$55s  & 10775 & A.\,Sarajedini \\
 ACS/WFC   & F814W  & Jul 01 2006     & 3s$+$4$\times$65s  & 10775 & A.\,Sarajedini \\
 WFC3/UVIS & F814W  & Sep 23 2010     & 2$\times$50s  & 12311 & G.\,Piotto    \\
 WFC3/UVIS & F814W  & Mar 17-18 2011  & 2$\times$50s  & 12311 & G.\,Piotto    \\
 WFC3/UVIS & F395N  & May 18 2011     & 2$\times$631s$+$2$\times$697s  & 12193 &  J.-W.\,Lee   \\
 WFC3/UVIS & F467M  & May 18 2011     & 2$\times$361s$+$2$\times$367s  & 12193 &  J.-W.\,Lee   \\
 WFC3/UVIS & F547M  & May 18 2011     & 74s$+$3$\times$75s  & 12193 &  J.-W.\,Lee   \\
 WFC3/UVIS & F438W  & Jul 17 2014     & 2$\times$141s  & 13297 &  G.\,Piotto   \\
 \hline
  &   &      &       $\omega\,$Centauri           &          &    \\
\hline
 ACS/WFC   & F435W  & Jun 27 2002     & 12s$+$3$\times$340s  & 9442 & A.\,Cool     \\
 WFC3/UVIS & F438W  & Jul 15 2009     &           35s  & 11452 &  J.\,Kim Quijano   \\
 WFC3/UVIS & F814W  & Jul 15 2009     &           35s  & 11452 &  J.\,Kim Quijano   \\
 WFC3/UVIS & F814W  & Jan 12 2010     &  8$\times$40s  & 11911 &  E.\,Sabbi   \\
 WFC3/UVIS & F438W  & Jan 14 2010     & 9$\times$350s  & 11911 &  E.\,Sabbi   \\
 WFC3/UVIS & F814W  & Jan 14 2010     &           40s  & 11911 &  E.\,Sabbi   \\
 WFC3/UVIS & F438W  & Apr 29 2010     & 7$\times$350s  & 11911 &  E.\,Sabbi   \\
 WFC3/UVIS & F814W  & Apr 29 2010     &  9$\times$40s  & 11911 &  E.\,Sabbi   \\
 WFC3/UVIS & F438W  & Jun 30 2010     & 9$\times$350s  & 11911 &  E.\,Sabbi   \\
 WFC3/UVIS & F814W  & Jun 30 2010     &  4$\times$40s  & 11911 &  E.\,Sabbi   \\
 WFC3/UVIS & F438W  & Jul 04 2010     &          350s  & 11911 &  E.\,Sabbi   \\
 WFC3/UVIS & F814W  & Jul 04 2010     &  5$\times$40s  & 11911 &  E.\,Sabbi   \\
 WFC3/UVIS & F438W  & Feb 15 2011     &          350s  & 12339 &  E.\,Sabbi   \\
 WFC3/UVIS & F438W  & Mar 24 2011     & 8$\times$350s  & 12339 &  E.\,Sabbi   \\
 ACS/WFC   & F435W  & Aug 18 2012     & 9$\times$6s$+$9$\times$339s  & 13066 &  L.\,J.\,Smith   \\
 ACS/WFC   & F435W  & Aug 27 2019     & 42s$+$3$\times$647s  & 15594 & V.\,Kozhurina-Platais    \\
    \hline\hline
\end{tabular}
  \label{tab:data}
 \end{table*}

\subsection{Selection of cluster members}\label{sec:selection}
To explore the internal kinematics of the stellar populations from Gaia DR2 data we identified the sample of stars with accurate astro-photometric measurements following the method described in our previous papers \citep{milone2018, cordoni2018, cordoni2019}. 
In a nutshell, we first selected only stars with accurate proper motions measurements, by using both the \texttt{astrometric\_gof\_al} (\texttt{As\_gof\_al}) and the Renormalized Unit Weight Error (RUWE) parameters \citep[see][]{lindegren2018}.
We then selected cluster members from the proper motion vector-point diagram (VPD).
We refer to \citet{cordoni2019} for a detailed description of the procedure.  
Finally, we corrected the photometry of cluster members for differential reddening using the method in \cite[][see their Section~3.1]{milone2012}.  The final CMDs are shown in the left panels of Figure \ref{fig:cmds}. 

In the case of the {\it HST} dataset, the photometric catalogues by \citet{milone2017, milone2018} already distinguished cluster members and field stars, based on stellar proper motions. Hence, we included in the analysis only those stars that, according to Milone and collaborators, belong to M\,22 and $\omega\,$Centauri.

\section{Multiple populations along the color-magnitude diagrams}\label{sec:cmds}

As shown in \cite{marino2019}, Type II GCs exhibit multimodal SGBs and RGBs in the photometric diagrams made with $U-V$ and $U-I$ colors that correspond to stellar populations with different metallicities. Hence, we exploit the $I$ vs.\,$U-I$ CMD of $\omega\,$Centauri and the $V$ vs.\,$U-V$ CMD of M\,22 to separate the stellar populations with low content of iron and s-process elements (Fe-poor) from the chemically enriched ones (Fe-rich). 

\begin{figure*}
  \centering
  \includegraphics[width=8.9cm,trim={0.5cm 0.cm 0.2cm 0.25cm},clip]{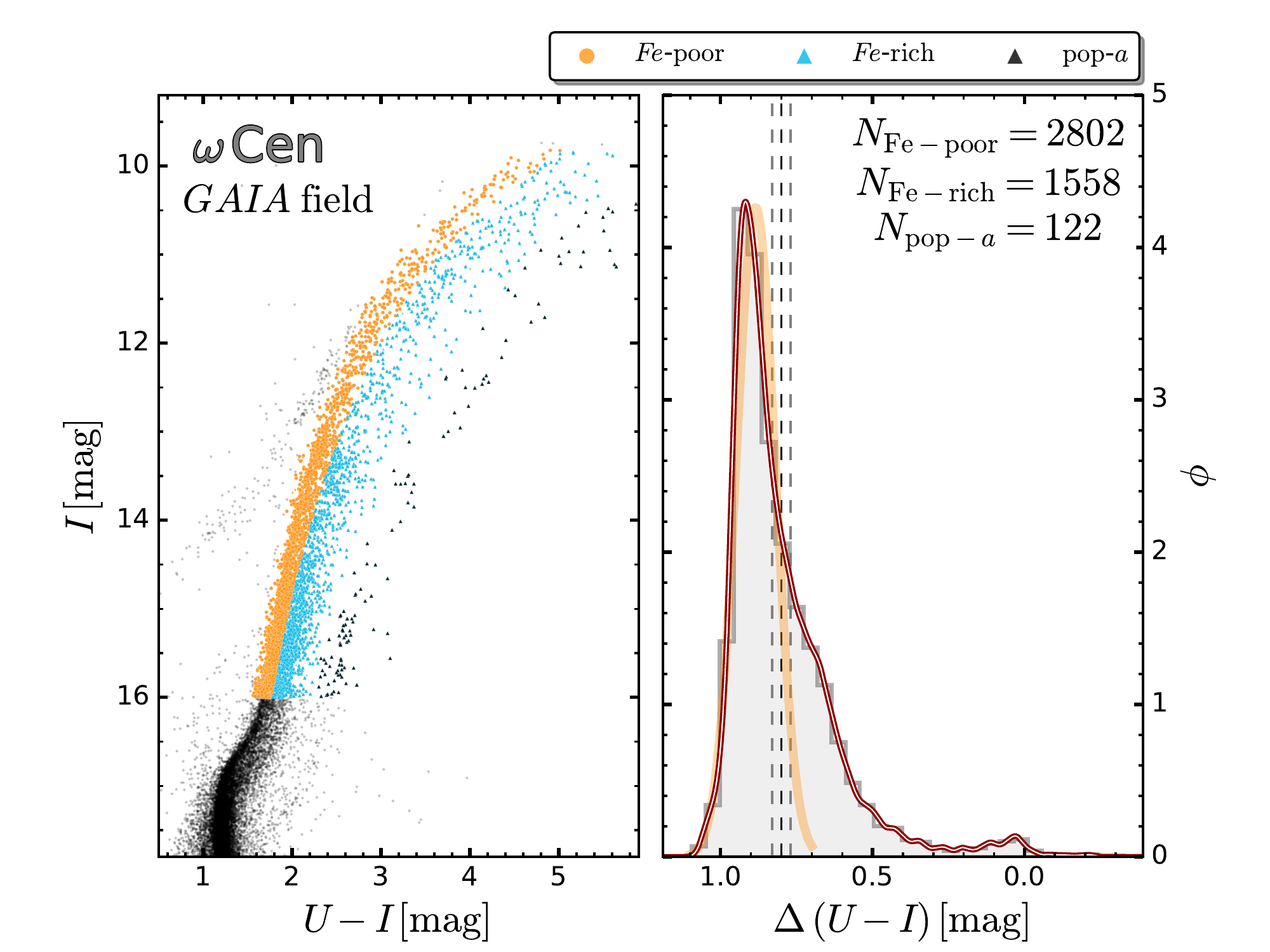}
  \includegraphics[width=8.9cm,trim={0.5cm 0.cm 0cm 0.25cm},clip]{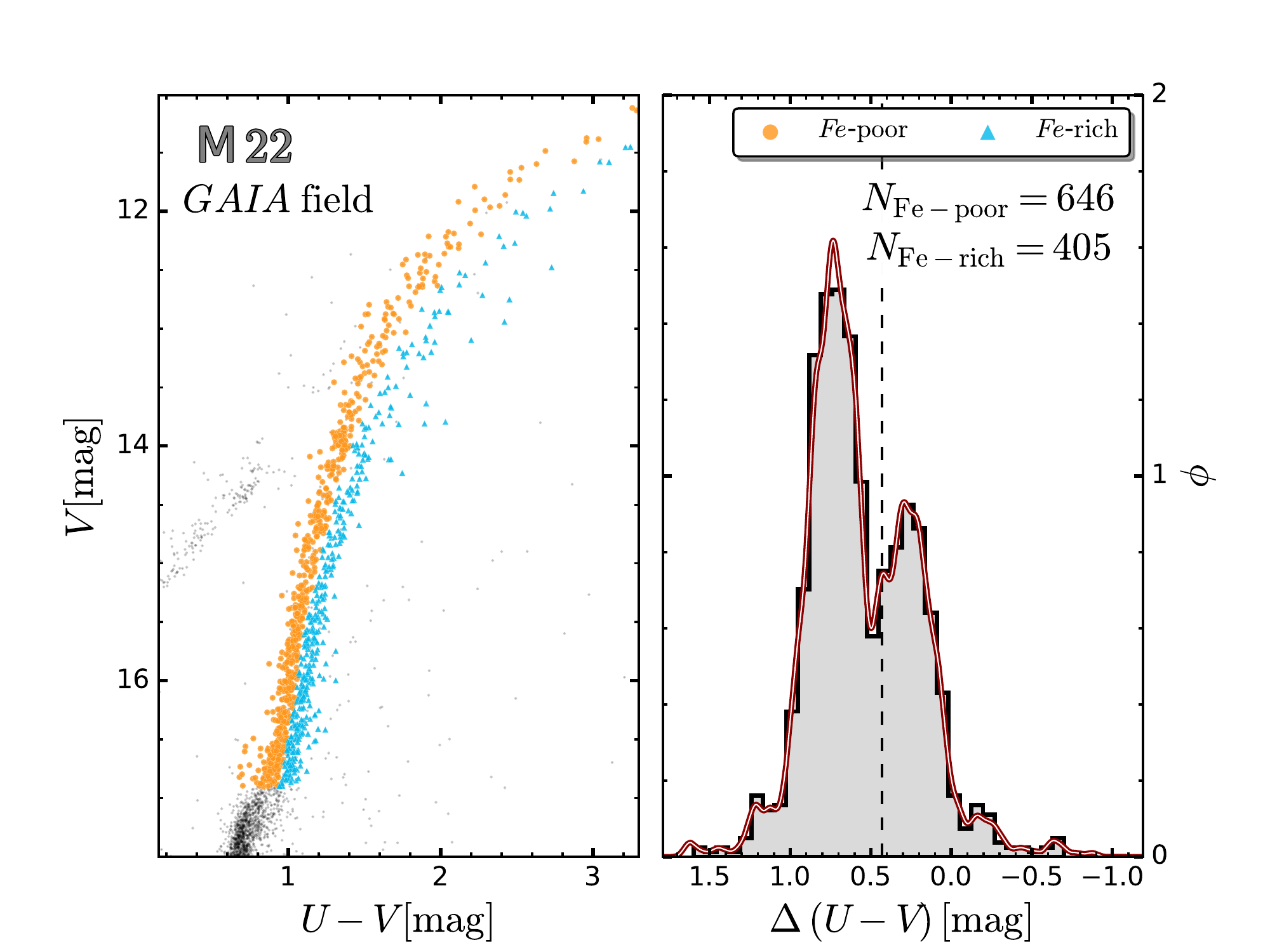}
\includegraphics[width=8.9cm,trim={0.5cm 0.cm 0.2cm 0.25cm},clip]{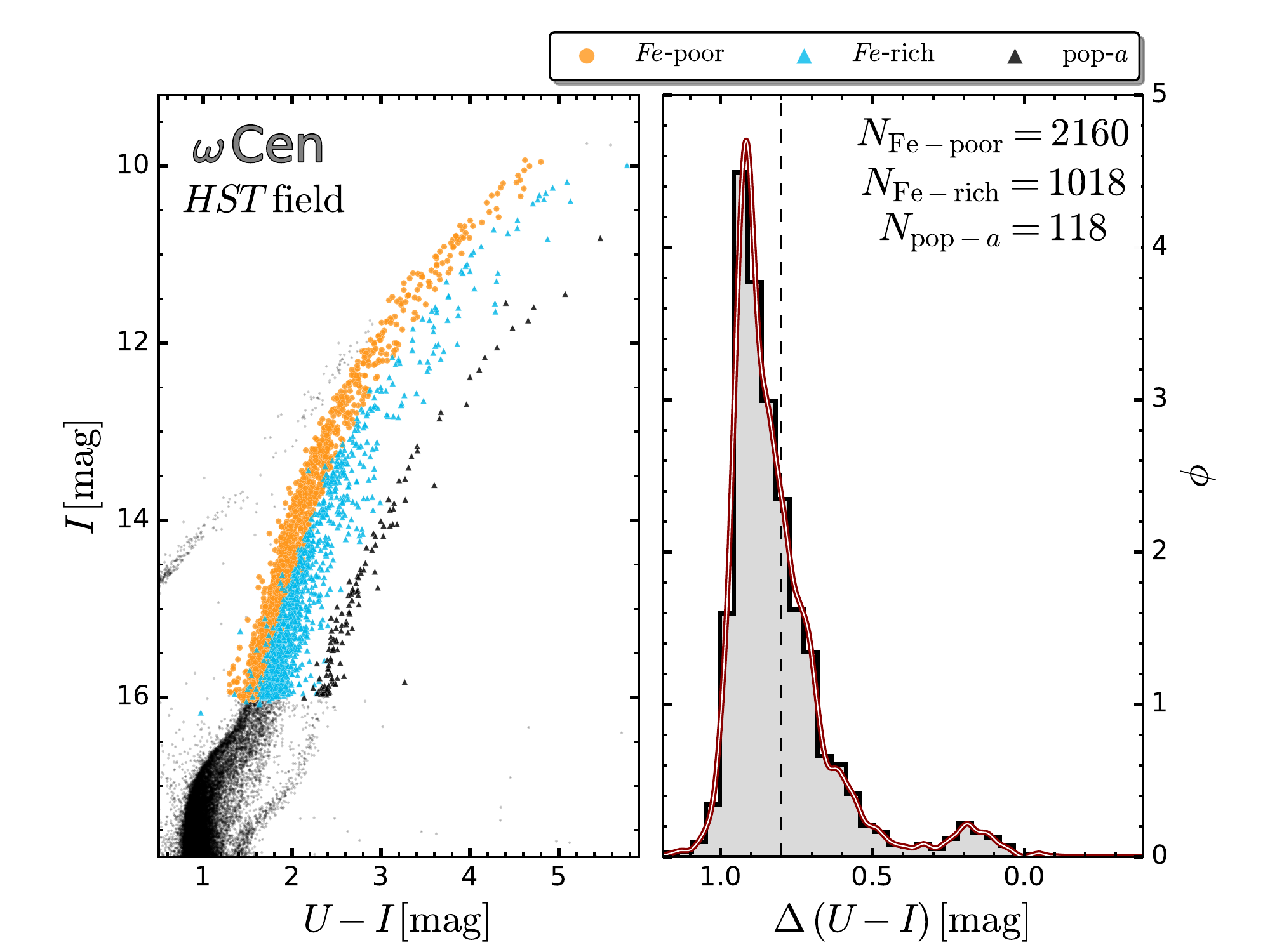}
  \includegraphics[width=8.9cm,trim={0.5cm 0.cm 0cm 0.25cm},clip]{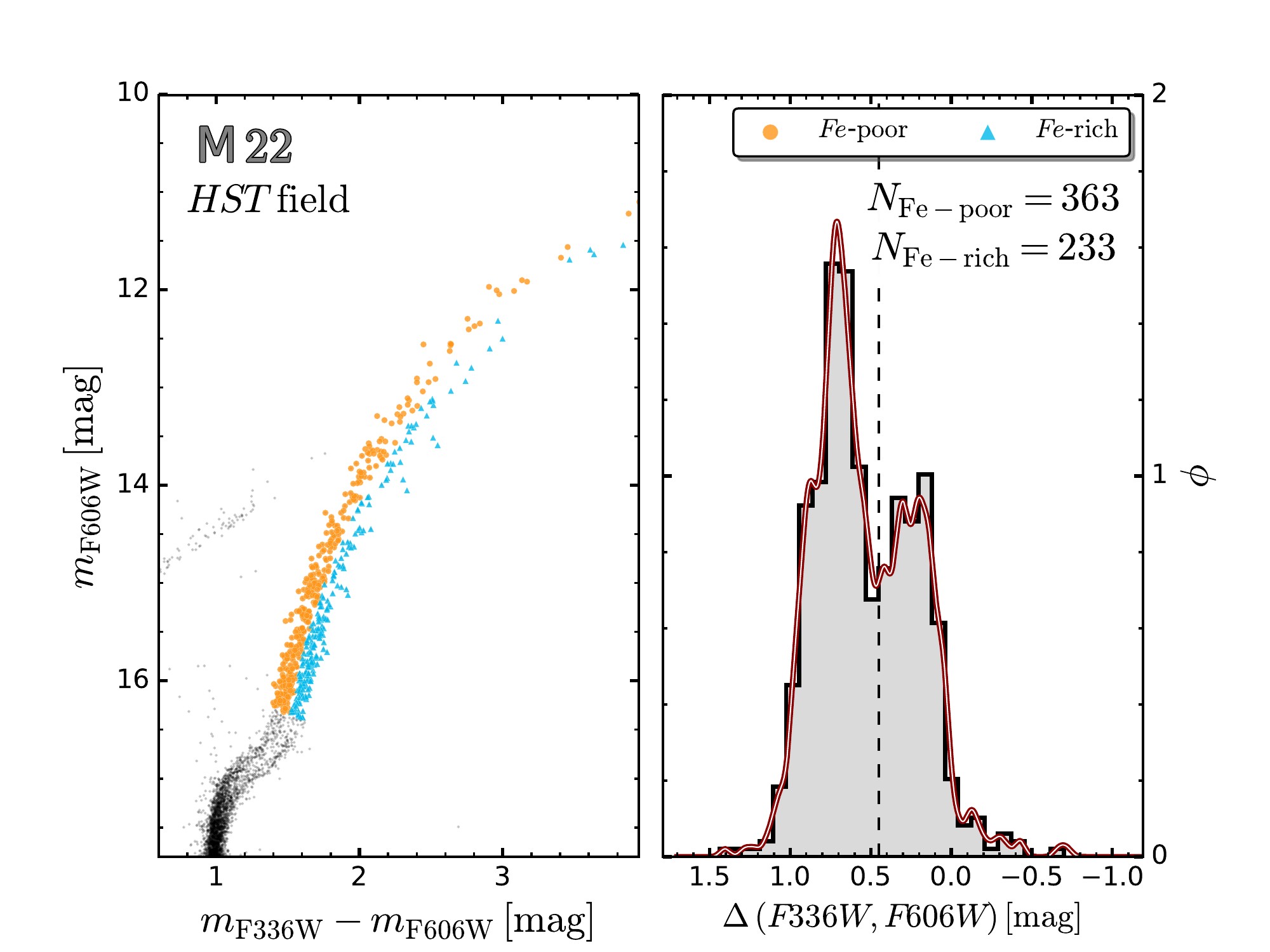}

  \caption{$I$ vs.\,$U-I$ and $V$ vs.\,$U-V$ diagrams for the selected cluster members of $\omega\,$Centauri (upper-left) and M\,22 (upper-right) from ground-based photometry (upper panels). We also show the histogram and the kernel-density distributions of $\Delta(U-I)$ and $\Delta(U-V)$ for the selected RGB stars of $\omega\,$Centauri and M\,22, respectively. The vertical black dashed lines are used to select Fe-poor and Fe-rich stars in the procedure I, marked with orange circles and cyan triangles, respectively. The orange Gaussian and the gray dashed lines in the top-left panels  are adopted to identify alternative groups of stars with different metallicities of $\omega$\,Centauri  from procedures II and III. See text for details.
  Lower panels show the corresponding diagrams from {\it HST} photometry of stars in the internal fields. To ensure consistency between the selection of Fe-poor/rich stars in the two fields of $\omega\,$Centauri, we converted \textit{HST} magnitudes $m_{\rm F336W}$ and $m_{\rm F814W}$ into $U$ and $I$ magnitudes of the photometric system by \citet{landolt1992}.}
  \label{fig:cmds}
\end{figure*}

The main procedure (I) to identify Fe-poor and Fe-rich stars  is similar to that used in \citet[][see their Figure 2]{cordoni2019}. Briefly, we determined the RGB boundaries as the 4$^{\rm th}$ and 96$^{\rm th}$ percentile of the color distributions, and we verticalized the CMD following the procedure described in \citet[][see their Section~3]{milone2017}. Finally, we derived the kernel-density distributions of stars in the verticalized I vs $\Delta$($U-I$) CMDs (red lines in the right panels of Figure~\ref{fig:cmds}) and identified by eye the groups of Fe-poor (orange dots) and Fe-rich (cyan triangles) RGB stars, which are located on the left and right side of the vertical dashed line, respectively. 
In the case of $\omega\,$Centauri we adopted an intermediate step, before identifying Fe-poor and Fe-rich stars in the \textit{HST} inner field. Specifically, to ensure consistency between the two fields and data sets, we converted the $m_{F336W}$ and $m_{F814W}$ magnitudes into $U$ and $I$ magnitudes.  The same process was redundant in the simpler case of M\,22, as revealed by the right panels of Figure~\ref{fig:cmds}. \\
To verify the impact of the adopted selection of $\omega$\,Centauri stars with different metallicities on the conclusions of the paper, we adopted two additional procedures (II and III). Procedure II consists of excluding stars with $U-I$ colors within $\pm$0.03 mag from vertical dashed line from the metal-poor and metal-rich sample defined above. Procedure III is based on the Monte-Carlo method for selection of metal-rich and metal-poor stars. We fitted the $\Delta$($U-I$) distributions of stars with $\Delta$($U-I$)$>0.85$ with a Gaussian function by means of least squares (orange transparent line in the top-left panel of Figure~\ref{fig:cmds}). Then, we randomly associated each star with a probability  to belong to the metal-poor and metal-rich sample based on ratio between the value of the best-fit Gaussian and the histogram distribution corresponding to its $\Delta$($U-I$) value. In the following, we present results based on the selection from procedure I, while in Section~\ref{sec:summary} we compare the results from the procedures I, II and III to demonstrate that the conclusions of the paper do not depend on the selection criterion.

\begin{table*}
\centering
\begin{tabular}{l|lllllllllll}
\toprule
\toprule
ID & RA (J2000)\footnote{from \citet[][updated as in 2010]{harris1996}} & DEC (J2000)$^{\rm a}$ & mass$^{\rm b}$ & $ d_{\rm sun}$\footnote{from \cite{baumgardt2018}} & $R_{\rm Gal}^{\rm a}$ & $R_{\rm c}^{\rm b}$ & $R_{\rm h }^{\rm b}$ & $R_{\rm t }^{\rm a}$ & $\log t_{\rm h}^{\rm b}$ & $\bar{V}_{\rm LoS}^{\rm b}$ \\ 
          &               &             & [$M_{\rm \odot}$]   & [kpc] & [kpc] & [kpc] & [arcmin] & [arcmin] & [yr] & [km/s] \\ 
\midrule
$\omega\,$Centauri       &   13 26 47.24 & $-$47 28 46.5  & $3.36\times 10^6$ & 5.2  & 6.40  &  2.37  &  5.00 & 48.4 & 10.39 &  234.28  \\ 
M\,22       &   18 36 23.94 & $-$23 54 17.1  & $4.16\times 10^5$ & 3.2  & 4.90  &  1.33  &  3.36  & 31.9 & 9.45 &  $-$147.76  \\ 
\bottomrule
\bottomrule
\end{tabular}
\caption{Identification, positional data and adopted structural parameters for the analyzed type-{\rm II} clusters. For each cluster we list position (RA, DEC), mass, distance from the Sun, galactocentric radius ($R_{\rm Gal}$), core radius ($R_{\rm c}$), half-light radius ($R_{\rm h}$), tidal radius ($R_{\rm t}$), logarithm of the half-mass relaxation time ($t_{\rm h}$), line-of-sight mean velocity ($\bar{V}_{\rm LoS}$).}
\label{tab:parameters}
\end{table*}

\subsection{Multiple populations with different light-element abundances in $\omega\,$Centauri}
Work based on high-resolution spectroscopy reveals that stellar groups in different metallicity bins of $\omega\,$Centauri host sub populations with different light-element abundances \citep[e.g.\,][]{marino2010, marino2011, johnson2010}.

\begin{figure*}[h!]
  \centering
  \includegraphics[width=0.8\textwidth, 				    trim={0cm 0cm 0cm 0cm},		clip]{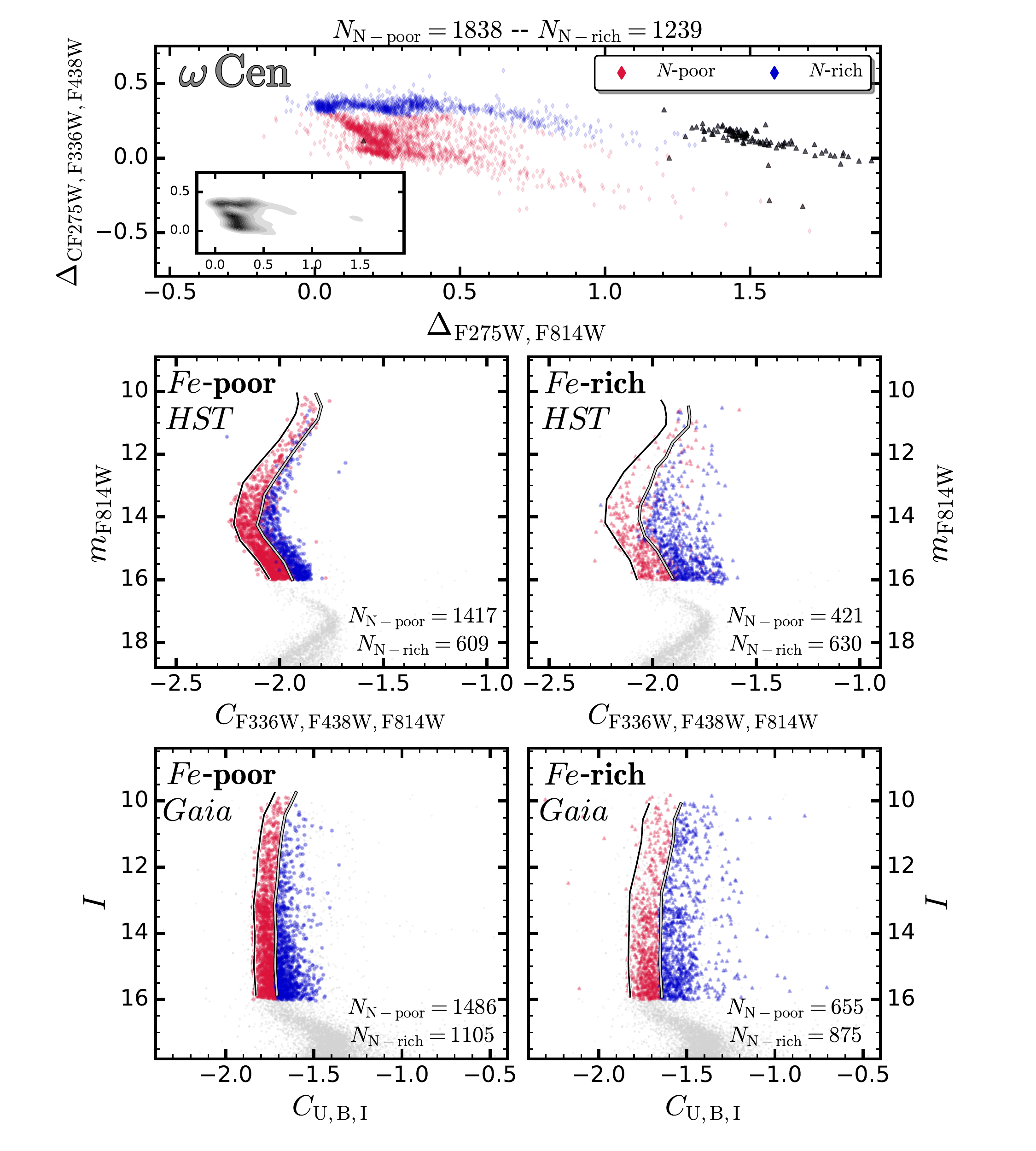}
  \caption{\textit{Top panel.} Pseudo two-color diagram, i.e. the Chromosome Map (ChM) of $\omega\,$Centauri. We adopted the selection criteria of \citet{marino2019} to identify stars with different nitrogen abundances. Specifically, N-poor stars (lower and middle stream) are represented with red diamonds, while N-rich stars (upper stream) are marked with blue diamonds. pop-$a$ stars are marked with solid black triangles. The bottom-left inset shows the 2D-KDE of the same ChM. \textit{Middle panels.} $m_{\rm F814W}$ vs.\,$C_{\rm F336W,F438W,F814W}$ for Fe-poor (left) and Fe-rich (right) stars as selected in Section~\ref{sec:cmds}. The fiducial lines in the CMDs are derived with the purpose of disentangling stellar populations with different nitrogen abundances among stars with different iron. \textit{Lower panels.} $I$ vs.\,$C_{\rm U,B,I}$ again for metal-poor (left) and metal-rich stars (right). The separation among N-poor and N-rich stars is derived shifting the bluest RGBs from the quantity derived in the \textit{HST} CMDs, as discussed in Section~\ref{sec:1P2P}.}
  \label{fig:stream}
\end{figure*}

An efficient tool to disentangle the distinct populations in GCs is provided by the pseudo two-color diagram called chromosome map, which revealed that $\omega\,$Centauri hosts at least 16 sup-populations \citep[ChM,][]{milone2015, milone2017}.
Based on multi-band {\it HST} photometry of $\omega\,$Centauri, \citet{marino2019} identified three main groups of stars that define distinct streams in the ChM. The stars of each stream span similar intervals of [Fe/H] but different content of He, C, N, O and Na. Specifically, the upper stream is composed of stars with extreme nitrogen abundances (hereafter N-rich sample), while middle- and the lower-stream stars have low and intermediate [N/Fe], respectively and will be called N-poor sample in the following.  N-rich and N-poor stars are represented with blue and red points, respectively, in the chromosome map of $\omega\,$Centauri plotted in the upper panel of Figure~\ref{fig:stream}.

Unfortunately, the {\it HST} photometry required to build the ChM is only available for stars with radial distances smaller than $\sim$2.5 arcmin. Hence, alternative photometric diagrams are needed to disentangle N-rich and N-poor stars in the external region of $\omega\,$Centauri.

As shown in the middle panels of Figure~\ref{fig:stream}, Fe-rich and Fe-poor stellar populations with different nitrogen abundances also populate different RGB regions in the $m_{\rm F814W}$ vs.\,$C_{\rm F336W,F438W,F814W}$=($m_{\rm F336W}-m_{\rm F438W}$)$-$($m_{\rm F438W}-m_{\rm F814W}$) pseudo-CMD. Specifically, N-poor stars exhibit lower $C_{\rm F336W,F438W,F814W}$ values than N-rich stars with the same luminosity. 

The lower panels of Figure~\ref{fig:stream}, show the $I$ vs.\,$C_{\rm U,B,I}=(U-B)-(B-I)$ pseudo CMD for stars with radial distances larger than $\sim 2.5\,\rm arcmin$, from ground-based photometry. Since the F336W, F438W and F814W filters are the {\it HST}-analogous of $U$, $B$ and $I$, the $C_{\rm F336W,F438W,F814W}$ and $C_{\rm U,B,I}$ have similar sensitivity to stellar populations with different chemical composition, we exploited \textit{HST} photometry to disentangle stellar populations with different nitrogen abundances, and then translated the separation into ground-based photometry. 

The black solid lines superimposed on the $m_{\rm F814W}$ vs.\,$C_{\rm F336W,F438W,F814W}$ and the $I$ vs.\,$C_{\rm U,B,I}$ diagrams are derived as in Section~\ref{sec:selection} and mark the bluest boundary of the RGBs. 
We determined the gray lines in the the $m_{\rm F814W}$ vs.\,$C_{\rm F336W,F438W,F814W}$ diagrams with the criteria to separate most of N-poor stars from N-rich stars, as selected in the ChM shown in the top panel.
 
To separate the bulk of N-rich and N-poor stars in the ground-based CMD we first estimated the 
$C_{\rm F336W,F438W,F814W}$ distance between the gray and black fiducials of each CMD for stars with different luminosities ($\Delta  C_{\rm F336W,F438W,F814W}$). We then determined the bluest RGB boundaries in the  $I$ vs.\,$C_{\rm U,B,I}$ for both Fe-poor and Fe-rich stars (lower panels). Finally, we shifted these fiducial lines by a certain amount, $\Delta C_{\rm U,B,I}$ that corresponds to the $\Delta  C_{\rm F336W,F438W,F814W}$ quantity derived from \textit{HST} photometry. \\
The selected N-poor and N-rich stars are shown with red and blue circles and triangles, respectively, for both metal-poor and metal-rich stars. 

$\omega\,$Centauri hosts a sample of metal-rich stars ([Fe/H]$\gtrsim -0.9$) that are called population-$a$ stars and define a distinct RGB sequence, on the red side of the bulk of RGB stars \citep[e.g.][]{lee1999, pancino2000}.
In the left panels of Figure~\ref{fig:cmds} we identified population-$a$ stars in the $I$ vs.\,$U-I$ (lower-left panel) and $I$ vs.\,$U-I$ CMDs (lower-right panel). The position of population-$a$ stars in the ChM  (see upper panel of Figure~\ref{fig:stream}) corroborate previous conclusion by \citet{marino2011, marino2019} that the majority of population-$a$ stars belong to the N-rich sample, and a small group of population-$a$ stars are N-poor. Specifically, $\sim 92\%$ of population-$a$ are enhanced in nitrogen, while only $\sim 8\%$ are N-poor.


We find that all the aforementioned stellar populations exhibit average proper motions consistent with each other within 1$\sigma$.

\section{Kinematics of stellar populations with different metallicities}\label{sec:fe subpop}
 
\subsection{Spatial distribution of Multiple Populations}\label{subsec:spatial}

In the following, we extend to the sample of Fe-poor and Fe-rich stars of $\omega\,$Centauri and M\,22 the procedure based on the two-dimensional Binned Kernel Density Estimate \citep{wand2015} used by \citet{cordoni2019} to analyse the spatial distributions of stellar populations in seven Type I GCs.

Due to crowding, stellar proper motions from GAIA DR2 are not available for most of the stars in the central regions of $\omega\,$Centauri.   
To increase the number of studied stars of $\omega\,$Centauri cluster members, we included in the analysis those stars that have membership probabilities larger than $90\,\%$ according to \citet{bellini2009}. 

\begin{figure}[h!]
  \centering
  \includegraphics[height=5.5cm, 				    trim={9cm 0cm 9cm 0cm},		clip]{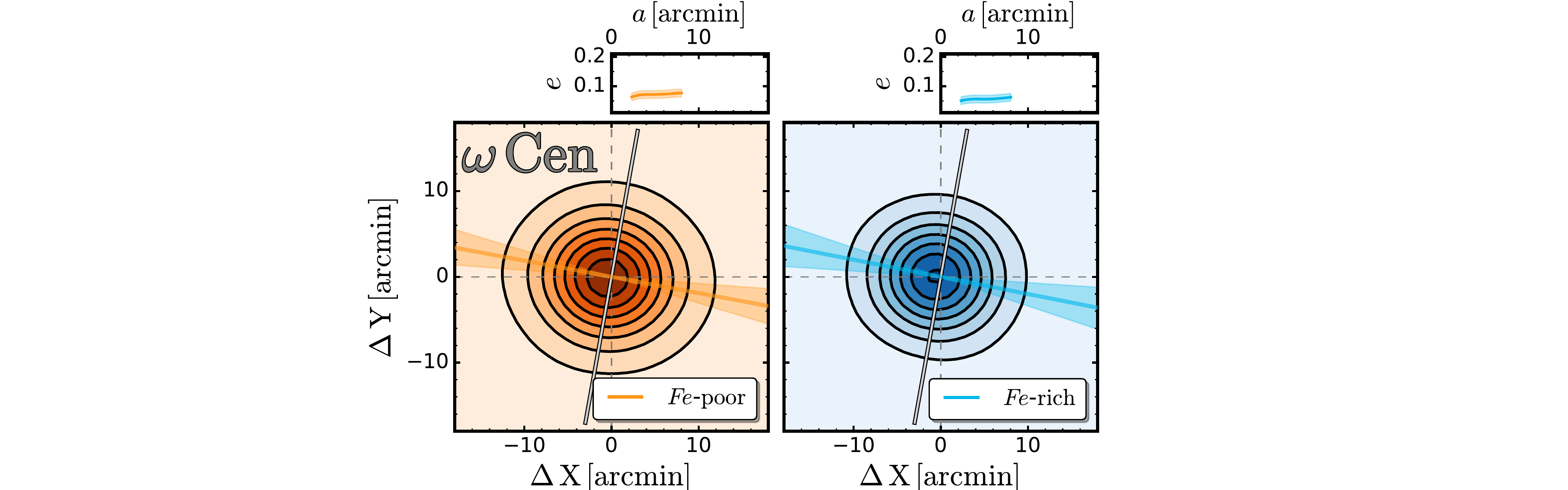}
  \includegraphics[height=5.5cm, 				    trim={9cm 0cm 9cm 0cm},		clip]{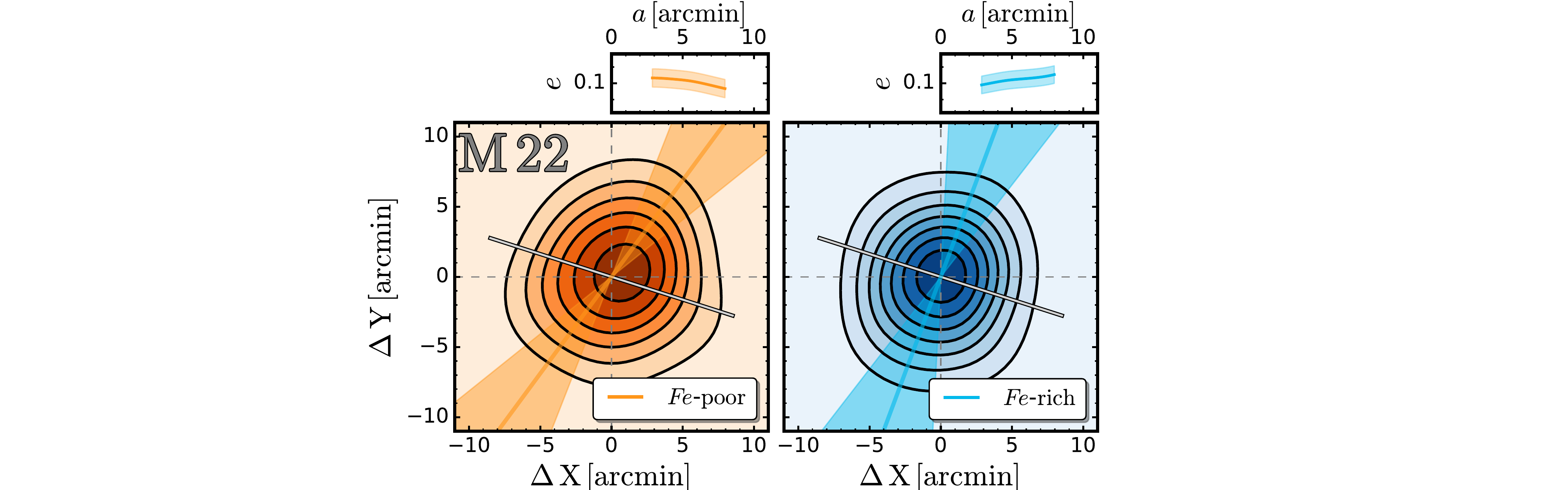}
  
  \caption{Density maps of metal-rich and metal-poor populations in $\omega\,$Centauri (top panels) and M\,22 (bottom panels). 
   The quantities on the abscissa and ordinate are the projected stellar coordinates relative to the cluster center. The color levels are indicative of stellar density and are based on the 2D Binned Kernel Density Estimate \citet{wand2015}. The iso-density contours are shown in each panel together with the rotation axis as determined in \citet{sollima2019} (solid black-grey line). 
   Red and blue lines in the top-insets show the ellipticity against the major axis, while the shaded regions represent the 1-$\sigma$ confidence bands.
  }
  \label{fig:spatial}
\end{figure}

Results for $\omega\,$Centauri are illustrated in the upper panels of Figure~\ref{fig:spatial} where we show the density plots of the Fe-poor (left) and Fe-rich (right) stars by using orange and blue colors, respectively. The contours are determined by smoothing the data-points with a Gaussian kernel of fixed size. The size is chosen with the criterion of minimizing the small scale structure without loosing the information on the global spatial distribution.
To properly compare the results, we adopted the same kernel size for all the populations of both clusters. 

\begin{figure*}[h!]
  \centering
  \includegraphics[width=8.9cm, 				    trim={4cm .0cm 6cm 0cm},		clip]{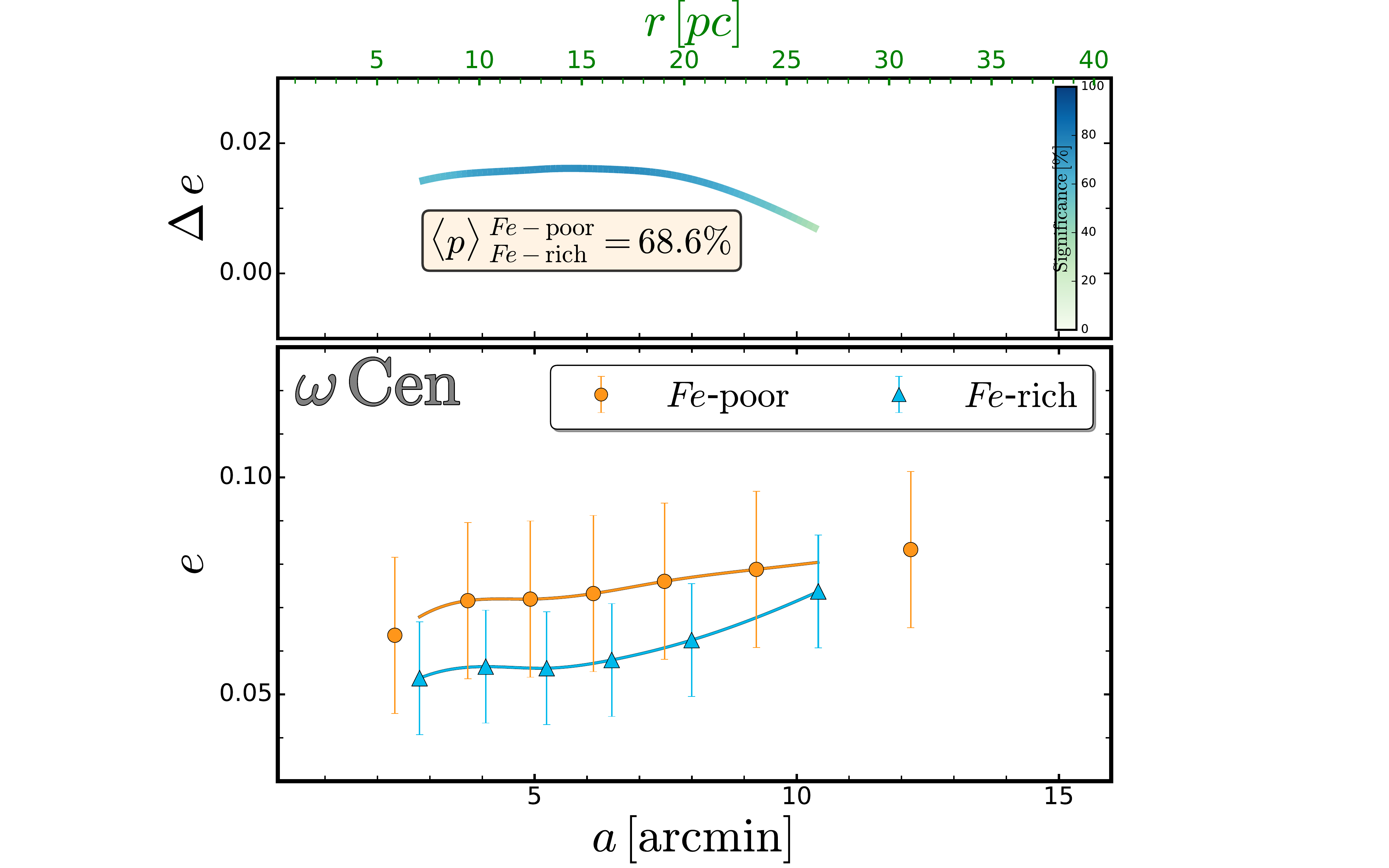}
  \includegraphics[width=8.9cm, 				    trim={4cm .0cm 6cm 0cm},		clip]{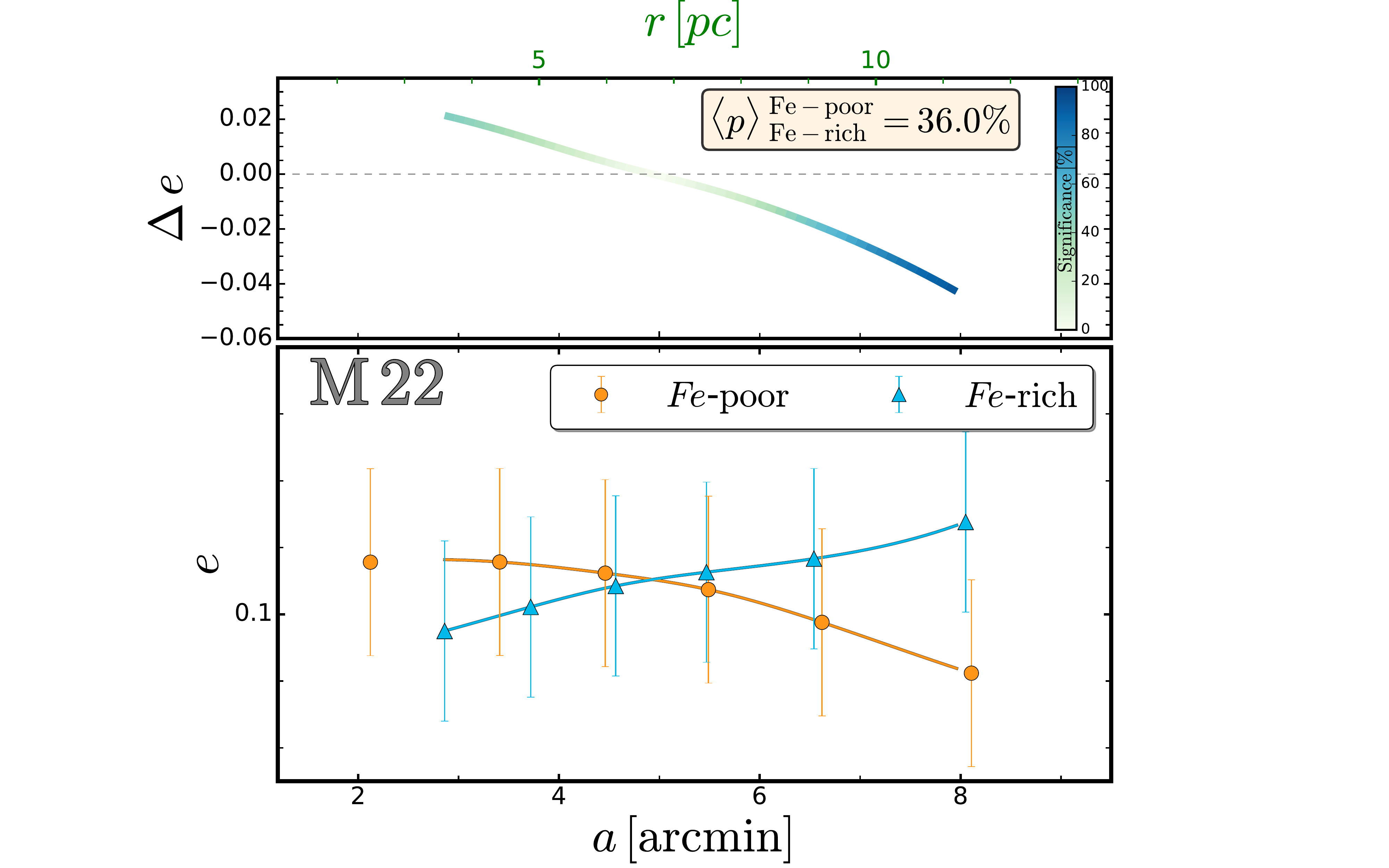}
  \caption{Ellipticity, $e$, of ellipses that best fit the spatial distribution of the different stellar populations of $\omega\,$Centauri (left) and M\,22 (right) against the semi-major axis, $a$ (lower panels). Upper panels show the difference between the ellipticity profiles as a function of $a$. The level of colors indicate the statistical significance of the difference, as indicated by the colorbar. See text for details.}
  \label{fig:ell}
\end{figure*}

We calculated six iso-density contours for each population and used the least square method to fit each contour with an ellipse as in \citet{radim1998}. The directions of the resulting minor and major axes are plotted in each panel. 
The resulting ellipticity, $e$, is plotted as a function of the semi-major axis, $a$ in Figure~\ref{fig:ell}. 

We confirm that $\omega\,$Centauri has an elliptical shape \citep[e.g.][]{harris1996}. 

The ellipticity of Fe-poor and Fe-rich stars slightly increases from e$\sim$0.05 to 0.07 and from    e$\sim$0.06 to 0.08, respectively, when moving from $a \sim 3\,\rm arcmin$ to $15\,\rm arcmin$. As a consequence, the ellipticity difference is $\Delta e \sim 0.015$ with statistical significance of $\sim$70\%.   

The major axis of the best-fit ellipses are coincident within one-sigma in both populations.   

In the case of M\,22, both Fe-poor and Fe-rich stars have similar ellipticities ($e \simeq 0.1$) over the entire analyzed field of view and their major axis have similar directions. \\
Finally, as expected for oblate rotators, we find that the rotation axis \citep[from][solid black-gray line in Figure~\ref{fig:spatial}]{sollima2019} is coincident with the semi-minor axis of the best fit-ellipses, i.e. perpendicular to the semi-major axis plotted in Figure~\ref{fig:spatial}. 

\subsection{Rotation in the plane of the sky}\label{subsec:rot}

\subsubsection{Rotation profile}\label{subsec:vprofile}

In this subsection, we analyze the average internal motions of Fe-poor and Fe-rich stars as a function of the radial distance from the cluster center. As a first step, we converted the $\mu_{\rm \alpha}\cos \delta$  and $\mu_{\rm \delta}$ components of proper motions into a radial ($\mu_{\rm RAD}$) and a tangential ($\mu_{\rm TAN}$) motions on the plane of the sky, correcting for the effect of perspective expansion/contraction as in \citet{vandenven2006}. 
We divided the cluster fields of $\omega\,$Centauri and M\,22 into 17 and 6 circular annulii, respectively,  determined with the naive estimator method \citep{silverman1986}. 
To account for the different star densities at different distances from the cluster center we used wider bins in the outskirt of the cluster (in the case of $\omega\,$Centauri: 5 arcmin for the innermost bin, 3.3 arcmin for $1.2R_{\rm h}< r < 3.2R_{\rm h}$,  and 4.2 arcmin for $3.2R_{\rm h}< r < 4.2R_{\rm h}$ and 5.9 arcmin for $r > 4.2R_{\rm h}$).

For each annulus, we used the routines provided by \citet{vasiliev2019} to compute the median radial ($\Delta \mu_{\rm RAD}$) and tangential ($\Delta \mu_{\rm TAN}$) motions, thus accounting for systematic errors in Gaia DR2 proper motions. Furthermore, \textit{Gaia} DR2 uncertainties on proper motion are underestimated by a factor of $\sim 10\%$  and $\sim 30\%$ for stars with magnitude $G>16$ and $G<13$, respectively. 

We therefore artificially increased the observed uncertainties on the proper motions as in \citet{bianchini2019}. \\

We then estimated the uncertainty relative to each point by bootstrapping with replacements performed 1,000 times. 

\begin{figure*}
  \centering
  \includegraphics[width=8.5cm, 				trim={0cm 0.95cm 0.cm 0cm},		clip]{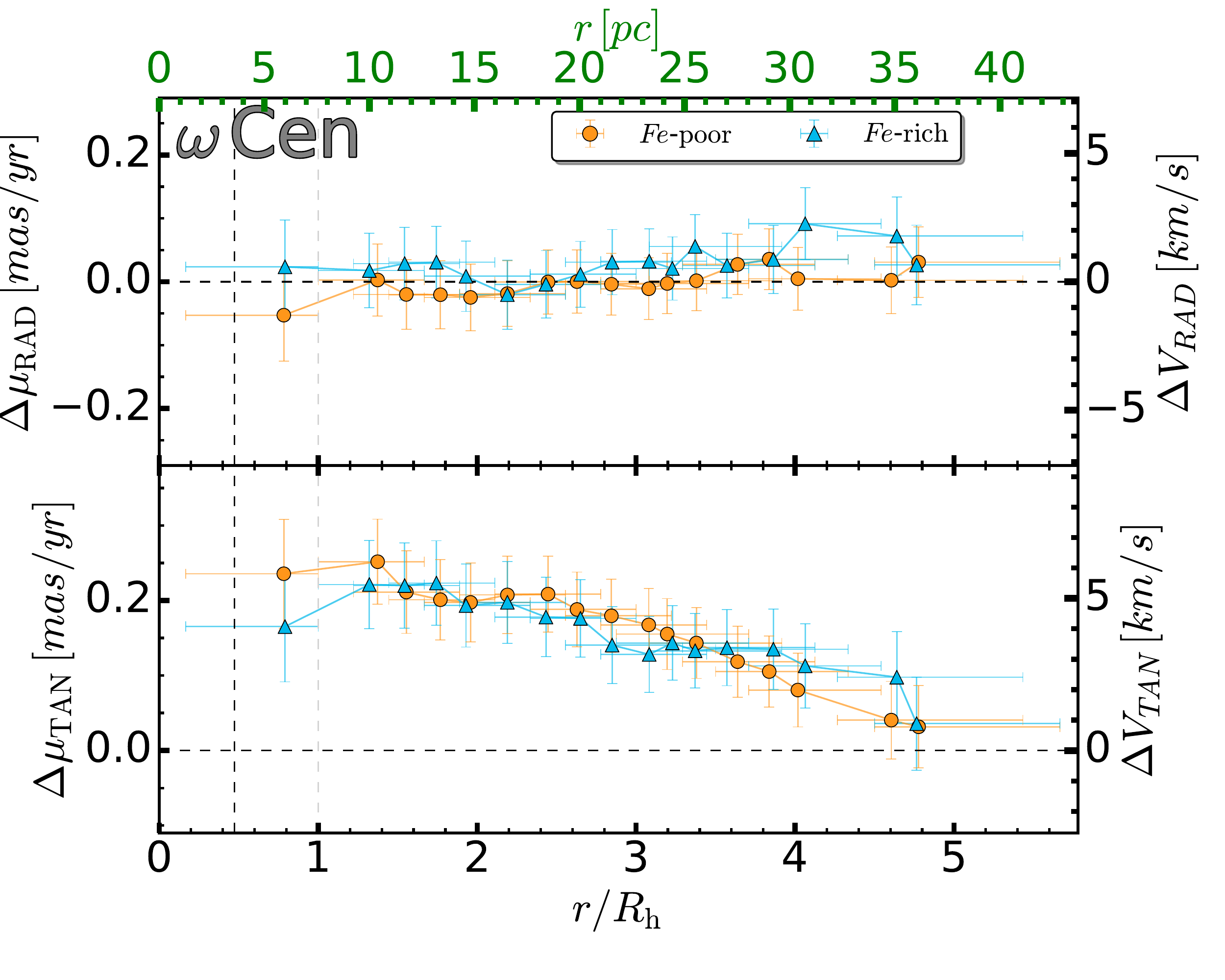}
  \includegraphics[width=8.5cm, 				trim={0cm 0.95cm 0.cm 0cm},		clip]{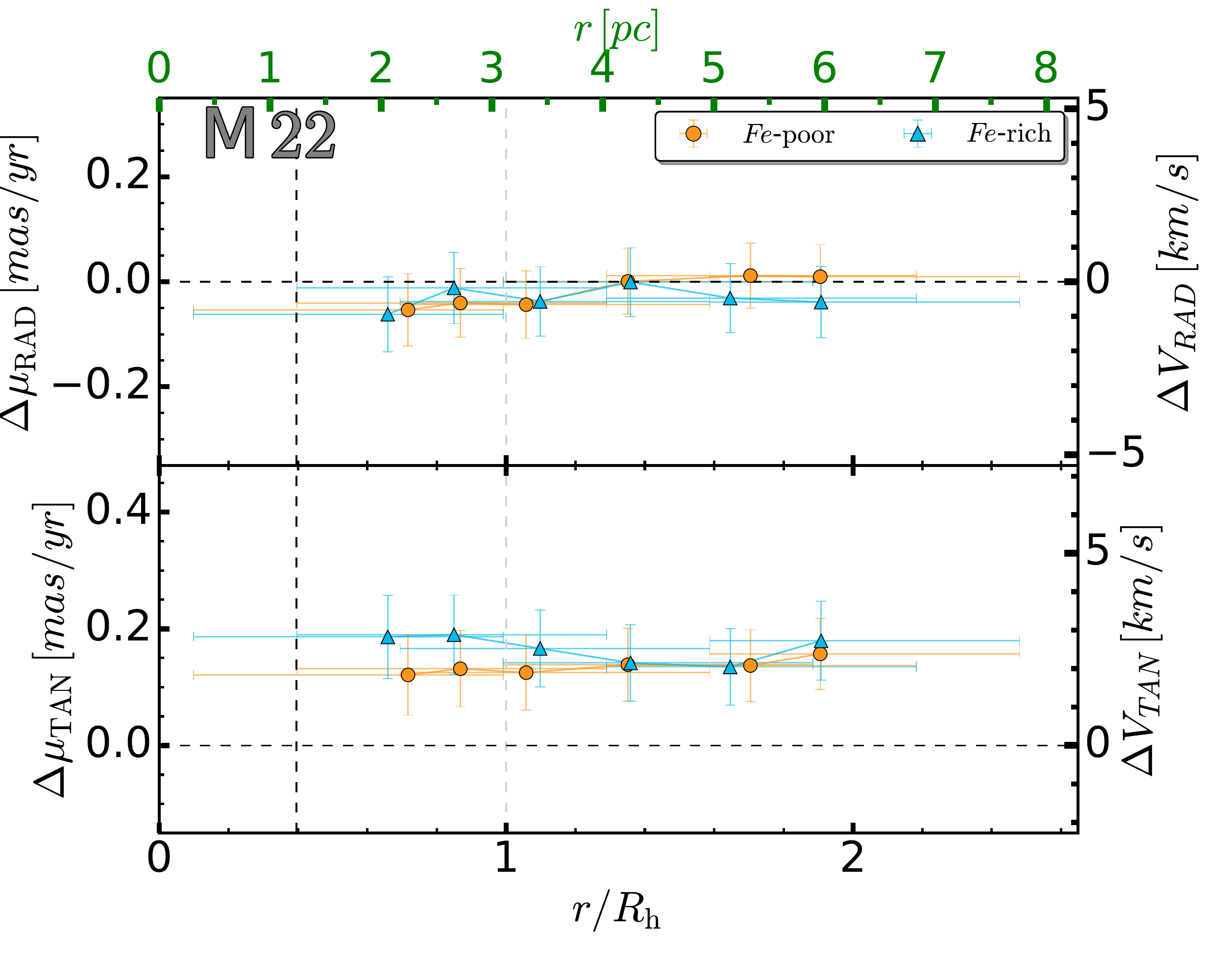}
  \caption{Average radial and tangential profile as a function of the radial distance from the cluster centers, for $\omega\,$Centauri (left) and M\,22 (right).  The radial quantity is normalized over the half-light radius from \cite{baumgardt2018}. Orange circles and cyan triangles represents Fe-poor and Fe-rich stars, respectively.}
  \label{fig:profiles}
\end{figure*}

The radial and tangential proper motions have been converted from mas/yr to km/s, ($\Delta V_{\rm RAD}$ and $\Delta V_{\rm TAN}$) adopting the distances listed in Table~\ref{tab:parameters}, derived in \citet{baumgardt2018}.
The radial and tangential velocity vs.\,the distance from the cluster center are shown in Figure~\ref{fig:profiles}. The radial coordinate has been normalized to the half-light radius, from \cite{baumgardt2018} (see  Table~\ref{tab:parameters}).  
A visual inspection of Figure~\ref{fig:profiles} reveals that the radial profiles of $\omega\,$Centauri and M\,22 are consistent with a zero-velocity up to at least $\sim$4 and $\sim$2 $R_{\rm h}$ respectively. 
A possible exception is provided by the Fe-rich population of $\omega\,$Centauri, which seems to exhibit larger radial velocities for distances greater than 4 $R_{\rm h}$.

Concerning the tangential profiles, we find positive values of $\Delta V_{\rm TAN}$ over the entire analyzed radial interval, thus favouring rotation among all the studied populations of $\omega\,$Centauri and M\,22. 
The tangential profiles of Fe-poor and Fe-rich stars in $\omega\,$Centauri exhibit their maximum amplitudes of about 6 km s$^{-1}$ at radial distances of about one half-light radii and steadily decrease towards larger distances from the cluster center.
In the case of M\,22 both populations are consistent with a flat tangential profile in the analyzed radial interval.  Our data do not allow to determine whether the rotation pattern of M\,22 strongly differs from that of $\omega$\,Centauri, or if the apparent flat rotation of M\,22 is due to the lack of observations in the external regions.

To compare the average velocity profiles of stellar populations we used the same procedure described in \citet[][see their Section 5]{cordoni2019}. 
We first used the Anderson-Darling (A-D) test to estimate the probability, $p$, that the tangential and radial motions of Fe-poor and Fe-rich stars are drawn from the same parent population. 
Furthermore, we compared the observed velocity profiles of Fe-poor and Fe-rich stars with $N_{\rm sim}=1000$ velocity profiles of simulated Fe-poor and Fe-rich stars. We assumed that the simulated stellar populations share the same velocity profiles and exhibit the same errors as the observed stars. For each bin we calculated the observed difference and counted the number of simulations that resulted in a difference greater than the observed one.

The fraction of simulations, $N^{*}/N_{\rm sim}$, where $\Delta_{\rm chi}\geq \Delta_{\rm obs}$ is indicative of the significance of the difference of the observed profiles. To quantify the global significance we computed the median of the significance of each bin, as well as the maximum and minimum values along the observed profiles. The results are listed in Table~\ref{tab:Significance} for each pair of velocity profiles. 

The results from the Anderson-Darling test and the values of $N_{\rm sim}$ are listed in Table~\ref{tab:Significance} and reveal that we find no differences between observed velocity profiles of Fe-poor and Fe-rich stars, neither in $\omega\,$Centauri nor in M\,22.

\subsubsection{Global rotation}
To further quantify the rotation in the plane of the sky of both clusters we adopted the procedure described in \citet[][see their Section 3]{cordoni2019}.
 
Briefly, we divided the field of view of each cluster into 30 circular sectors with an arc length of 45$^{\circ}$. We computed the median $\mu_{\rm \alpha}\cos\delta$ and $\mu_{\rm \delta}$ motions for Fe-poor and Fe-rich stars in each sector, and then subtracted the cluster median motion. As previously done for the radial and tangential velocity profiles, we account for Gaia systematic errors as in \citet{vasiliev2019}.
 
\begin{figure*}
  \centering
  \includegraphics[width=8.5cm,trim={0cm 0.7cm 0cm 1cm},clip]{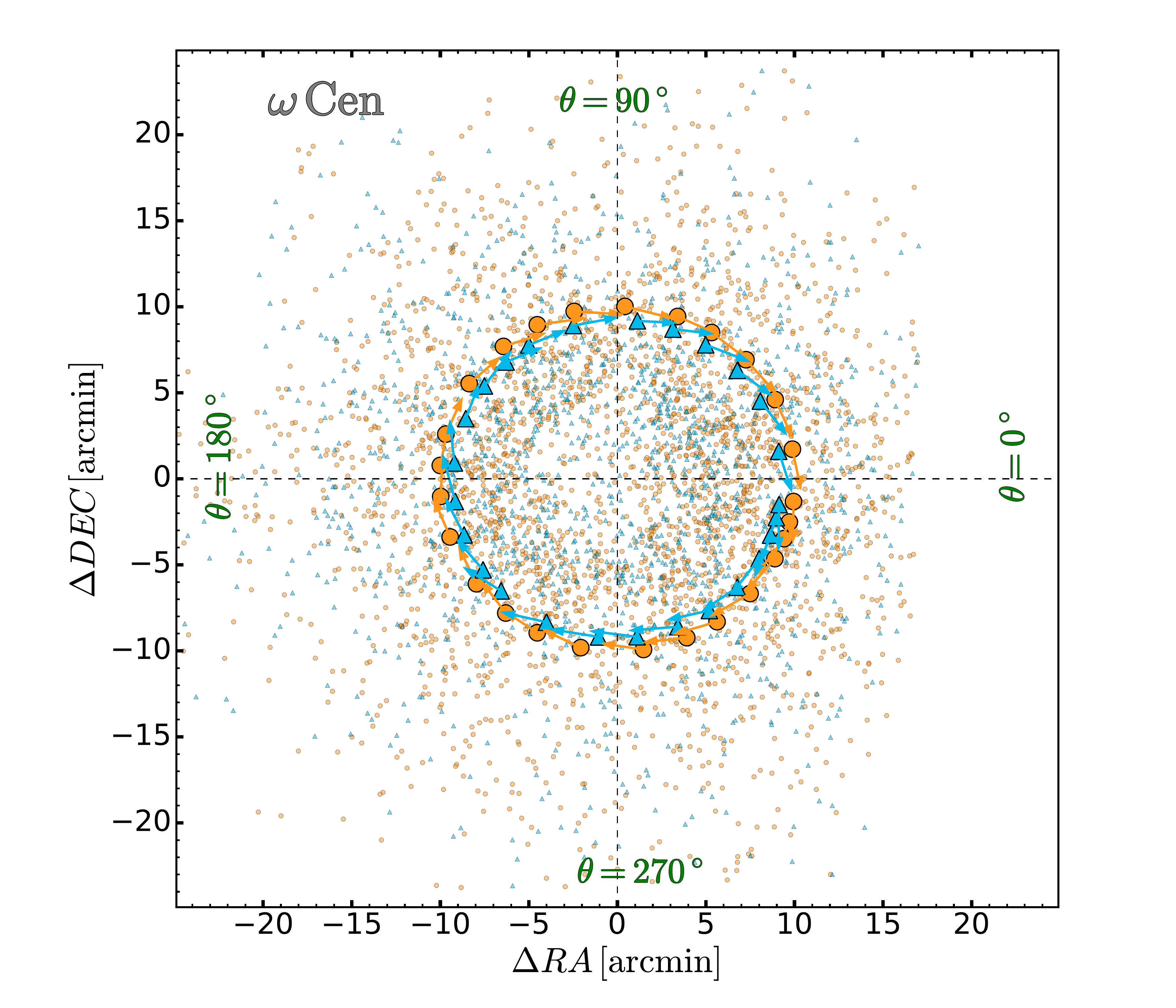}
  \includegraphics[width=8.5cm,trim={0cm 0.7cm 0cm 0cm},clip]{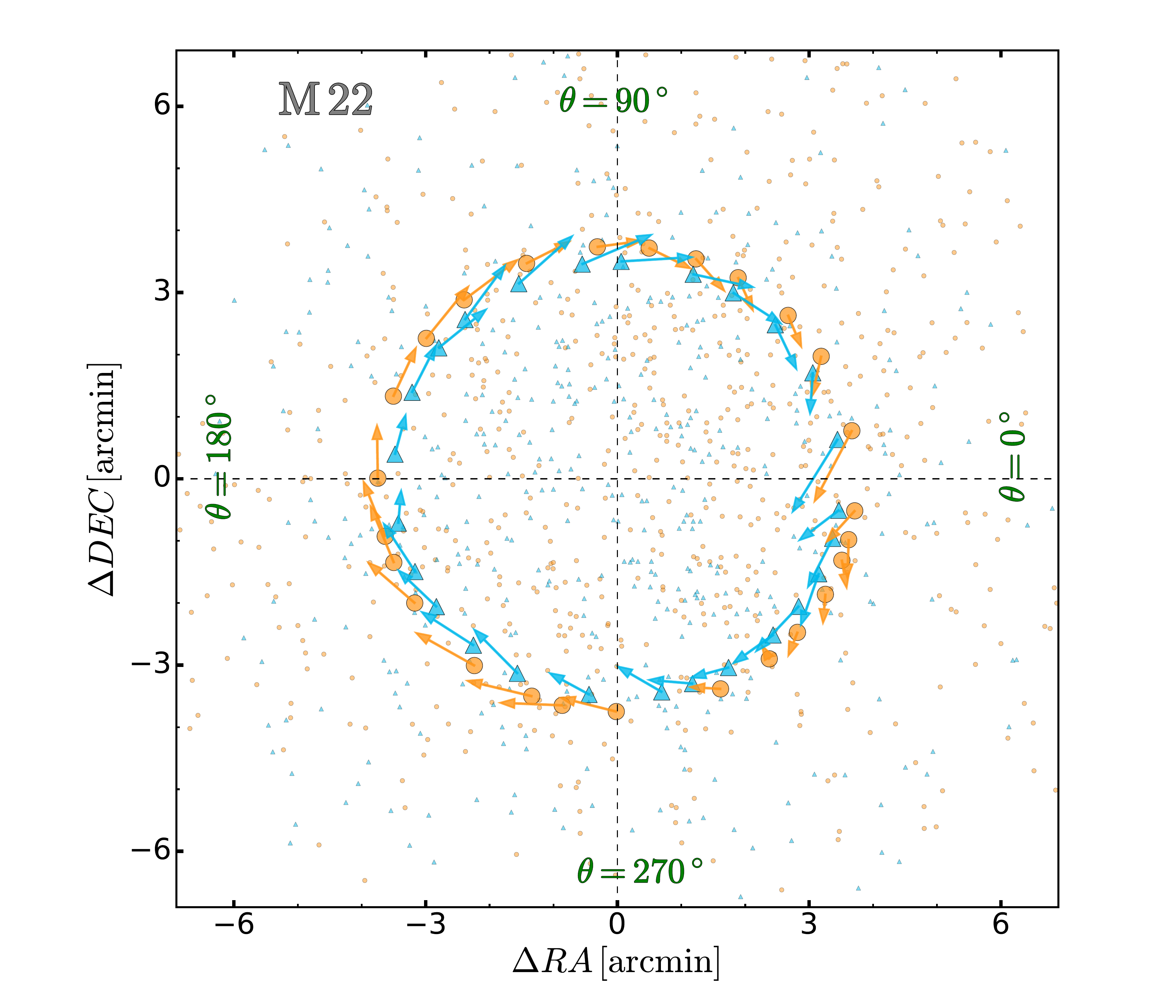}

  \includegraphics[width=8.5cm,trim={0cm 0cm 0cm 1.5cm},clip]{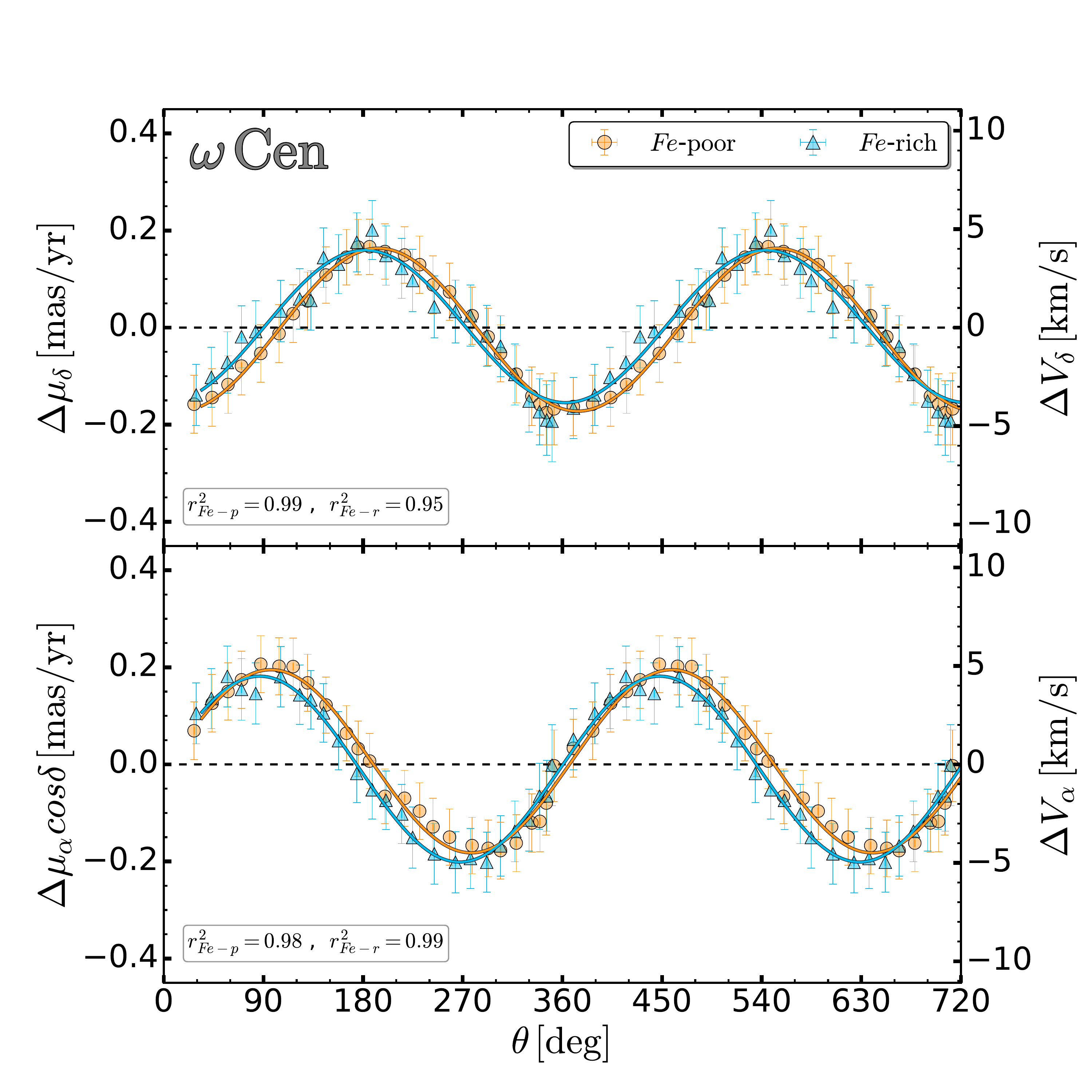} 
  \includegraphics[width=8.5cm,trim={0cm 0cm 0cm 1.5cm},clip]{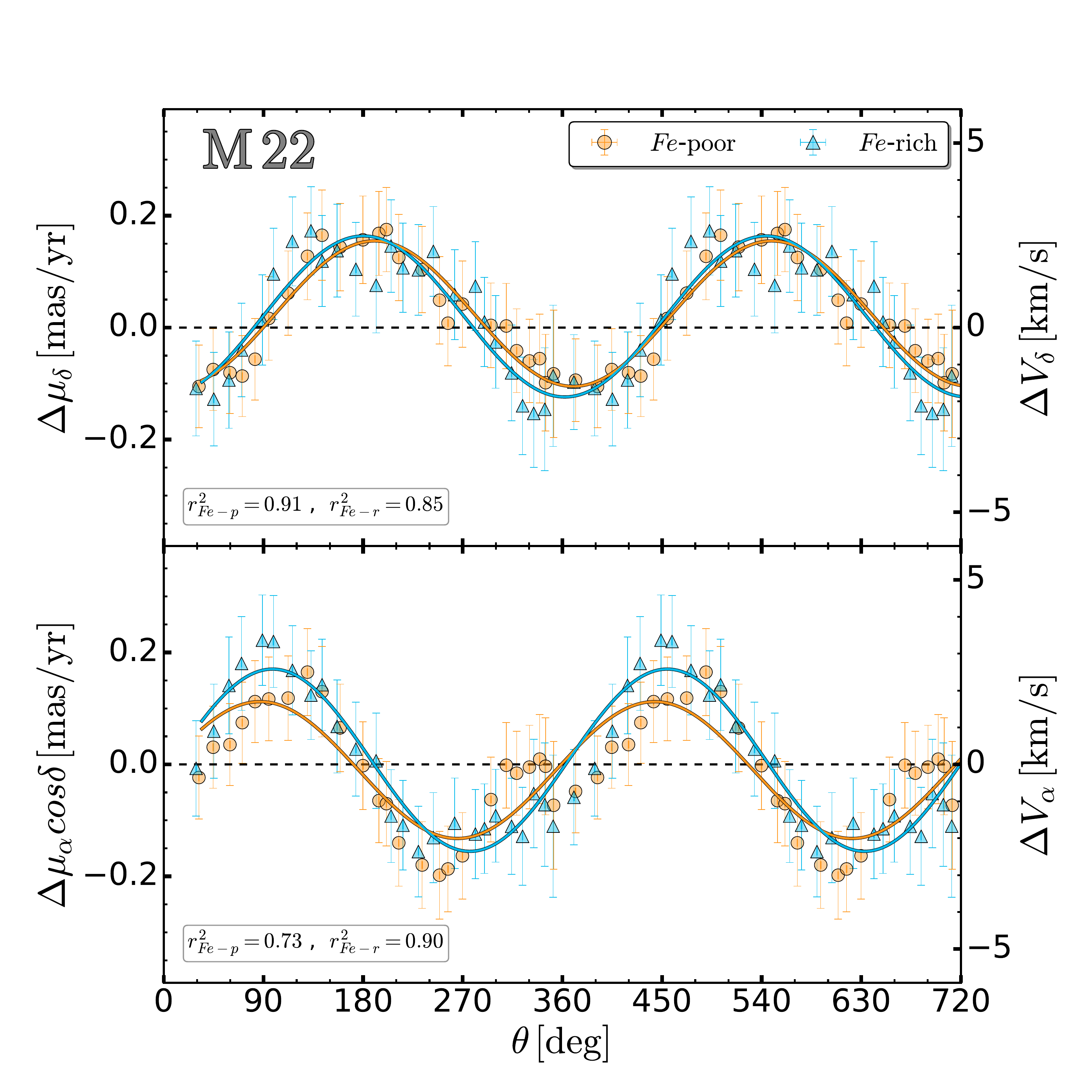} 

  \caption{\textit{Top panels.} Relative positions of Fe-poor and Fe-rich RGB stars in $\omega\,$Centauri and M\,22 with respect to the  cluster center. Orange and cyan arrows indicate the average motion computed in each circular sector, as described in Section~\ref{subsec:rot}, scaled by a factor of 5 for a better visual impact. The radial distances of the arrows correspond to the median radius of stars in each population. \textit{Bottom panels.} $\Delta\mu_{\rm \alpha}\cos\delta$ and $\Delta\mu_{\rm \delta}$ as a function of $\theta$ for the two analyzed clusters. The bottom-left insets show the value of the r$^2$ parameter, indicative of the quality of the fit.}
  \label{fig:rot}
\end{figure*}

The resulting quantities, $\Delta\mu_{\rm \alpha}\cos\delta$ and $\Delta\mu_{\rm \delta}$, are shown in the bottom panels of Figure \ref{fig:rot} as a function of the position angle $\theta$, defined counter-clockwise from the east. 
A visual inspection of this figure suggests that, consistently with the previous method, both populations of $\omega\,$Centauri and M\,22 rotate in the plane of the sky. 
This result is illustrated in the top panels of Figure~\ref{fig:rot}, where we show the positions of the stars relative to the cluster center, together with the vectors that indicate the median motions of Fe-poor and Fe-rich stars computed in each circular sector.

As in \cite{cordoni2019}, we used the least-squares technique to fit the sine function 
\begin{equation}\label{eq:1}
f(\theta)=M+A\cdot\sin(F\cdot\theta+\phi)     
\end{equation}
to all Fe-poor and Fe-rich stars in $\omega\,$Centauri and M\,22. The best fit functions to Fe-poor and Fe-rich stars are represented with orange and azure lines, respectively, in the bottom panels of Figure~\ref{fig:rot}. 
We estimate the goodness of the fit by means of the $r^2$ parameter:
\begin{equation}\label{eq:2}
r^2=1-\frac{\sum_i  (y_i-f(\theta,i))^2}{\sum_i(f (\theta, i)-\bar{y})^2}
\end{equation}
where $y_{i}$ is the value of $\mu_\alpha\cos\delta (\mu_\delta$) for each star, $i$, $\theta$ is the corresponding position angle, $\bar{y}$ is the average value of $y$, and $f$ is the best-fit function.
The $r^{2}$ values for Fe-poor and Fe-rich are listed in bottom left insets of Figure~\ref{fig:rot}. The fact that the motions of $\omega\,$Centauri and M\,22 provide a good match between the data and the sine interpolation ($r^{2}>0.7$), confirms the visual impression that both populations rotate in the plane of the sky.
 
Figure~\ref{fig:rot} shows that the sine functions that provide the best fit of Fe-poor and Fe-rich stars exhibit slight different rotation patterns.
To investigate whether these differences are significant or not, we followed the procedure introduced by \citet[see their Section~4.1]{cordoni2019}.
Specifically, we ran 1,000 Monte Carlo simulations of Fe-poor and Fe-rich stars where we assumed that the two populations exhibit the same proper motion distribution corresponding to the sine function that best fits the observed Fe-poor stars. We assumed that the two populations host the same number of stars as the observed Fe-rich and Fe-poor stars and added the corresponding observational errors to the simulated proper motions of each star.
We used Equation~2 to fit the proper motion distributions of Fe-poor and Fe-rich stars and  calculated the difference between their phases ($\Delta \phi^{\rm abs}$) and amplitudes ($\Delta A^{\rm abs}$). The number of simulations where $(|\Delta {\rm A}^{\rm sim}| \geq |\Delta {\rm A}^{\rm obs}|)$ and $(|\Delta \phi^{\rm sim}| \geq |\Delta \phi^{\rm obs}| )$ are indicative of the probability that the observed phase and amplitude differences between the corresponding rotation curves is due to observational errors alone.
Results are listed in Table~3 and show that there is no evidence for significant differences between the amplitudes and phases of the sine functions that best matches the proper motion distributions of Fe-poor and Fe-rich stars of $\omega\,$Centauri and M\,22. 
 
For completeness, we plot in Figure~\ref{fig:rot} the values of $\mu_\alpha\cos\delta$ and $\mu_\delta$ inferred for population-$a$ stars against $\theta$. The small sample of population-$a$ stars does not allow to properly fit the data with a sine function ($r^{2}=0.35$) and to understand whether this population shares the same rotation pattern as the bulk of $\omega\,$Centauri stars or not.  

\begin{table*}
\centering
\begin{tabular}{ll|ll|ll|cc}
\toprule
\toprule
 ID & & Region & motion component & $\Delta {\rm A}^{\rm obs}$ & $\Delta \phi^{\rm obs}$ 
 & P$(|\Delta {\rm A}^{\rm sim}| \geq |\Delta {\rm A}^{\rm obs}|)$ & P$(|\Delta \phi^{\rm sim}| \geq |\Delta \phi^{\rm obs}| )$ \\ 
 & & & & mas/yr & rad & & \\
\midrule
\textbf{$\omega\,$Centauri} & Fe-poor $-$ Fe-rich & all & $\mu_{\alpha}\cos{\delta}$  & $0.006\pm0.020$ & $0.179\pm0.090$ & 0.756 & 0.403 \\ 
                            &                   &        & $\mu_\delta$                & $0.019\pm0.022$ & $0.192\pm0.103$ & 0.333 & 0.145 \\
                            &                   & $R_1$  & $\mu_{\alpha}\cos{\delta}$  & $0.026\pm 0.038$ & $0.113\pm 0.134$ & 0.501 & 0.507 \\ 
                            &                   &        & $\mu_\delta$                & $0.055\pm 0.039$ & $0.047\pm 0.125$ & 0.193 & 0.845 \\ 

                            &                   & $R_2$  & $\mu_{\alpha}\cos{\delta}$  & $0.006\pm 0.026$ & $0.010\pm 0.121$ & 0.848  & 0.941 \\ 
                            &                   &        & $\mu_\delta$                & $0.032\pm 0.028$ & $0.261\pm 0.133$ & 0.312  & 0.176 \\ 

                            &                    & $R_3$ & $\mu_{\alpha}\cos{\delta}$  & $0.004\pm 0.026$ & $0.399\pm 0.121$ & 0.899  &     0.112 \\ 
                            &                    &       & $\mu_\delta$                & $0.004\pm 0.028$ & $0.203\pm 0.129$ & 0.916  & 0.454 \\
\hline
                            & N-poor $-$ N-rich & all & $\mu_{\alpha}\cos{\delta}$  & $0.044\pm0.019$ & $0.031\pm0.092$ & 0.878 & 0.795 \\ 
                            &                 &     & $\mu_\delta$                  & $0.004\pm0.018$ & $0.051\pm0.110$ & 0.898 & 0.741 \\
\hline
        \textbf{Fe-poor}    & N-poor $-$ N-rich & all & $\mu_{\alpha}\cos{\delta}$  & $0.008\pm0.023$ & $0.038\pm0.155$ & 0.986 & 0.733 \\ 
                            &                 &     & $\mu_\delta$                  & $0.015\pm0.025$ & $0.114\pm0.180$ & 0.541 & 0.147 \\
\hline
        \textbf{Fe-rich}    & N-poor $-$ N-rich & all & $\mu_{\alpha}\cos{\delta}$  & $0.023\pm0.025$ & $0.107\pm0.101$ & 0.538 & 0.655 \\ 
                            &                 &     & $\mu_\delta$                  & $0.041\pm0.024$ & $0.112\pm0.140$ & 0.196 & 0.670 \\    
                            &                   & $R_1$  & $\mu_{\alpha}\cos{\delta}$  & $0.102\pm 0.035$ & $0.111\pm 0.168$ & 0.004 & 0.585 \\ 
                            &                   &        & $\mu_\delta$                & $0.093\pm 0.042$ & $0.127\pm 0.118$ & 0.009 & 0.477 \\ 

                            &                   & $R_2$  & $\mu_{\alpha}\cos{\delta}$  & $0.030\pm 0.025$ & $0.164\pm 0.186$ & 0.451  & 0.569 \\ 
                            &                   &        & $\mu_\delta$                & $0.043\pm 0.024$ & $0.067\pm 0.195$ & 0.233  & 0.854 \\ 
\hline
\hline
\textbf{M\,22}              &  Fe-poor $-$ Fe-rich & all   & $\mu_{\alpha}\cos{\delta}$  & $0.041\pm 0.020$ & $0.210\pm 0.195$ & 0.345 & 0.379 \\ 
                            &                   &       & $\mu_\delta$                & $0.014\pm 0.018$ & $0.219\pm 0.224$ & 0.745 & 0.231 \\ 
                            & N-poor $-$ N-rich   & all & $\mu_{\alpha}\cos{\delta}$  & $0.067\pm0.020$ & $0.273\pm0.153$ & 0.077 & 0.396 \\ 
                            &                 &     & $\mu_\delta$                & $0.041\pm0.022$ & $0.024\pm0.143$ & 0.344 & 0.930 \\
           
\bottomrule
\bottomrule
\end{tabular}
\caption{Comparison of the rotation curves in the $\mu_{\rm \alpha} cos{\delta}$ vs.\,$\theta$,  $\mu_{\delta}$ vs.\,$\theta$ vs.\,$\theta$ planes of metal-poor and metal-rich stars in the entire field of view of M\,22 and $\omega\,$Centauri and in the analyzed three sub-regions, $R_{\rm 1}$, $R_{\rm 2}$, and $R_{\rm 3}$ of the $\omega\,$Centauri field of view.
 We provide the A-D values from the  Anderson-Darling test and the corresponding probability that metal-poor and metal-rich stars come from the same parent distribution (p-val). We list the amplitude ($\Delta{A^{\rm obs}}$) and phase differences ($\Delta{\phi}^{obs}$) of the curves that provide the best-fit with metal-poor and metal-rich stars and the probability that the observed difference in phase and amplitude are due to observational errors as inferred from Monte-Carlo simulations.}
\label{tab:Significance}
\end{table*}


\subsubsection{Rotation of stellar populations in $\omega\,$Centauri as a function of radial distance}\label{subsec:RotWcen}

\begin{figure*}
  \centering
  \includegraphics[width=18cm, trim={0cm 0cm 0cm 0cm}, clip]{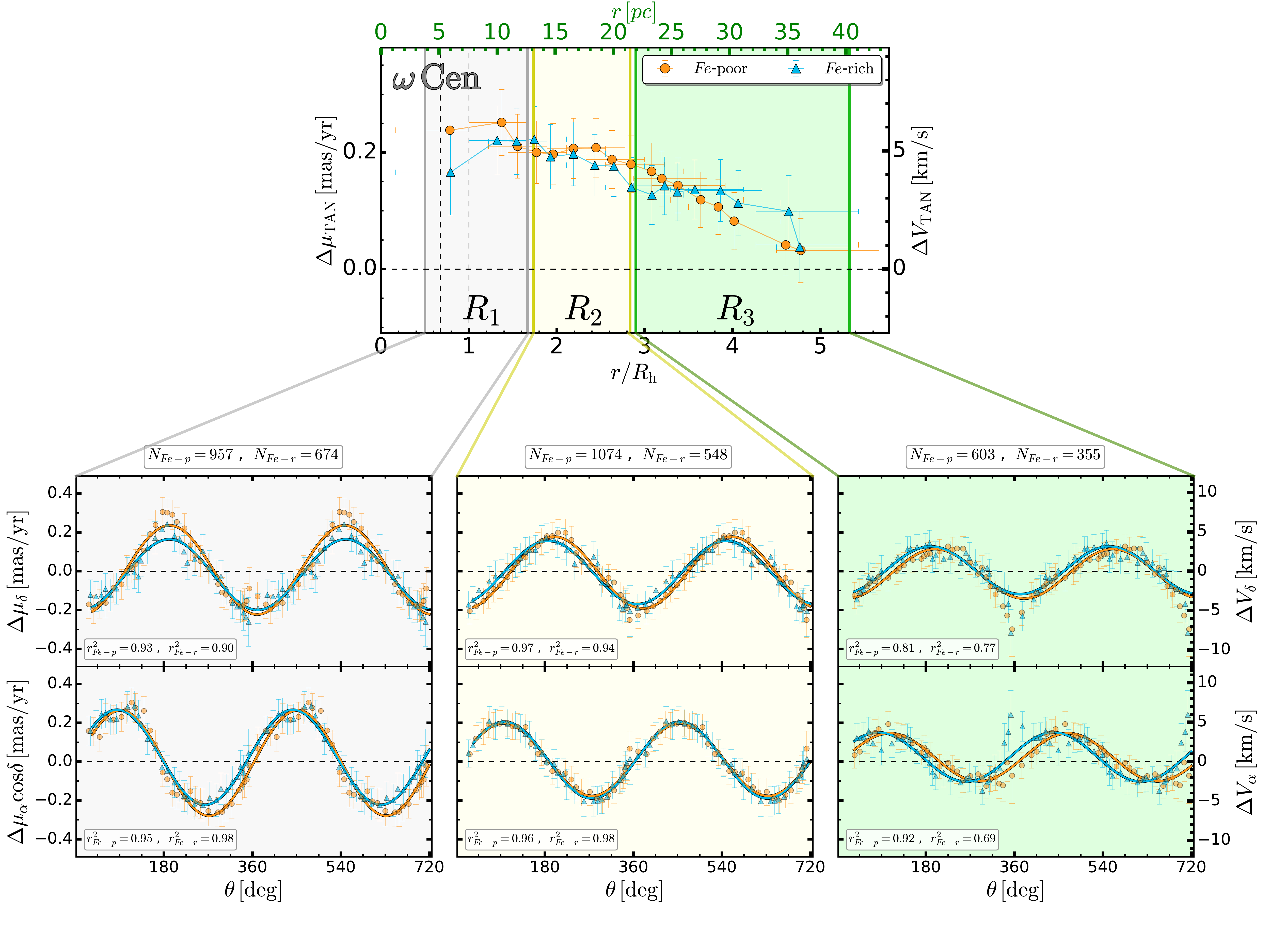}
  \caption{\textit{Top panel}. Reproduction of the $\Delta \mu_{\rm TAN}$ vs.\,r/$R_{\rm h}$ diagram of Fe-poor and Fe-rich stars in $\omega\,$Centauri plotted in Figure~\ref{fig:profiles}. \textit{Bottom panels.} 
  $\Delta\mu_{\rm \alpha}\cos\delta$ and $\Delta\mu_{\rm \delta}$ against the position angle $\theta$ for stars in regions $R_{\rm 1}$--$R_{\rm 3}$  defined in the upper panel. The sine functions that provide the best fit with the observations of Fe-poor and Fe-rich stars are represented with orange and azure lines, respectively.}
  \label{fig:regions}
\end{figure*}

The large number of stars available in $\omega\,$Centauri, together with the wide field of view where Gaia DR2 and $UBI$ ground based photometry from \citet{stetson2019} offer us the opportunity to investigate the rotation of Fe-poor and Fe-rich stars at different radial distances from the cluster center. 
We therefore exploit the median tangential velocity profile, re-proposed in the top panel of Figure~\ref{fig:regions} to select three circular annuli ($R_{\rm 1}$--$R_{\rm 3}$ in Figure~\ref{fig:regions}) 
with significantly different values of $\Delta \mu_{\rm TAN}$. 
The three regions are selected so that they contain comparable number of stars.
The individual numbers of Fe-poor and Fe-rich stars is indicated in the insets of the lower panels of Figure~\ref{fig:regions}. \\
We'd like to point out that the ratio between Fe-poor and Fe-rich stars increases from the innermost region to the middle region, and remains almost constant between the middle region and the outermost one. Thus confirms previous findings by \citet{bellini2009}. 

Concerning the rotation in the plane of the sky, we find that the amplitudes of the best fit sinusoidal functions for both Fe-poor and Fe-rich populations decrease from about six to three km s$^{-1}$ when moving from $r\sim1\,R_{\rm h}$ to $r\sim4\,R_{\rm h}$. 

The sine functions that provide the least-squares best fit with the observed proper motion distributions of Fe-poor and Fe-rich stars have slightly different values of amplitude $A$ and phase $\phi$. 
We followed the procedure introduced by \citet[][see their Section~4.1]{cordoni2019} and described above to quantify the statistical significance of the observed phase and amplitude differences. Results are listed in Table~\ref{tab:Significance} and show that Fe-poor and Fe-rich stars are consistent with sharing the same rotation pattern in the three analyzed regions.

\subsection{Velocity dispersion profiles} \label{subsec:disp}

\begin{figure*}[h!]
  \centering
  \includegraphics[width=8.5cm, 				trim={0cm 2.95cm 0.cm 0cm}, 	clip]{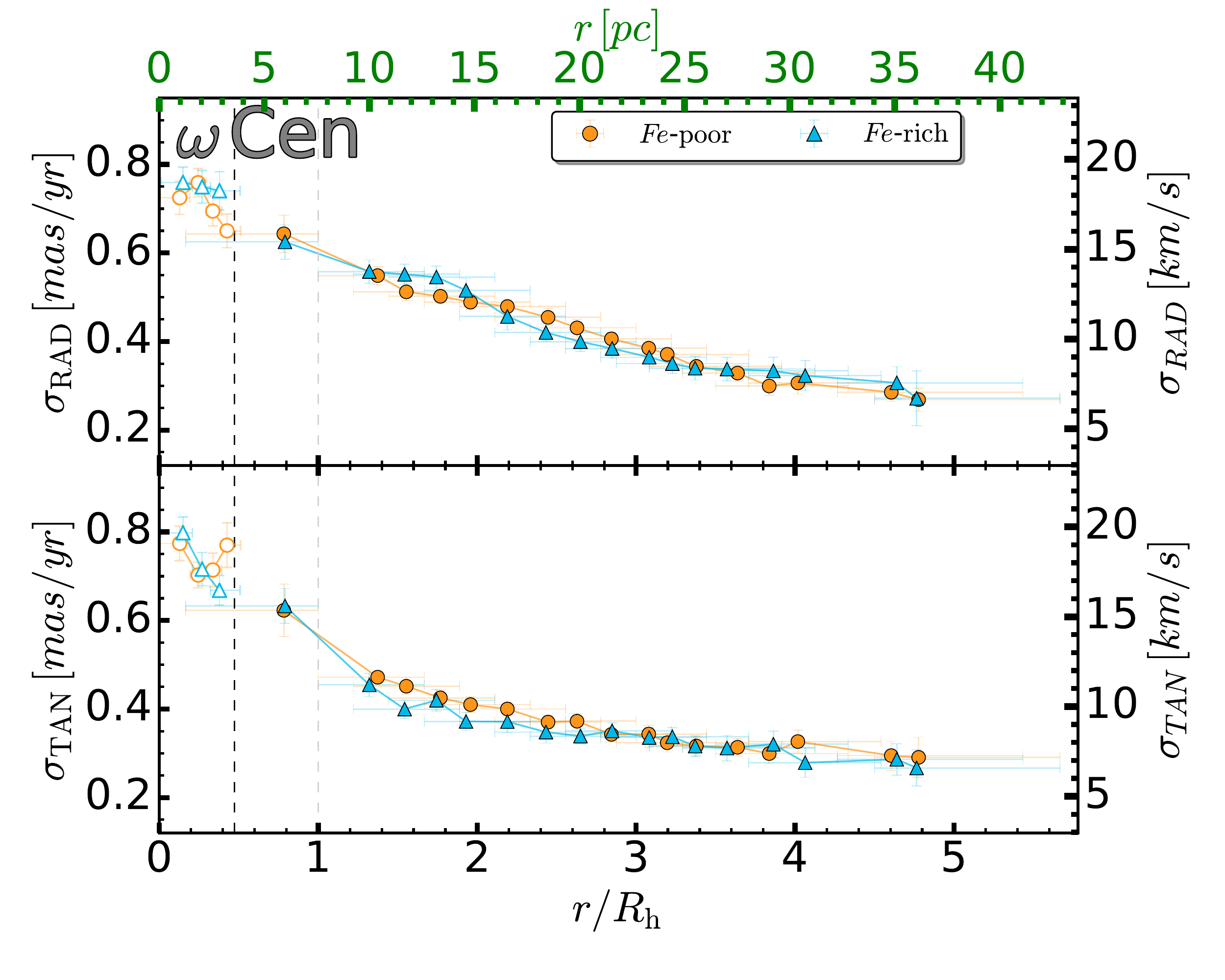}
  \includegraphics[width=8.5cm, 				trim={0cm 2.95cm 0.cm 0cm}, 	clip]{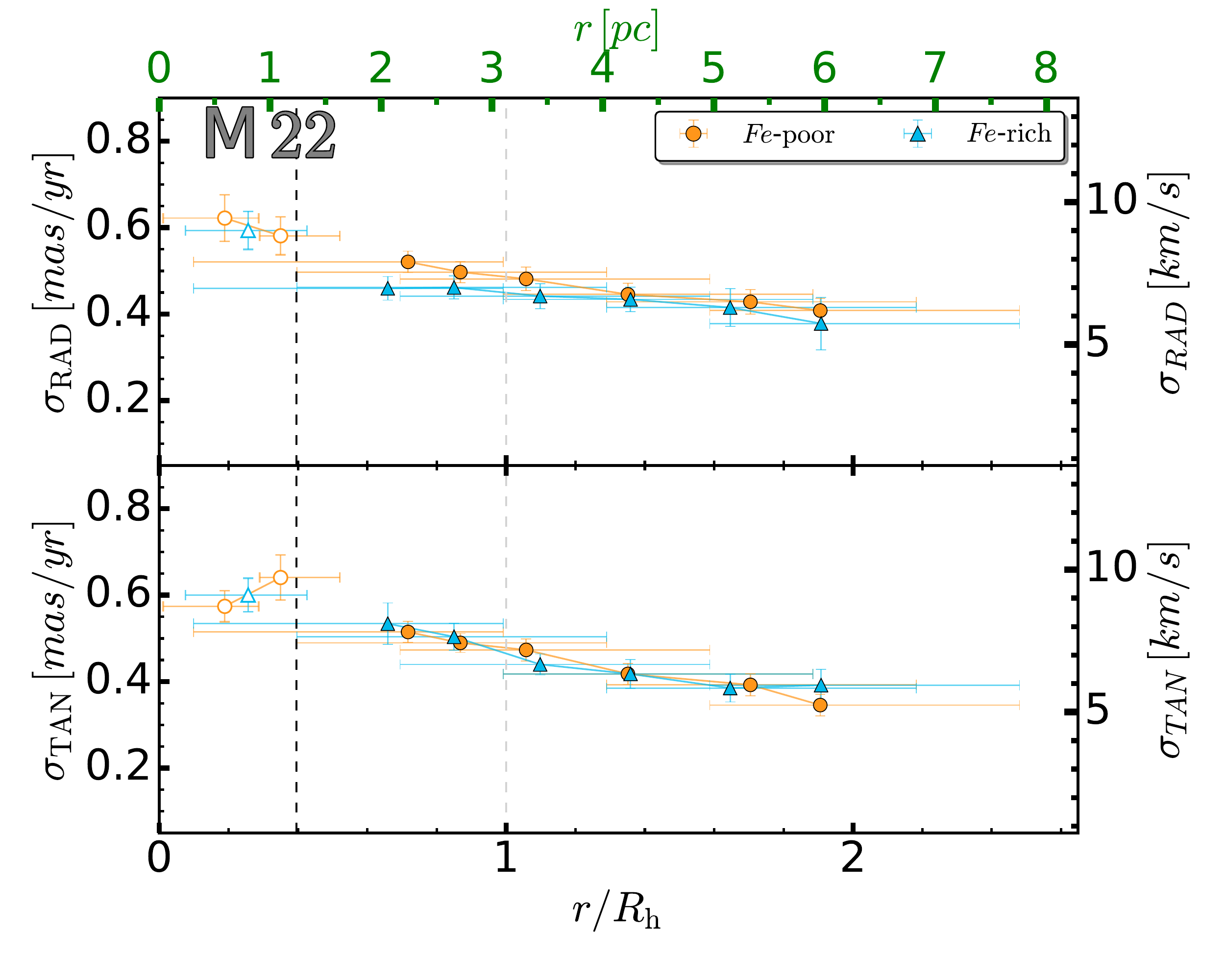}

  \includegraphics[width=8.5cm, 				trim={0.cm 0.cm 0.cm 1.75cm},    clip]{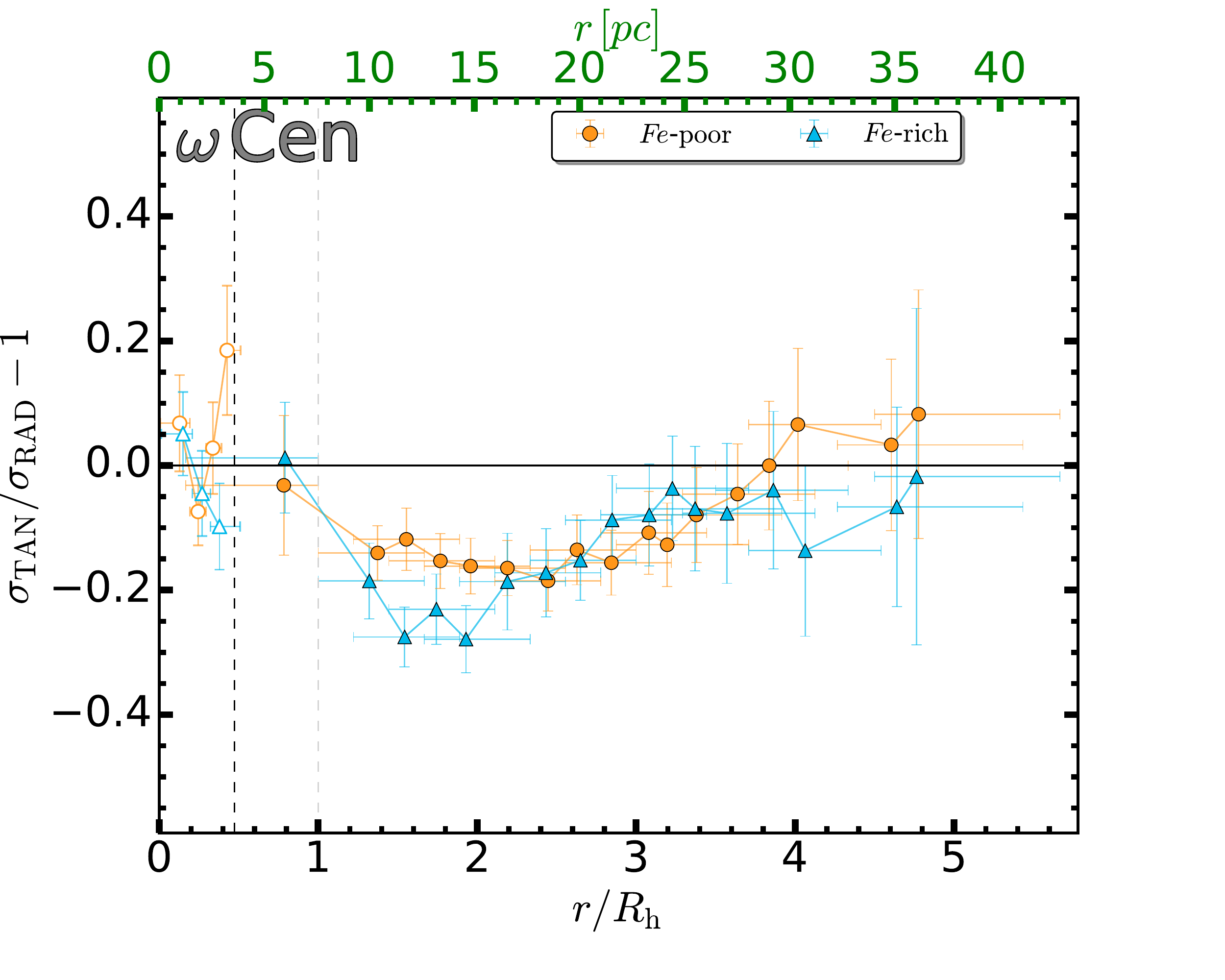}
  \includegraphics[width=8.5cm, 				trim={0.cm 0.cm 0.cm 1.75cm}, 	
  clip]{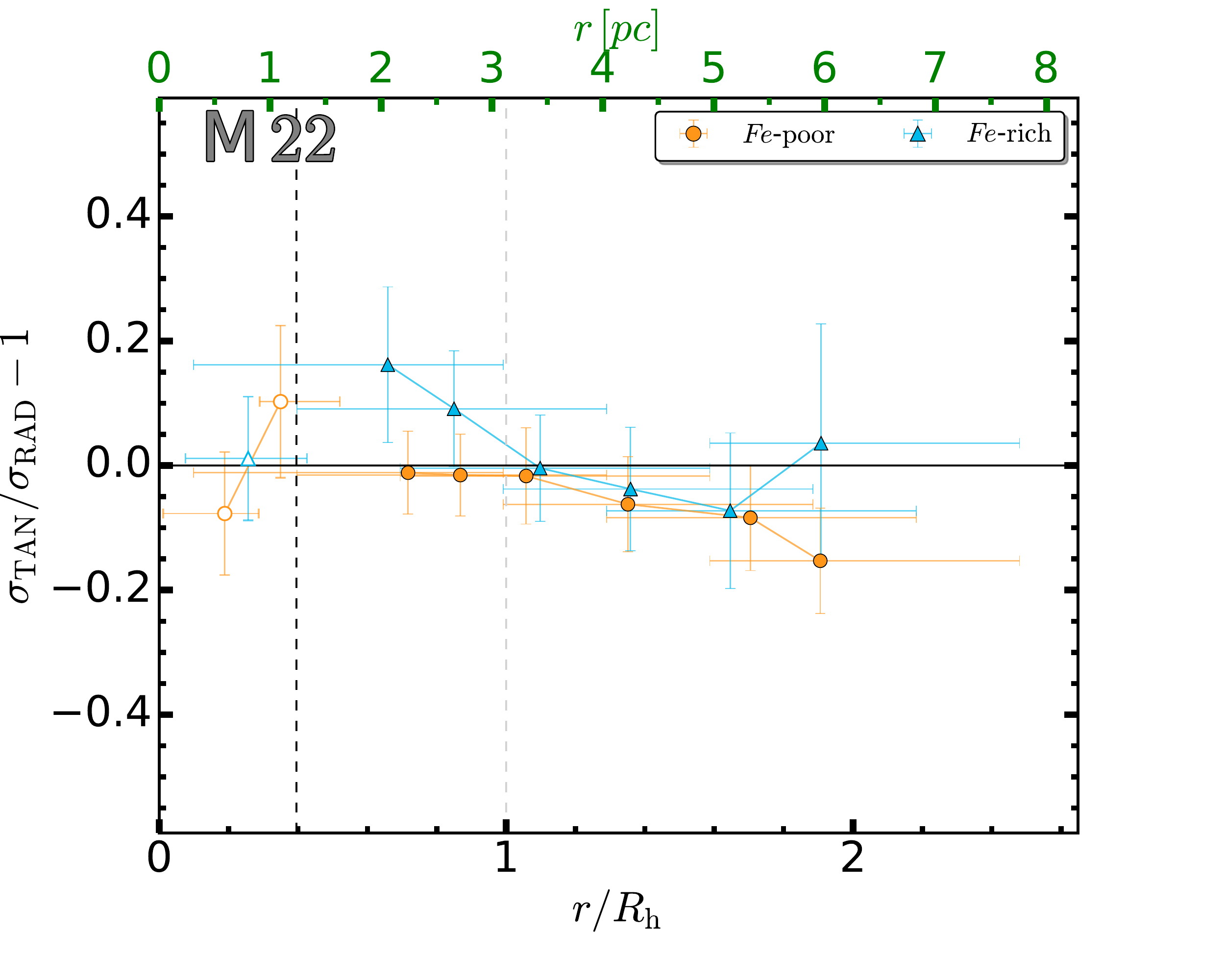}
  \caption{Velocity dispersion profiles for $\omega\,$Centauri (top-left) and M\,22 (top-right).
  Empty markers are {\it HST} results, while filled markers correspond to Gaia DR2 proper motions.
  Bottom panels show the corresponding anisotropy radial profiles. The horizontal lines plotted in the bottom panels correspond to isotropic stellar systems. Orange circles and cyan triangles represent Fe-poor and Fe-rich stars, respectively. The black and gray vertical dashed lines highlight the core radii and the half-light radii of each cluster from \cite{baumgardt2018}. The radial quantity is normalized over the half-light radius. }
  \label{fig:anisotropia}
\end{figure*}

To estimate the radial and tangential velocity dispersion profiles of Fe-poor and Fe-rich stars of $\omega\,$Centauri and M\,22 we followed the procedure described in \cite{mackey2013, marino2014}  and \citet{bianchini2018}. Briefly,  we maximized the likelihood function 
$$
\lambda = \prod_{\rm i=1}^Np(v_{\rm i},\epsilon_{\rm i})
$$  
with the probability of finding a star with velocity $v_{\rm i}$ and uncertainty $\epsilon_{\rm i}$ defined by Equation \ref{eq:3}. 
The corresponding uncertainties have been computed by bootstrapping with replacement a 1000 times.
\begin{equation}
\label{eq:3}
p(v_i,\epsilon) = \frac{1}{2\pi\sqrt{(\sigma^2+\epsilon^2_{\rm i})}}\exp \left(-\frac{(v_{\rm i}-v)^2}{2(\sigma^2+\epsilon^2_{\rm i})}\right)
\end{equation}
The radial and tangential velocity dispersion profiles as a function of the radial distance from the cluster center are plotted in the upper panels of Figure~\ref{fig:anisotropia}, where we used filled and open symbols to represent results from Gaia DR2 and {\it HST} proper motions, respectively.

The velocity dispersions of $\omega\,$Centauri and M\,22 reach their maximum values of $\sim$18 km s$^{-1}$ and $\sim$9 km s$^{-1}$, respectively, in the innermost analyzed regions and decline to $\sim$7 and $\sim$6 km s$^{-1}$, respectively, in the cluster outskirts.

We quantified the anisotropy of cluster stars as  $\beta = \sigma_{\rm TAN}/\sigma_{\rm RAD}-1$, with $\beta<0$ indicating radial anisotropy and $\beta>0$ tangential anisotropy. $\beta=0$ is characteristic of an isotropic stellar system. \\
$\beta$ is plotted as a function of the radial distance from the cluster center in the bottom panels of Figure~\ref{fig:anisotropia}.
Finally, we assessed the statistical significance of the observed differences in the kinematical profiles with the same procedure described in \citet[][]{cordoni2019}. The average ``p-value'', together with the its maximum and minimum are listed in Table~\ref{tab:vProfileSignificance} for all the analyzed populations and sub-populations. \\
Our results show that the studied populations of $\omega\,$Centauri are radially anisotropic in the central regions, with the Fe-rich population being more radially anisotropic than the Fe-poor ones. In the outermost region of the cluster both populations are consistent with an isotropic system. 
Furtermore, the p-values listed in Table~\ref{tab:vProfileSignificance} show that the observed differences are statistically not significant.

In M\,22 the radial profiles of $\beta$ for Fe-poor and Fe-rich stars are consistent with each other and are both approximately isotropic.  \\

\begin{table*}[h]
\centering
\begin{tabular}{l|l|cc|ccccc}
\toprule
\toprule
 Cluster & &  $\Delta \mu_{\rm RAD}$ & $\Delta \mu_{\rm TAN}$&  $\Delta \mu_{\rm RAD}$ & $\Delta \mu_{\rm TAN}$ & $\Delta\sigma_{\rm RAD}$ & $\Delta\sigma_{\rm TAN}$ & $\Delta\beta$ \\
       &  & A-D & A-D & $P$ & $P$ & $P$ & $P$ & $P$ \\
\midrule
$\mathbf{\omega}\,$\textbf{Centauri} & Fe-poor $-$ Fe-rich          & 0.100 & 0.110 & $0.690_{0.999}^{0.230}$ & $0.750_{0.949}^{0.660}$ & $0.230_{0.989}^{0.026}$ & $0.449_{0.946}^{0.012}$ & $0.600_{0.923}^{0.030}$ \\
                                     & N-poor $-$ N-rich            & 0.120 & 0.090 & $0.619_{0.945}^{0.172}$ & $0.866_{0.995}^{0.583}$ & $0.412_{0.977}^{0.011}$ & $0.151_{0.927}^{0.007}$ & $0.555_{0.988}^{0.065}$ \\
\textbf{Fe-poor}                     & N-poor $-$ N-rich            & 0.180 & 0.130 & $0.600_{0.970}^{0.186}$ & $0.745_{0.980}^{0.237}$ & $0.478_{0.957}^{0.004}$ & $0.132_{0.909}^{0.021}$ & $0.638_{0.979}^{0.040}$ \\
\textbf{Fe-rich}                     & N-poor $-$ N-rich            & 0.160 & 0.060 & $0.378_{0.970}^{0.003}$ & $0.186_{0.825}^{0.000}$ & $0.400_{0.959}^{0.005}$ & $0.280_{0.874}^{0.012}$ & $0.514_{0.998}^{0.043}$ \\
                                     & N-poor  $-$ $\rm pop$-$a$    & 0.023 & 0.020 & $0.600_{0.884}^{0.416}$ & $0.144_{0.809}^{0.066}$ & $0.295_{0.447}^{0.216}$ & $0.428_{0.385}^{0.026}$ & $0.682_{0.976}^{0.346}$ \\
                                     & N-rich  $-$ $\rm pop$-$a$    & 0.014 & 0.040 & $0.650_{0.802}^{0.472}$ & $0.201_{0.472}^{0.064}$ & $0.850_{0.959}^{0.778}$ & $0.446_{0.697}^{0.290}$ & $0.388_{0.680}^{0.058}$ \\
\midrule
\textbf{M\,22} & Fe-poor $-$ Fe-rich           & 0.201 & 0.015 & $0.810_{0.988}^{0.587}$ &             $0.749_{0.976}^{0.509}$ & $0.312_{0.644}^{0.009}$ & $0.501_{0.997}^{0.104}$ & $0.609_{0.920}^{0.231}$ \\ \\
               & N-poor $-$ N-rich             & 0.250 & 0.026 & $0.772_{0.953}^{0.387}$ & $0.709_{0.896}^{0.536}$ & $0.460_{0.963}^{0.059}$ & $0.504_{0.863}^{0.240}$ & $0.587_{0.999}^{0.177}$ \\ \\
\bottomrule
\bottomrule
\end{tabular}
\caption{Third and fourth columns indicate the probability (p-value) that the two populations come from the same parent distribution, according to the Anderson-Darling (A-D) test.
We considered the radial distributions of the quantities listed in the first line: $\Delta \mu_{\rm RAD}$, $\Delta \mu_{\rm TAN}$, $\Delta\sigma_{\rm RAD}$, $\Delta\sigma_{\rm TAN}$ and $\Delta\beta$.
 The test has been carried out independently in the radial and tangential component. The remaining columns list the probability, $P$, that the velocity distributions come from the same parent distribution ($P=1$) or not ($P=0$), determined as described in Section~\ref{subsec:vprofile}. We provide the average value of $P$ and the minimum and maximum $P$ values.}
\label{tab:vProfileSignificance}
\end{table*}

\section{Multiple stellar populations with different light-element abundances}\label{sec:1P2P}

In this section, we investigate the stellar populations of $\omega\,$Centauri and M\,22 selected on the basis of their content of light elements.
In $\omega\,$Centauri, we analyzed the entire groups of N-rich and N-poor stars identified in Figure~\ref{fig:stream}. Moreover, we separately  compared the spatial distributions and the kinematics of the sub-sample of N-poor and N-rich stars that belong to the Fe-poor population alone and of the sub-sample of N-poor, N-rich and population-a stars among Fe-rich stars. 
In M\,22, which has a smaller number of RGB stars than $\omega\,$Centauri, we limited the analysis to entire sample of N-rich and N-poor stars. 
The spatial distributions of the stellar populations with different light-elements of both M\,22 and $\omega\,$Centauri are analyzed in Section~\ref{sub:spatialN}, while Section~\ref{sub:kinematicsN} is focused on their  internal kinematics.

\subsection{Spatial distribution of N-rich and N-poor populations}\label{sub:spatialN}
\begin{figure*}[h!]
  \centering
 \includegraphics[height=6.0cm, 				    trim={4cm 0cm 2cm 0cm},		clip]{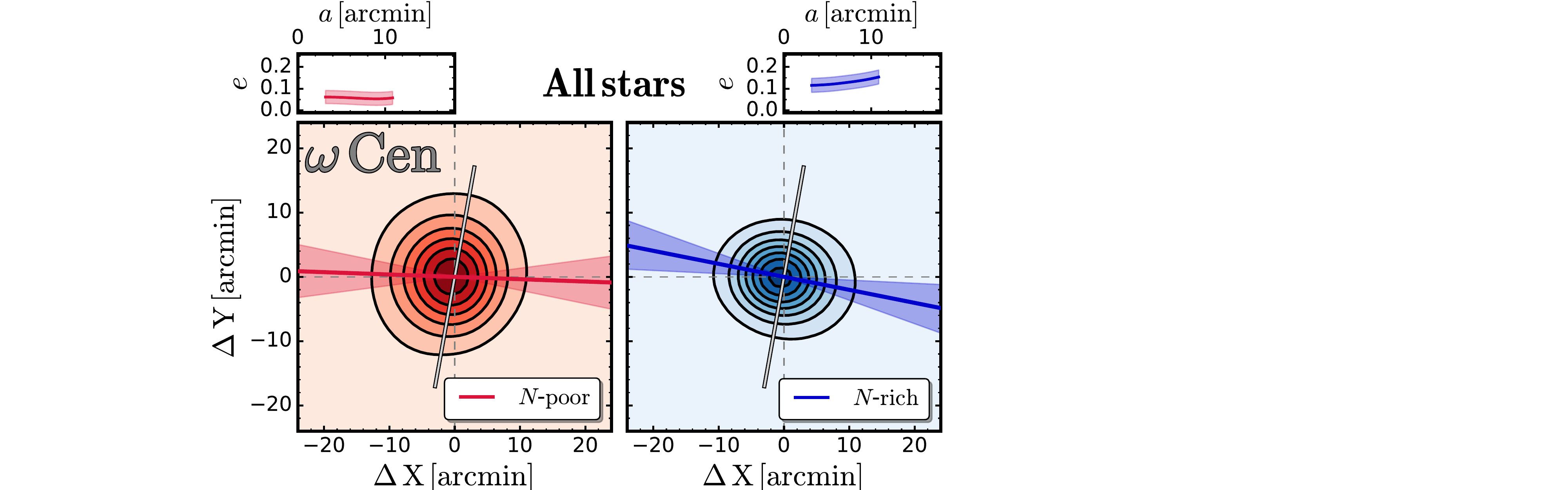}
  \includegraphics[height=6.0cm, 				    trim={4cm 0cm 2cm 0cm},		clip]{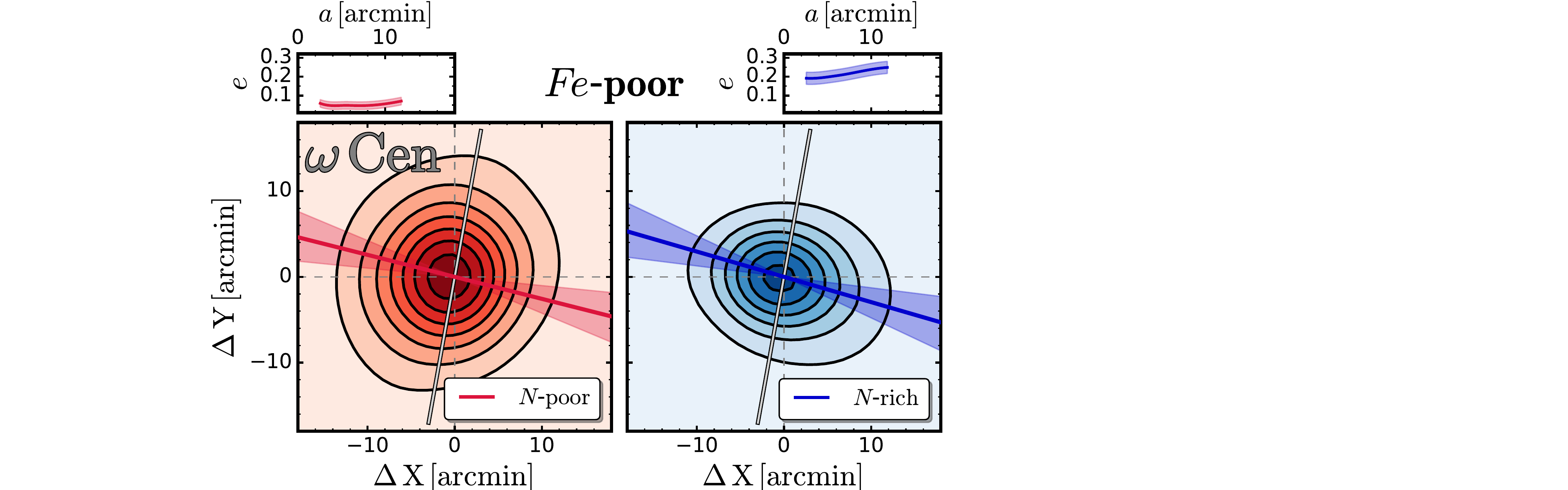}
  \includegraphics[height=6.0cm, 				    trim={4cm 0cm 2cm 0cm},		clip]{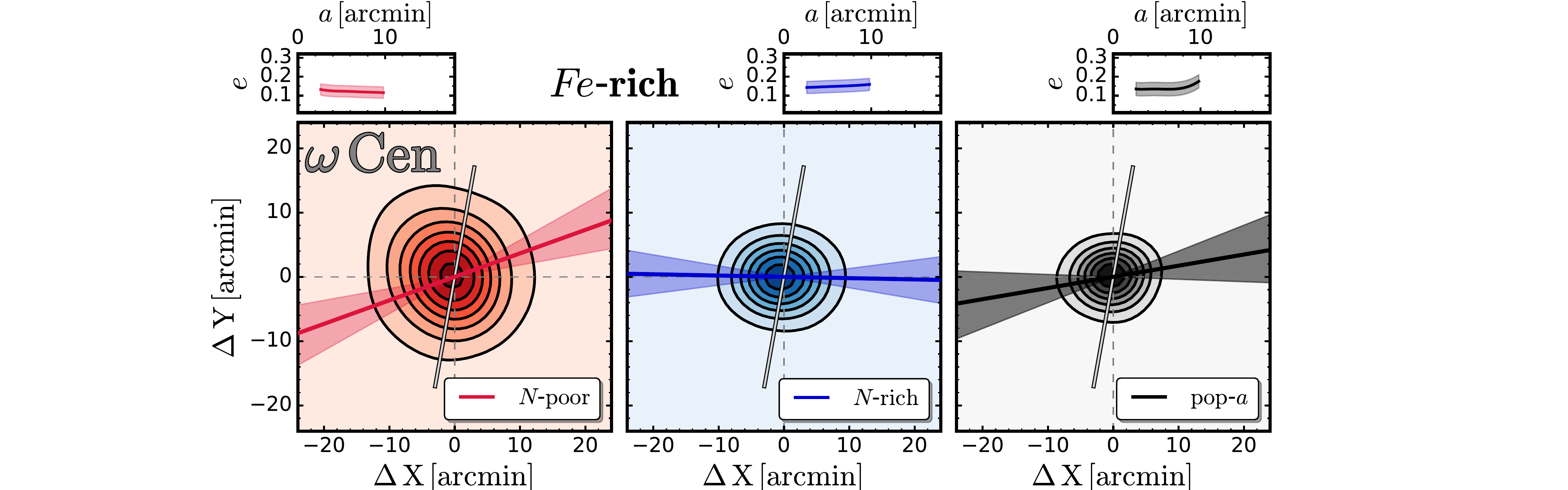}
   \caption{Density maps of stellar populations of $\omega\,$Centauri with different nitrogen abundances and of population-$a$ stars. 
  Top and center rows refer to Fe-poor and Fe-rich stars, respectively, while  the bottom panels show the spatial distribution of the entire sample of N-poor and N-rich stars. The ellipticities of the isodensity contours are plotted as a function of the semi-major axis.}
  \label{fig:spatial subpop}
\end{figure*} 

\begin{figure*}
  \centering
  \includegraphics[height=4.75cm, 				    trim={5.0cm 0.5cm 6.5cm 0cm},		clip]{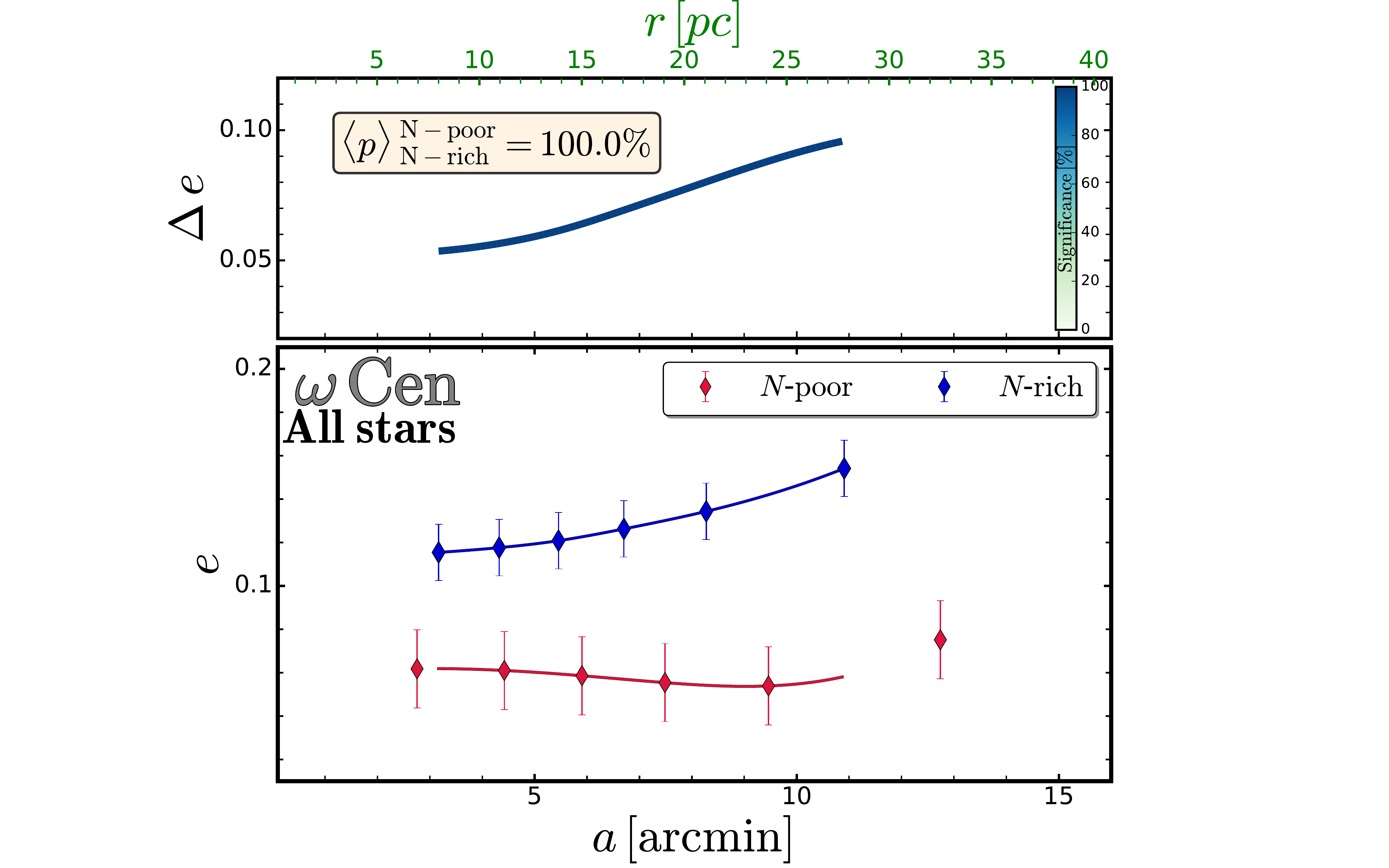}
  \includegraphics[height=4.75cm, 				    trim={6.6cm 0.5cm 6.5cm 0cm},		clip]{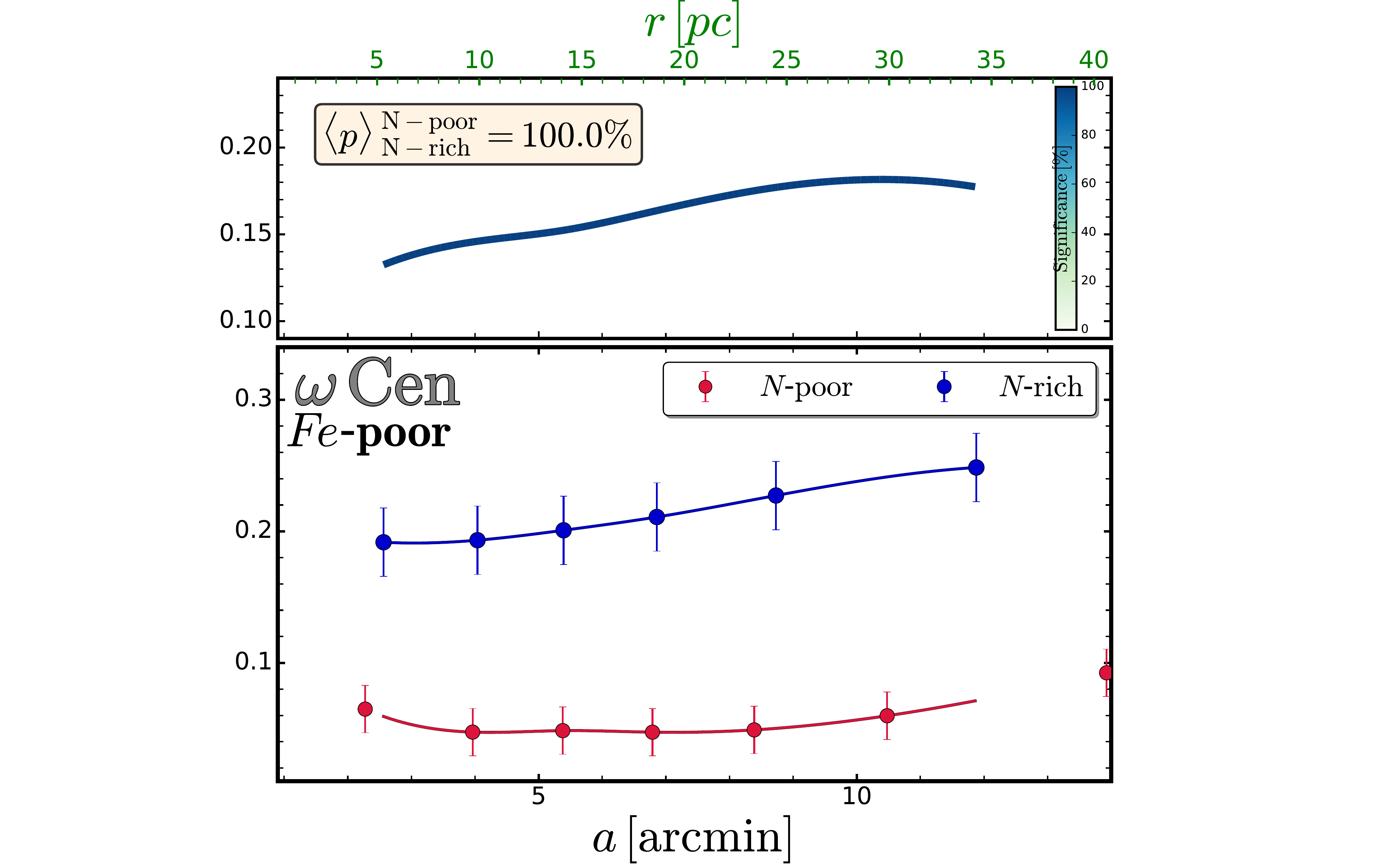}
  \includegraphics[height=4.75cm, 				    trim={6.6cm 0.5cm 6.5cm 0cm},		clip]{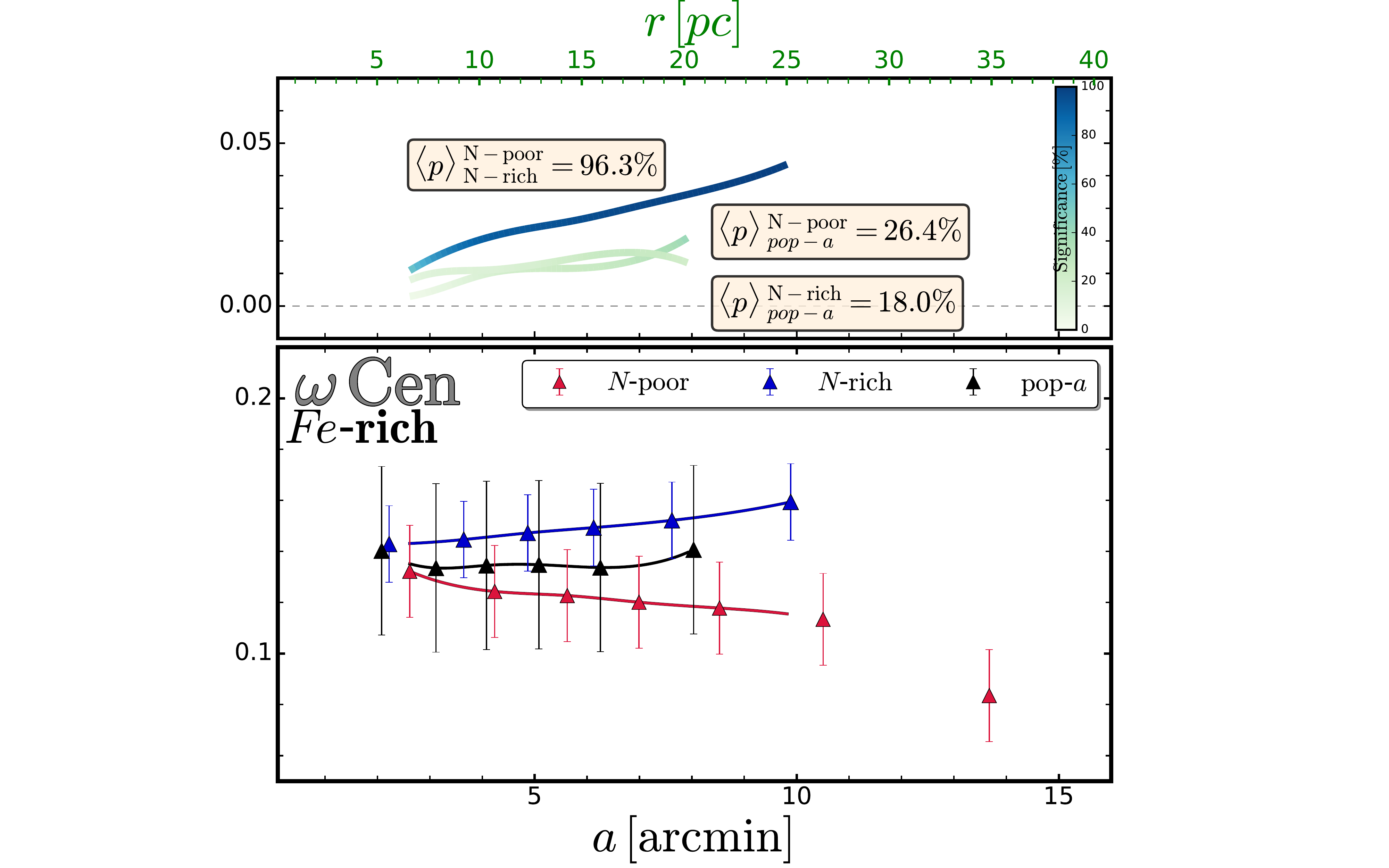}
  \caption{\textit{Lower panels.}  Ellipticity of N-rich and N-poor stellar populations as a function of the major semi-axis of the best-fit ellipses, $a$ for in the entire sample of analyzed $\omega\,$Centauri stars (left), among Fe-poor stars (middle) and Fe-rich stars (right). The latter panel also includes population-a stars. \textit{Upper panels.} Absolute values of ellipticity differences inferred from the populations quoted in each panel against $a$. The level of colors indicate the statistical significance of the difference as indicated by the colorbar. }
  \label{fig:ell subpop}
\end{figure*} 

To investigate the spatial distributions of stellar populations with different nitrogen content, we extended the analysis introduced in Sections~\ref{subsec:spatial} to the selected groups of N-rich and N-poor stars. 
Results on $\omega\,$Centauri are illustrated in Figures~\ref{fig:spatial subpop} and ~\ref{fig:ell subpop}. 

The upper panels of Figure~\ref{fig:spatial subpop} compare the density contours of the overall N-poor and N-rich samples. 
Clearly, N-rich stars, which have average ellipticity, $e \sim 0.13$, exhibit more elliptical distributions than N-poor stars ($e \sim 0.05$). 

The ellipticity difference between the spatial distributions of N-rich and N-poor stars is larger when we limit the analysis to the Fe-poor stars as shown in the middle panels of Figure~\ref{fig:spatial subpop}. Specifically,  N-rich Fe-poor stars exhibit more flattened distributions (ellipticity $e \sim$0.20) than N-poor Fe-poor stars, which have $e\sim 0.05$.  

Qualitatively, the spatial distributions of the Fe-rich sub-populations with different nitrogen abundances follow a similar behaviour as their Fe-poor counterparts, although the ellipticity differences among the various sub-populations are less pronounced. Indeed, as shown in the lower panels of Figure~\ref{fig:spatial subpop}, N-rich Fe-rich stars have an average ellipticity of $e \sim 0.15$ , which is slightly higher than that of N-poor Fe-rich stars ($e\sim 0.12$). The ellipticity difference is significant at the $\sim$2.1 $\sigma$-level.
On the other hand, population-$a$ stars have, on average, $e\sim 0.13$, and the small ellipticity difference with N-rich Fe-rich and N-poor Fe-rich is not statistically significant.  

Overall, Figure~\ref{fig:spatial subpop} reveals that the median semi-major axes of the best-fit ellipses of all populations are consistent with each other within one sigma and are almost perpendicular to the global rotation axis determined in \citet{sollima2019}.

For completeness, we plot the ellipticity of the various subpopulations as a function of the major axis of the best-fit ellipse, $a$, in the bottom panels of Figure~\ref{fig:ell subpop}. Upper panels show the absolute value of the ellipticity differences $|{\Delta e}|$ between the populations quoted in the figures against $a$. The color scale is indicative of the statistical significance of the difference.
All populations are consistent with having constant ellipticity in the analyzed interval of $a$. 

\begin{figure}[h!]
  \centering
  \includegraphics[width=14.0cm, 				    trim={5cm 0cm 2cm 0cm},		clip]{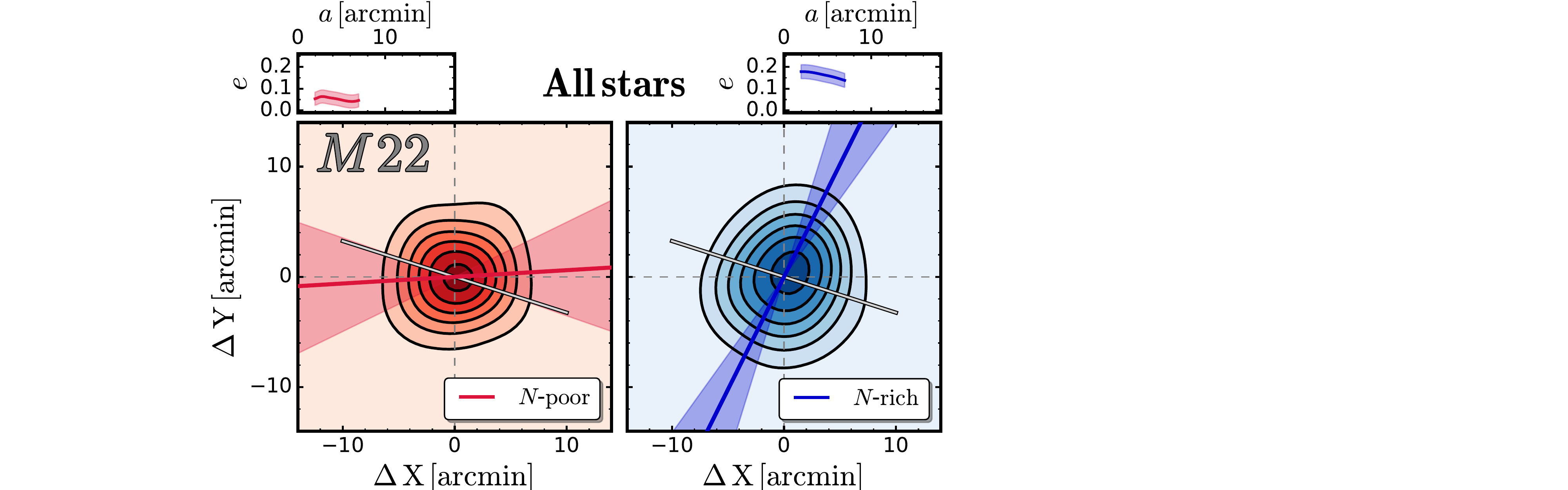}
  \includegraphics[width=8cm, 				    trim={4.0cm 0.5cm 6.5cm 0cm},		clip]{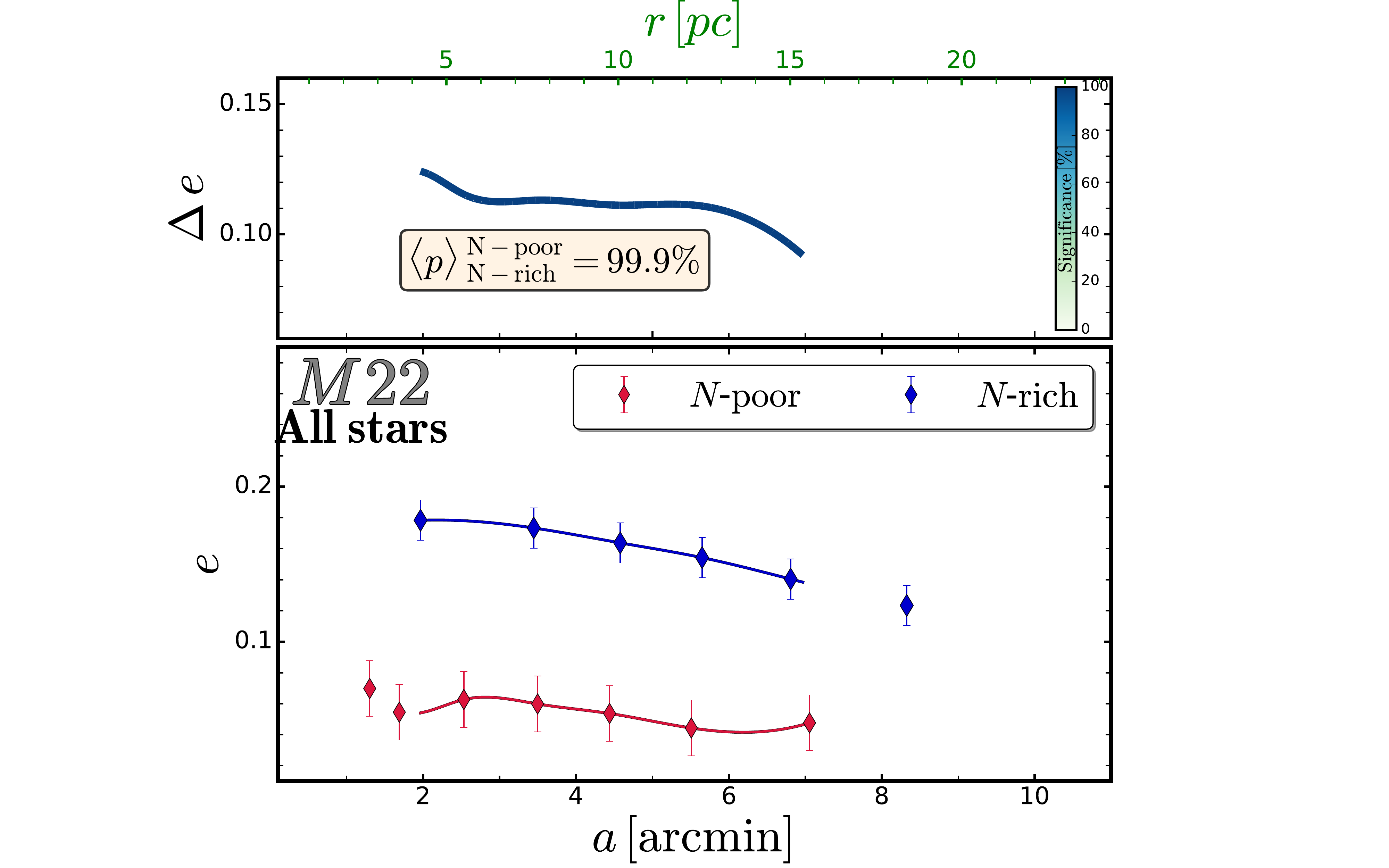}
  \caption{\textit{Top panels.} Same as Figure~\ref{fig:spatial} and \ref{fig:spatial subpop} for the N-poor (red, left panel) and N-rich (blue, right panel) in M\,22. \textit{Bottom panel.} Ellipticity profile as in Figure~\ref{fig:ell} and \ref{fig:ell subpop}.}
  \label{fig:spatial_subpopM22}
\end{figure}

Results on the spatial distributions of the N-poor and N-rich stellar populations of M\,22 are illustrated in Figure~\ref{fig:spatial_subpopM22}. N-rich stars have average ellipticity $e \sim 0.15$ and clearly exhibit a more-flattened distribution than N-poor stars, which have $e \sim 0.05$.

\subsection{Internal kinematics }\label{sub:kinematicsN}
The internal kinematics of the stellar populations with different nitrogen abundances and of population-a stars are derived by using the methods described in Section~\ref{subsec:disp}.

\begin{figure*}[h!]
  \centering
 \includegraphics[width=5.5cm, 				trim={0cm 2.95cm 0.cm 0cm},		clip]{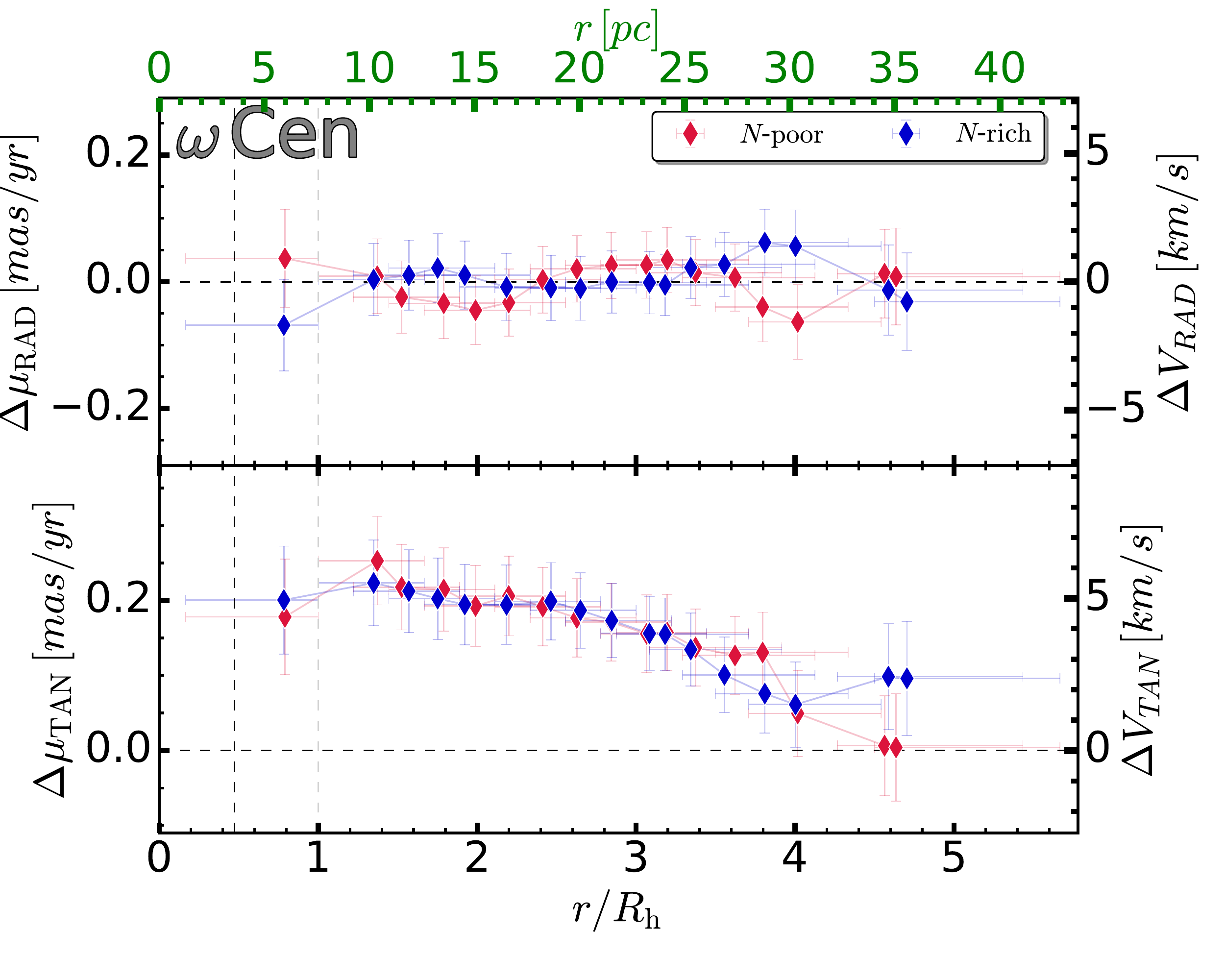}
  \includegraphics[width=5.5cm, 				trim={0cm 2.95cm 0.cm 0cm},		clip]{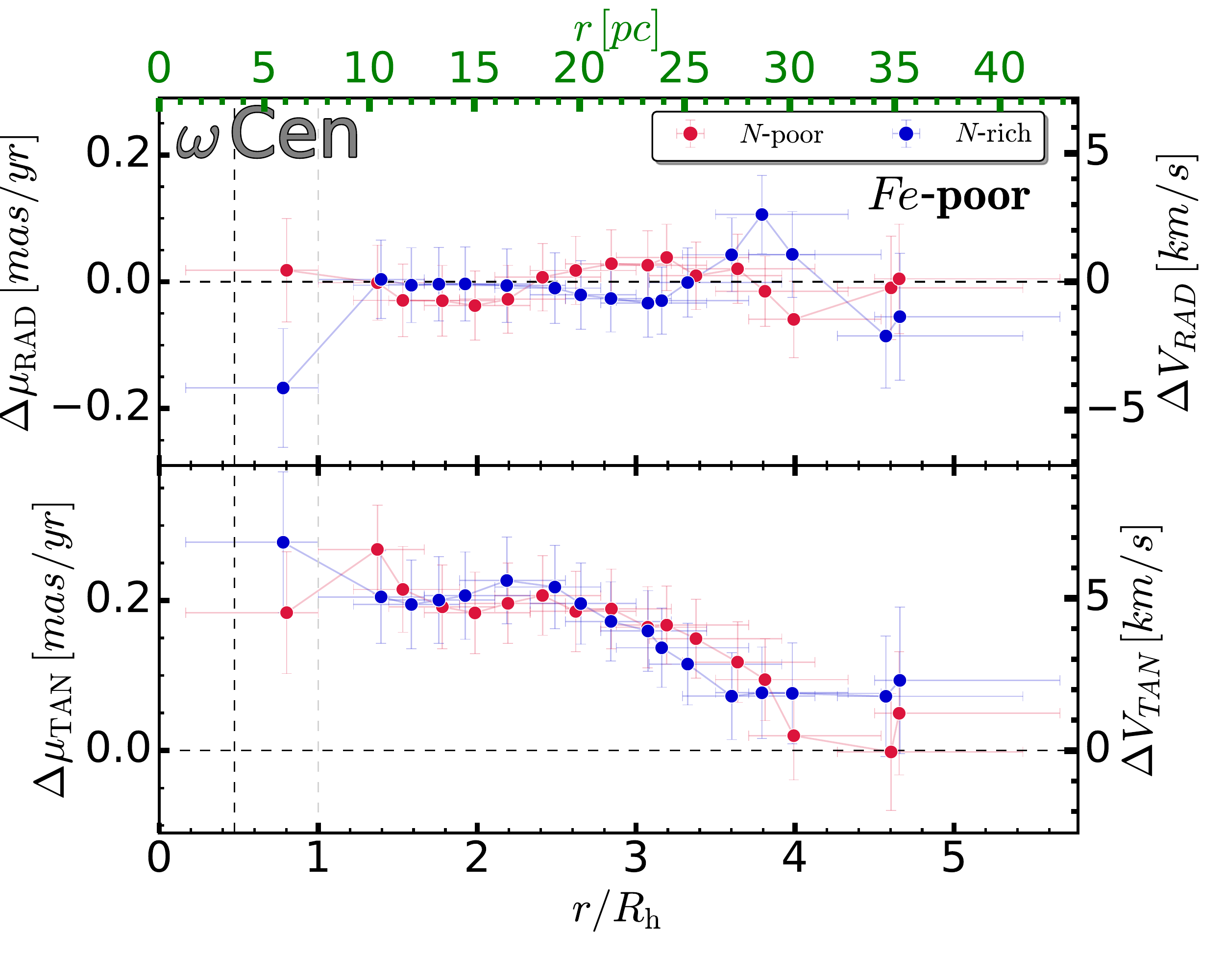}
  \includegraphics[width=5.5cm, 				trim={0cm 2.95cm 0.cm 0cm},		clip]{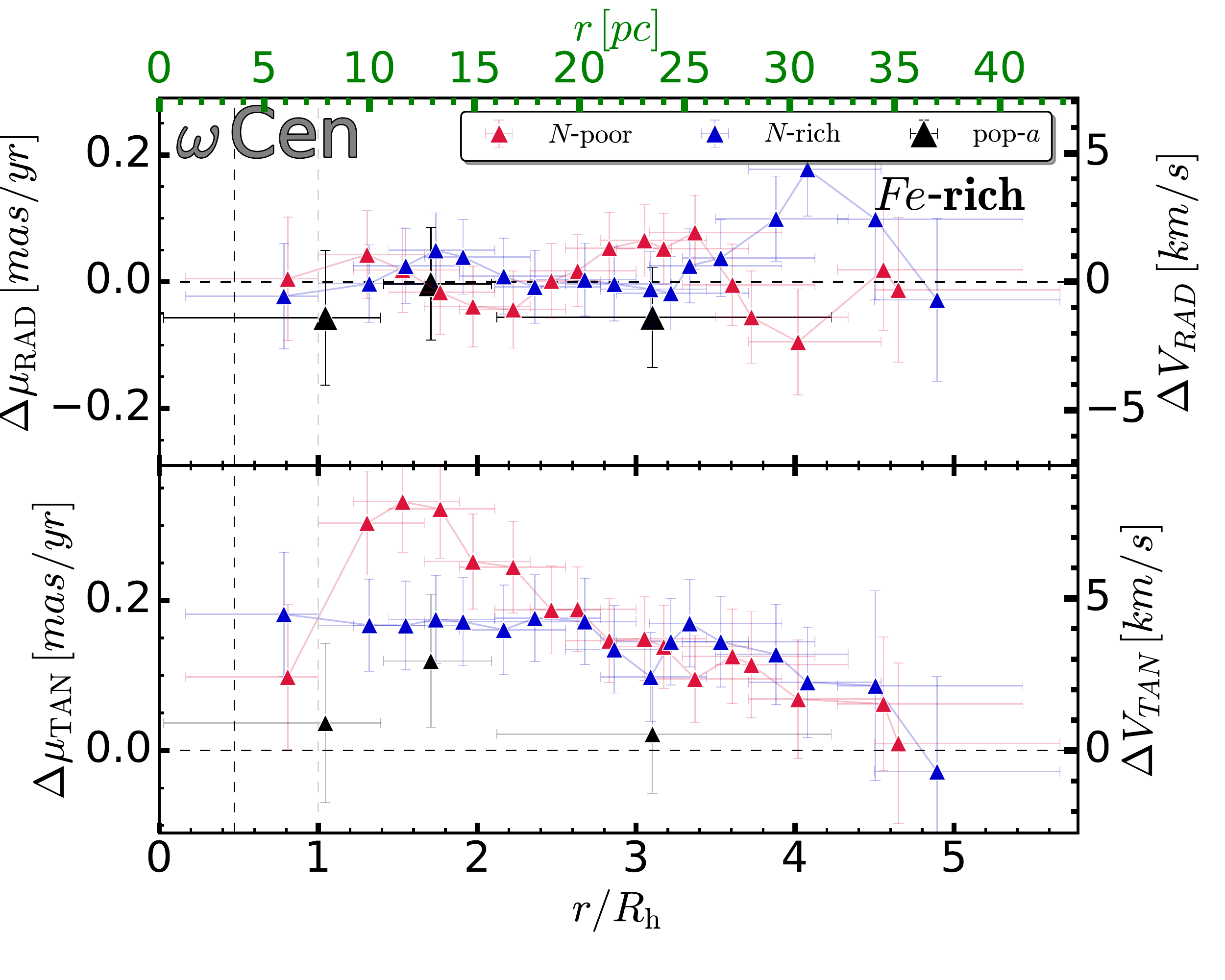}
 
  \includegraphics[width=5.5cm, 				trim={0cm 2.95cm 0.cm 1.75cm}, 	clip]{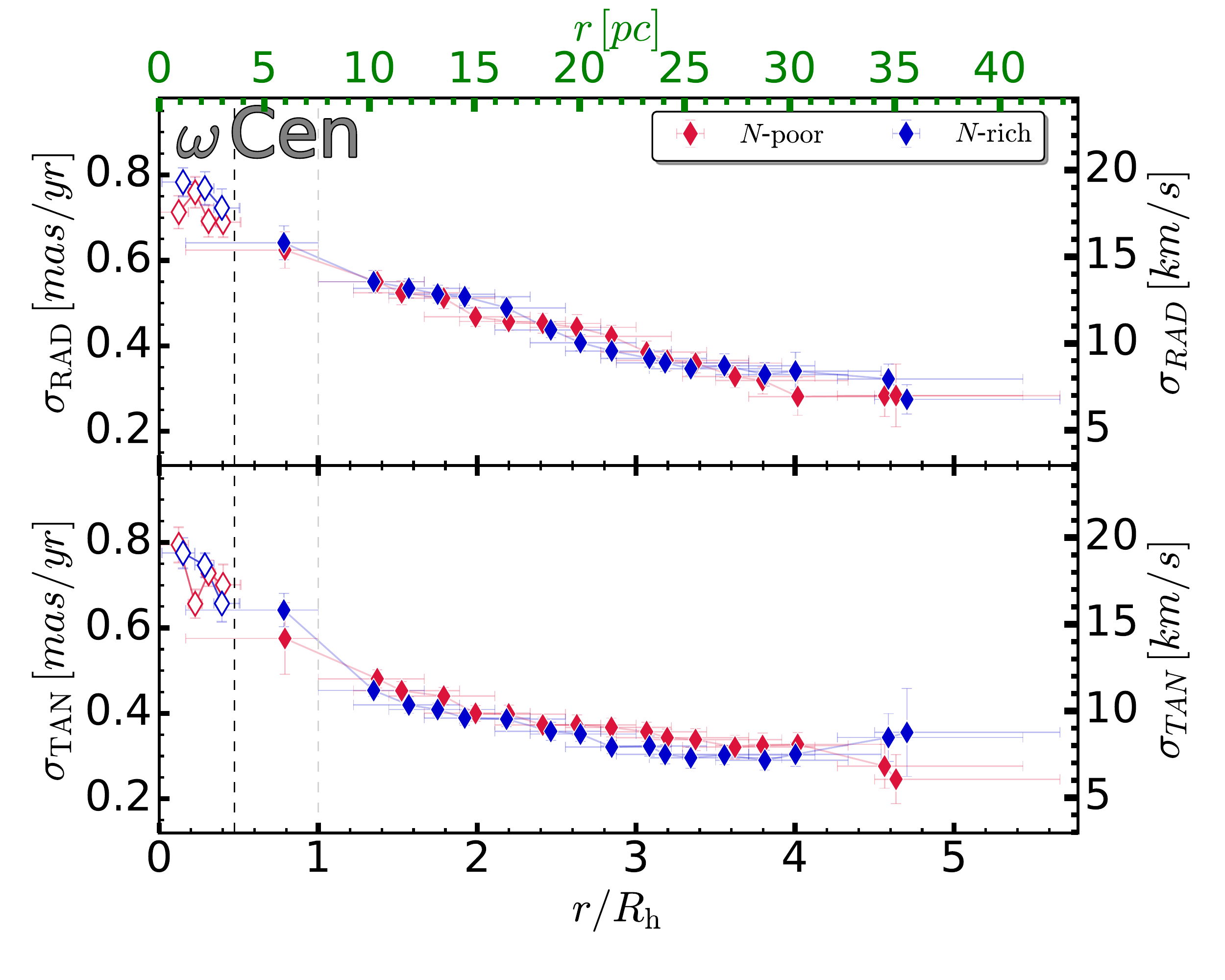}
  \includegraphics[width=5.5cm, 				trim={0cm 2.95cm 0.cm 1.75cm}, 	clip]{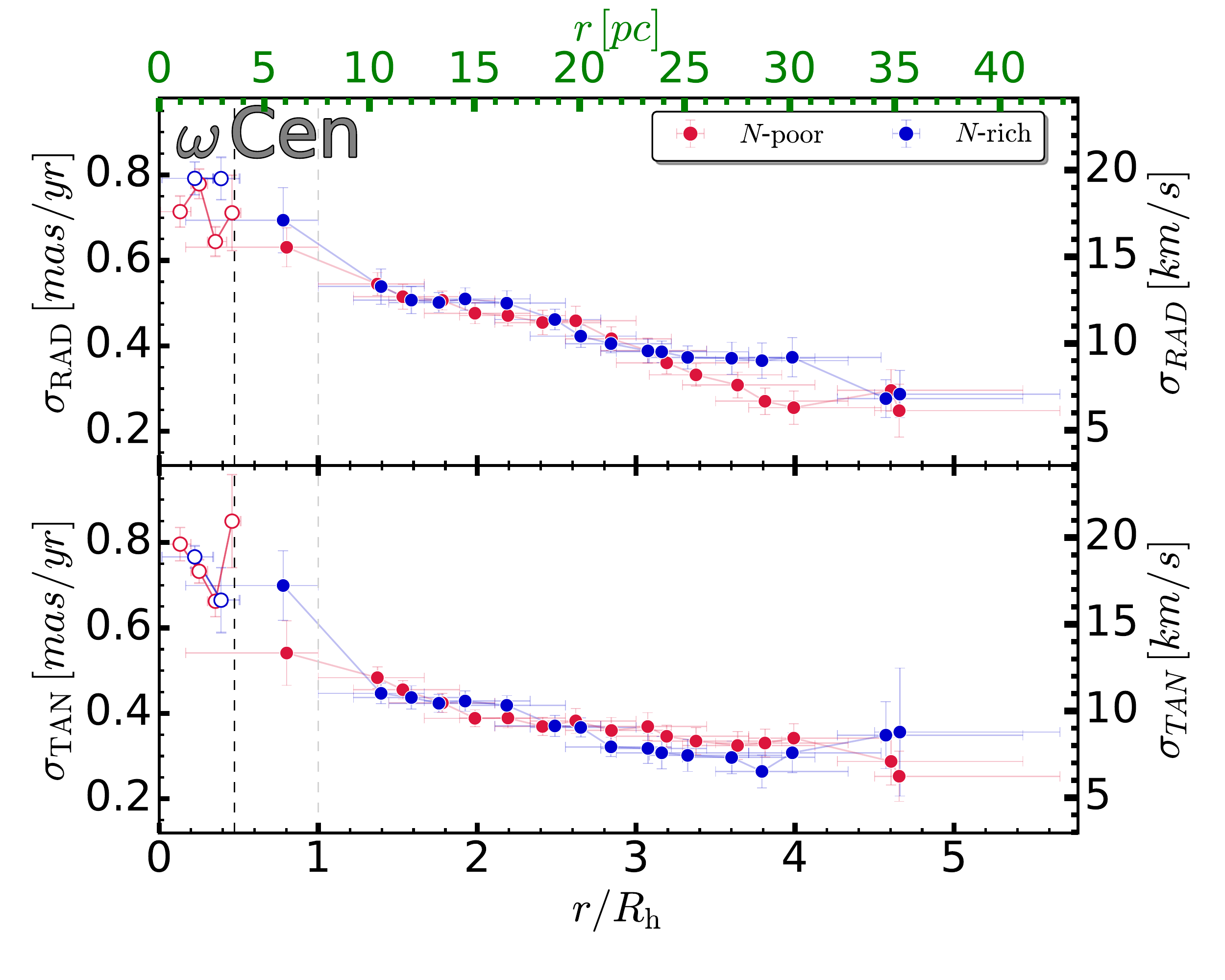}
  \includegraphics[width=5.5cm, 				trim={0cm 2.95cm 0.cm 1.75cm}, 	clip]{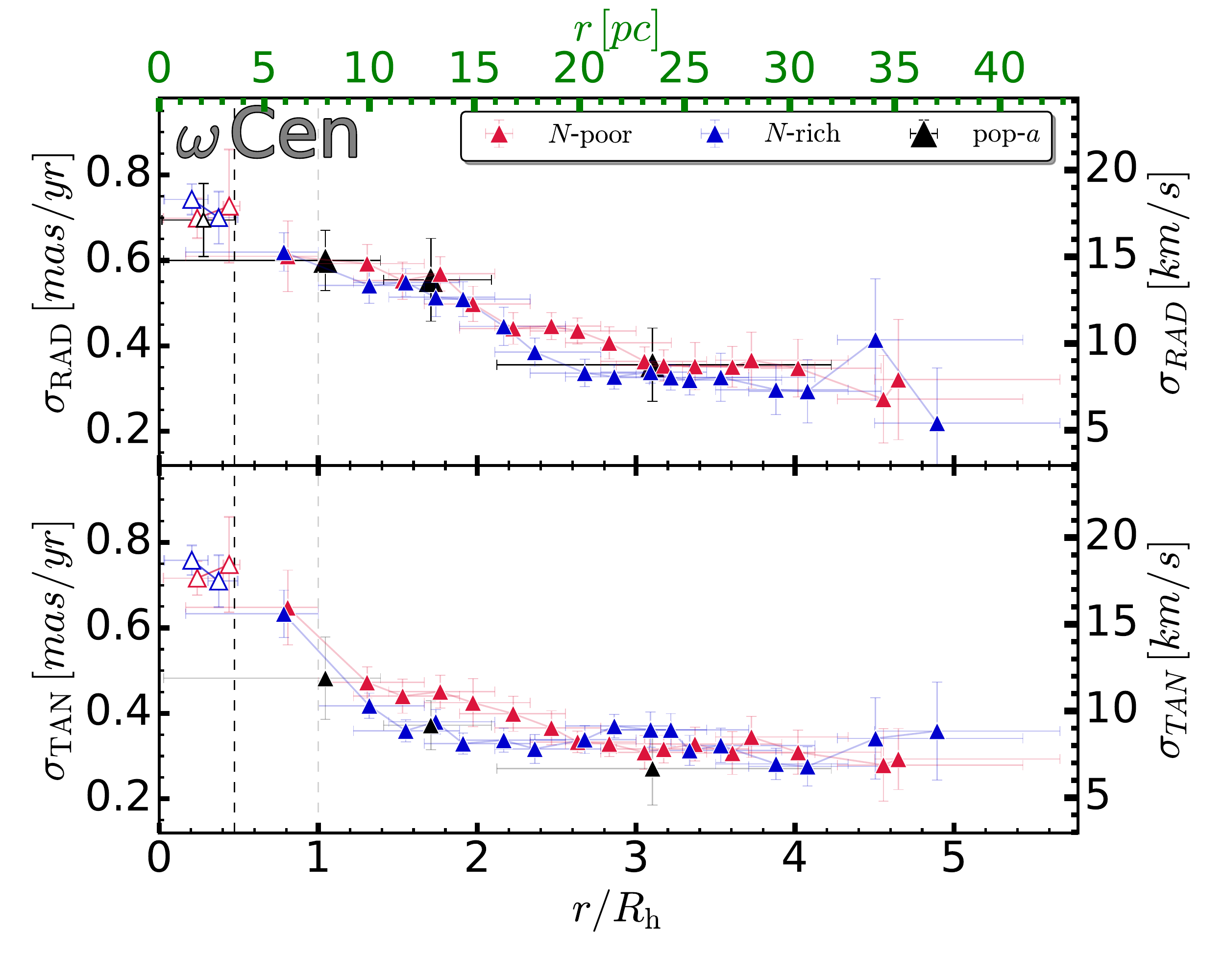}
 
   \includegraphics[width=5.5cm, 				trim={0.cm 0.cm 0.cm 1.75cm}, 	 clip]{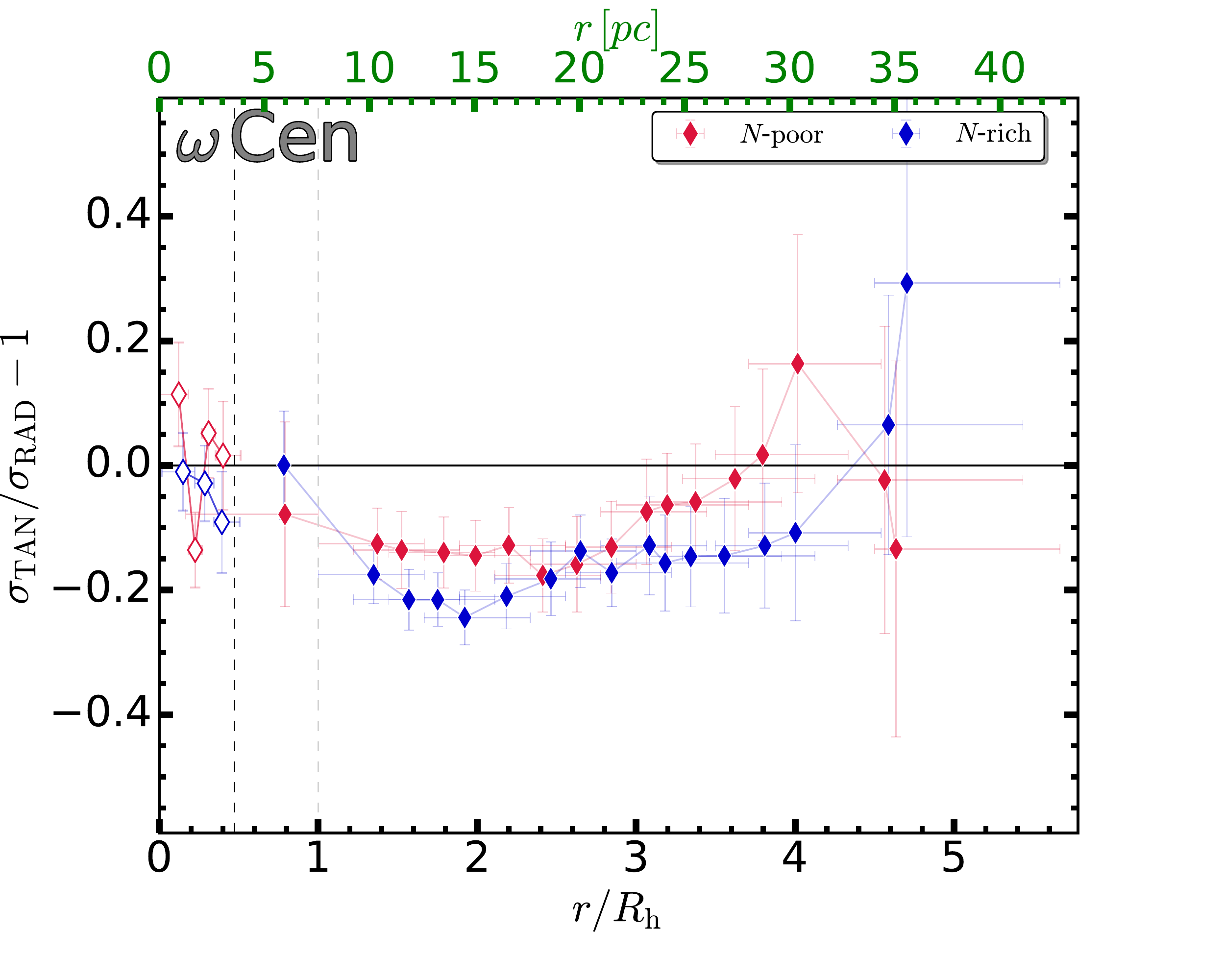}
    \includegraphics[width=5.5cm, 				trim={0.cm 0.cm 0.cm 1.75cm},    clip]{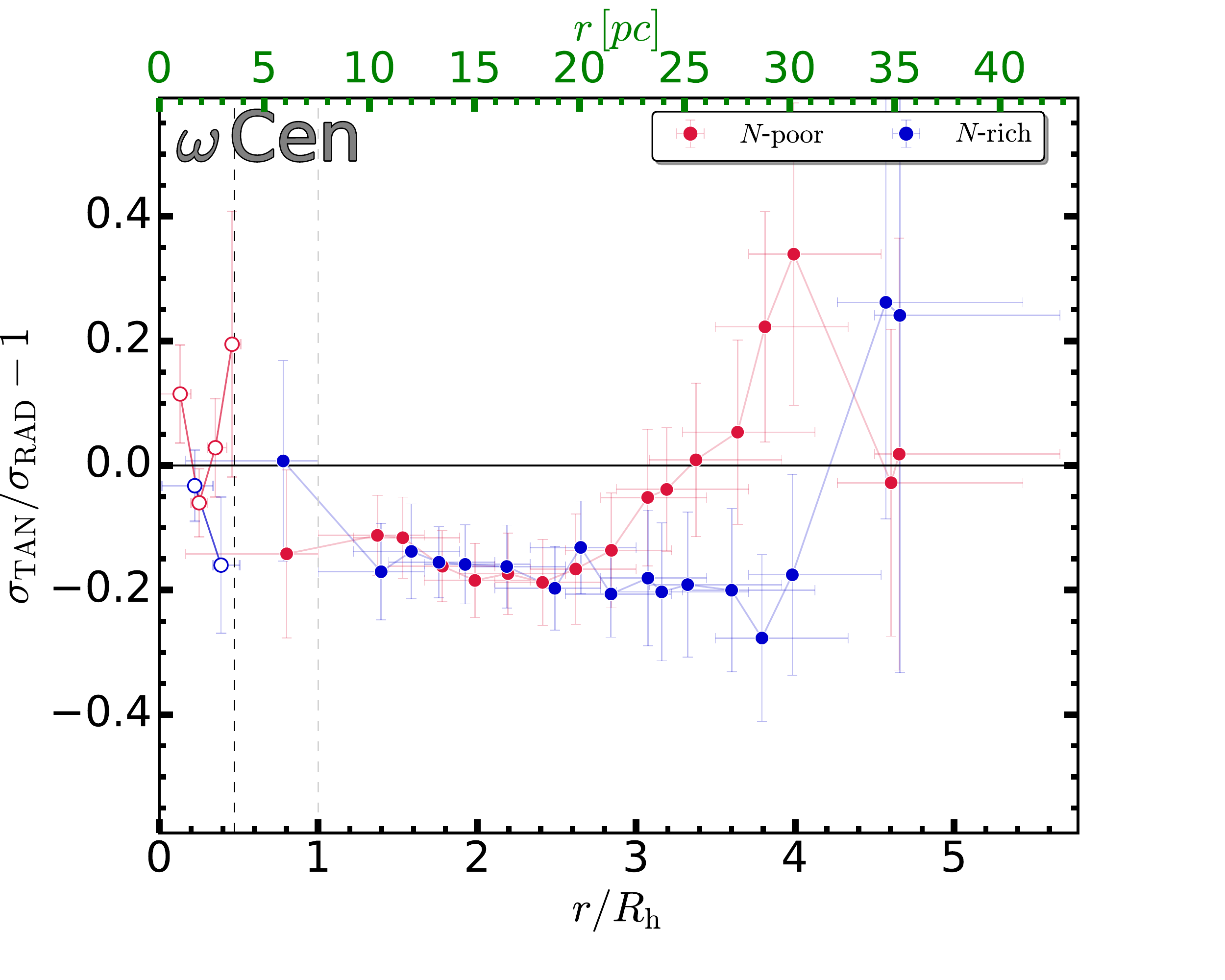}
  \includegraphics[width=5.5cm, 				trim={0.cm 0.cm 0.cm 1.75cm}, 	 clip]{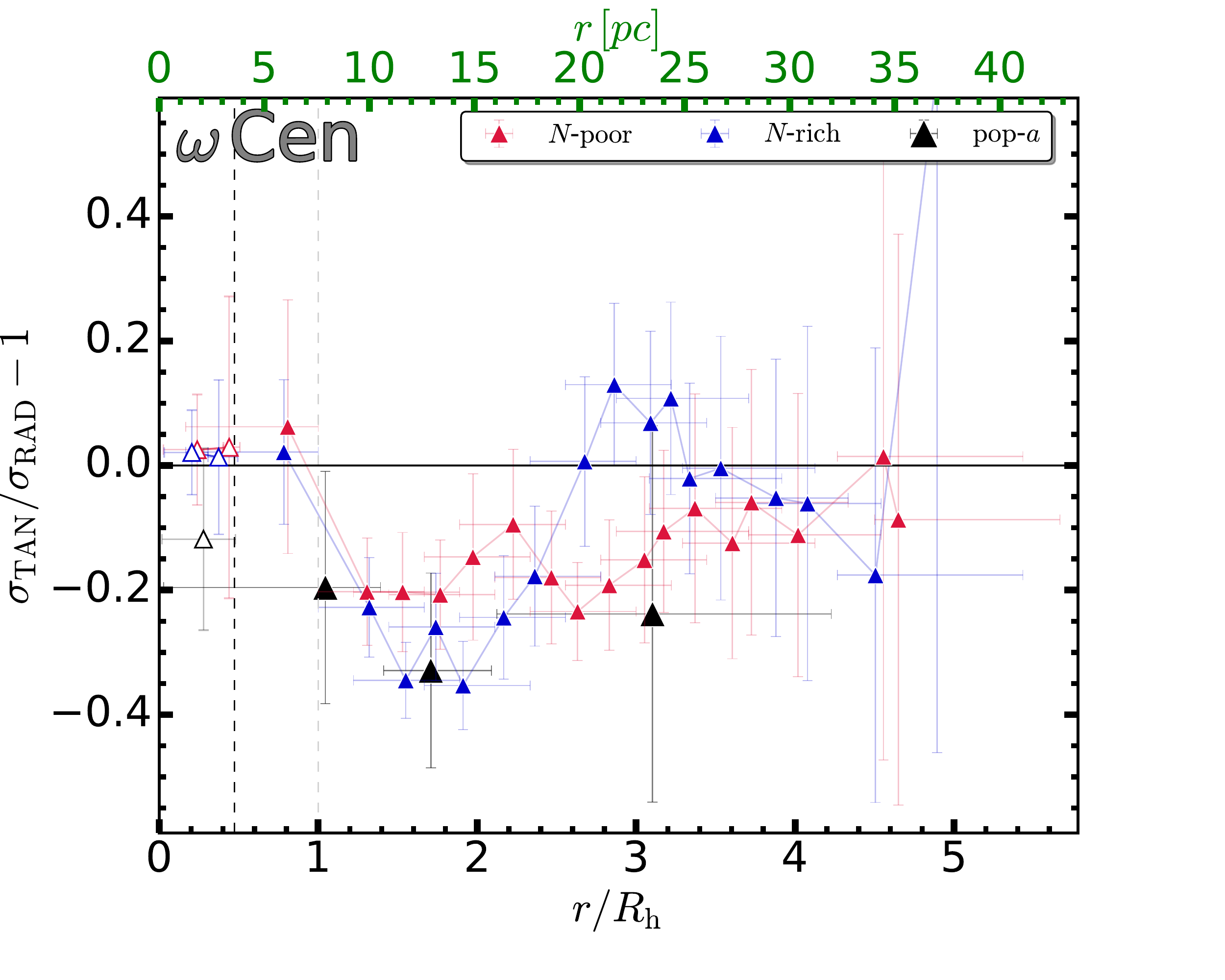}

   \caption{From top to bottom: average velocity, velocity dispersion and anisotropy profiles as a function of the distance from the cluster center for N-poor (red dots) and N-rich (blue triangles) stars.
   Left panels refer to the entire sample of $\omega\,$Centauri stars, while in the middle and right panels we considered the Fe-poor and Fe-rich populations, respectively.
   The velocity profiles of population-a stars are plotted with black triangles in the right panels.
  The black and gray vertical dashed lines highlight the core radius and the half-light radius from \cite{baumgardt2018}. The radial quantity is normalized over the half-light radius. }
  \label{fig:profile subpop}
\end{figure*}

The velocity profiles of the entire sample of N-rich and N-poor stars are plotted in the left panels of Figure~\ref{fig:profile subpop} and the corresponding results on the sub-populations of N-poor and N-rich populations among Fe-poor and Fe-rich stars are illustrated in the middle and right panels of Figure~\ref{fig:profile subpop}, respectively. The right panels also include the velocity profiles of population-$a$ stars.


A visual inspection of the top-left panels of Figure~\ref{fig:profile subpop} reveals that both N-rich and N-poor stars exhibit significant rotation in the plane of the sky, with $\Delta \mu_{\rm TAN}$ ranging from $\sim$6 km/s towards the cluster core to $\sim$2 km/s in the most-distant regions.

Stellar populations with different nitrogen abundances exhibit radial anisotropic motions between $\sim$1 and 3 half-light radii. Differences in the radial profile of $\beta$ are present the region with $r/R_{\rm h}\sim 1.5-2.5$, where N-rich stars have more-radially anisotropic motions and between $\sim$3 and $4.5$ half-light radii, where N-poor stars are consistent with having isotropic motions while N-rich stars have $\beta \sim -0.2$. 

When we consider the sample of Fe-poor stars alone, we find that N-poor and N-rich stars have similar rotation patterns. In contrast, stellar populations with different nitrogen abundances seem to exhibit different tangential-velocity profiles in the radial annulus between $\sim 0.8$ and $\sim 2.3$ half-light radii. 

The average values of $\Delta \mu_{\rm TAN}$ for N-poor and N-rich stars, estimated as in \citet{vasiliev2019}, are 
0.27$\pm$0.06 and 0.18$\pm$0.06, respectively.
However, these uncertainties, which account for systematic errors that affect Gaia DR2 proper motions, are upper limits to the true errors on the relative proper motions.
Indeed, Gaia DR2 systematic errors depend on stellar colors and positions. Hence, they mostly cancel out when we consider the relative motions of N-rich and N-poor stars that have similar colors and spatial distributions. 
The average $\Delta \mu_{\rm TAN}$ difference between N-poor and N-rich stars is 0.09$\pm$0.03 if we do not consider the contribution of Gaia DR2 systematics. In this case the difference would be significant to the 3$\sigma$-level.

\begin{figure*}
  \centering
  \includegraphics[width=13cm, trim={0cm 0cm 0cm 0cm}, clip]{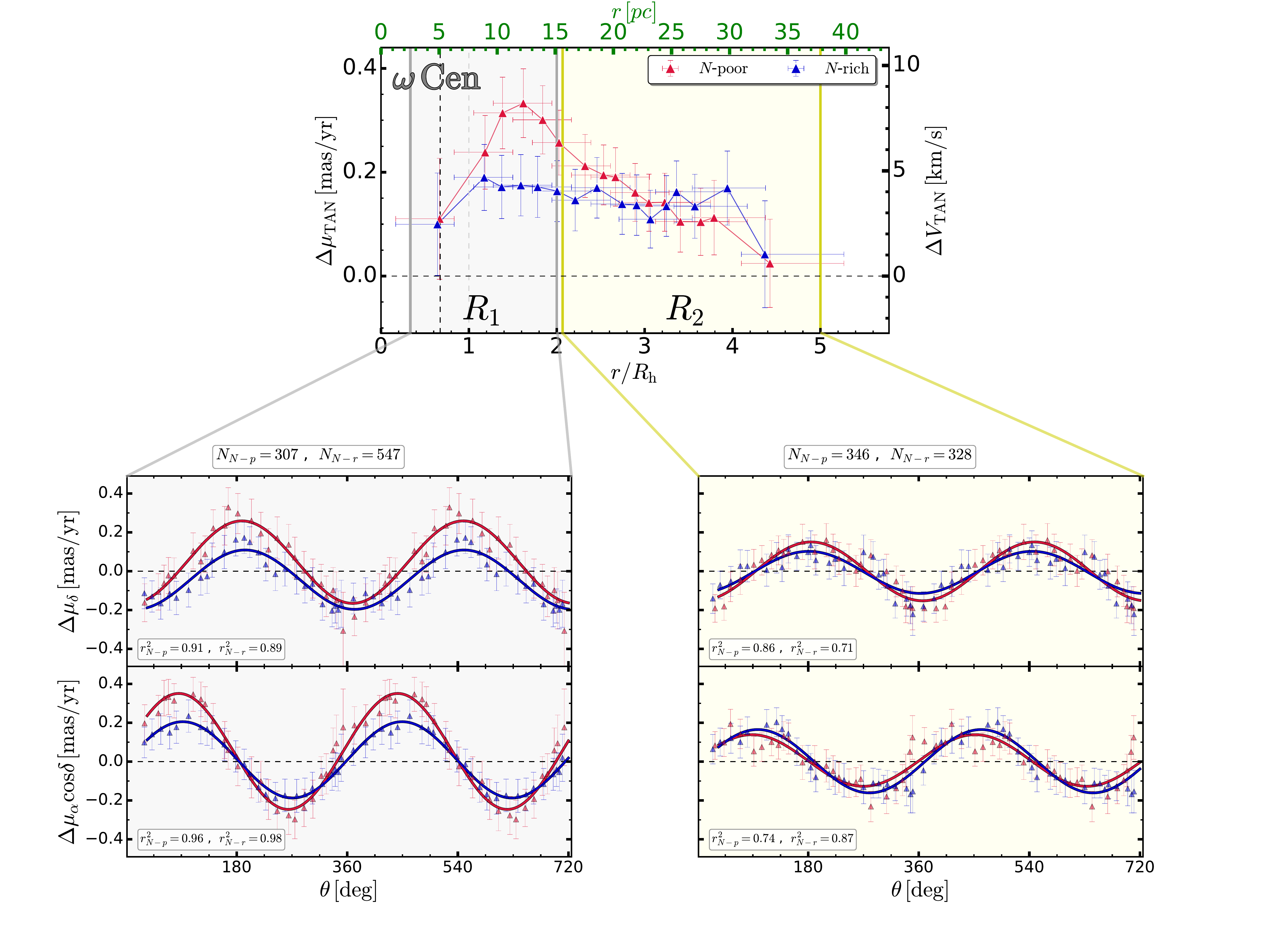}
  \caption{Reproduction of the radial profile of $\Delta \mu_{\rm TAN}$ for the N-rich and N-poor groups of Fe-rich stars (top) in $\omega\,\rm Cen$. 
  In the bottom panels, $\Delta\mu_{\rm \alpha}\cos\delta$ and $\Delta\mu_{\rm \delta}$ are plotted as a function of the position angle $\theta$ for stars in regions $R_{\rm 1}$ and $R_{\rm 2}$ defined in the top panel. The sine functions that provide the best fit with the observations of N-poor and N-rich stars are represented with red and blue lines, respectively. }
  \label{fig:regions ferich}
\end{figure*}

To further investigate the rotation of stellar populations with different nitrogen abundances among Fe-rich stars, we plot in  Figure~\ref{fig:regions ferich} $\Delta \mu_{\rm \alpha} cos{\delta}$ and $\Delta \mu_{\delta}$ as a function of $\theta$ for N-poor Fe-rich and N-rich Fe-rich stars in the two radial bins. We find that the rotation curves of N-poor stars within two half light-radii from the cluster center exhibit higher amplitudes than those of N-rich stars in the same radial bin. The amplitude differences derived from the $\Delta \mu_{\rm \alpha} cos{\delta}$ vs.\,$\theta$ and $\Delta \mu_{\delta}$ vs.\,$\theta$ planes are significant at 2.4-$\sigma$ and 2.1-$\sigma$, level respectively.   Hence, the probability that the amplitude differences observed in both components of proper motions are due to observational errors is smaller than 0.2\%. 
The rotation curves of the two populations are consistent with having the same amplitudes when we consider stars with $r>2 R_{\rm h}$.

\section{Summary and conclusions} \label{sec:summary}
We combined Gaia DR2 and {\it HST} proper motions with multi-band photometry from {\it HST} and ground-based facilities to  investigate the spatial distributions and the internal kinematics of multiple stellar populations in the Type II GCs M\,22 and $\omega\,$Centauri over a wide field of view, from the cluster center up to $\sim$2.5 and $\sim$5.5 half-light radii,   respectively.

We first identified stellar populations with different iron abundances along the RGB from differential-reddening corrected  CMDs build with appropriate combinations of $U, V, I$ magnitudes (or $m_{\rm F336W}, m_{\rm F606W}, m_{\rm F814W}$ magnitudes, in the case of {\it HST} data). Then, we identified and analyzed stellar populations with different nitrogen content. 
 The main results for stellar populations with different metallicities of M\,22 can be summarized as follows:

\begin{itemize}
    \item Fe-poor and Fe-rich stars exhibit the same average proper motions within 1-$\sigma$. The same result is true also for stellar populations with different Nitrogen abundance.
    \item Fe-poor and Fe-rich stars share similar spatial distributions with an average ellipticity, $e \sim 0.1$ (Figure~\ref{fig:ell}).
    \item Both populations exhibit significant rotation in the plane of the sky and their rotation curves are characterized by similar phases and amplitudes (A$\sim$2.5 km/s, Figure~\ref{fig:rot}).
    The tangential-velocity profiles of Fe-poor and Fe-rich stars are nearly flat in the analyzed radial interval with an average $\Delta \mu_{\rm TAN} \sim 2.5$ km/s (Figure~\ref{fig:profiles}).
    \item Fe-poor and Fe-rich stars share similar velocity-dispersion profiles, with both the radial and tangential component ranging from $\sim$9 to $\sim$6 km/s when moving from the cluster center to a radius of $\sim$2.5 half-light radii. 
    Both populations exhibit isotropic motions (Figure~\ref{fig:anisotropia}). 
\end{itemize}

The main findings on the Fe-poor and Fe-rich stars of $\omega\,$Centauri include:

\begin{itemize}
    \item The stellar populations with different metallicities share the same motions, thus confirming the conclusions by \cite{anderson2010, sanna2020}. Similarly, stellar populations with different N have the same average proper motions.
    \item The spatial distributions of both stellar populations have similar elliptical shapes with ellipticity, $e \sim 0.06$, and similar directions of the major axes (Figure~\ref{fig:ell}).
    \item The rotation pattern in the plane of the sky is similar for Fe-poor and Fe-rich stars. The tangential-velocity component decreases from $\sim 6$ km/s at a radial distance of about one half-light radius from the center to $\sim$2 km/s at $r/R_{\rm h} \sim 5$ (Figure~\ref{fig:profiles} and \ref{fig:rot}). 
    \item The rotation curves of both populations share similar amplitudes and phases. When we investigate regions with different radial distances from the cluster center, we find that the amplitude of the rotation decreases when moving away from the cluster center (Figure~\ref{fig:regions}).
    \item Both populations exhibit similar velocity-dispersion profiles in the plane of the sky, with the values of radial and tangential velocities ranging from $\sim$18 km/s, in the cluster center, to 7 km/s, at a distance of $\sim 5$ km/s (Figure~\ref{fig:anisotropia}).   
    \item The motions of the stellar populations with different metallicities are isotropic within about one half-light radii from the cluster center and radially anisotropic from $\sim 1$ up-to $\sim 4\,R_{\rm h}$ (Figure~\ref{fig:anisotropia}). 
    The motions become isotropic in the outermost regions.  
\end{itemize}

In addition, we identified two main groups of N-poor and N-rich stars of both $\omega\,$Centauri and M\,22 and studied their spatial distributions and internal kinematics. In the case of $\omega\,$Centauri we also investigated the population-$a$, which is composed of the most metal-rich stars of this cluster \citep[e.g.][]{johnson2010, marino2011}.

The main results on stellar populations with different nitrogen of $\omega\,$Centauri can be summarized as follows. 
\begin{itemize}
    \item N-rich stars of $\omega\,$Centauri exhibit a flatter spatial distribution than N-poor stars. The difference is more pronounced when we consider the sample of metal-poor stars alone, where the N-poor Fe-poor and N-rich Fe-poor sub-populations have average ellipticities of $\sim$0.06 and $\sim$0.22, respectively (Figure~\ref{fig:ell subpop}).
    Population-$a$ stars exhibit higher ellipticity ($e \sim 0.13$) than the bulk of $\omega\,$Centauri stars, which have $e \sim 0.07$.
    \item N-poor and N-rich stars of $\omega\,$Centauri exhibit similar rotation patterns. However, when we consider the Fe-rich population alone, we find that N-rich Fe-rich stars have lower tangential velocities than N-poor Fe-poor stars in the radial interval between $\sim 1$ and $\sim 3$ half-light radii (Figure~\ref{fig:profile subpop}). In this region, the amplitude of the rotation curve of N-rich Fe-rich stars of $\omega\,$Centauri seems smaller than that of N-poor Fe-rich stars, but the amplitude difference is significant at $\sim$2.4-sigma and 2.1 level only, when we consider the motions $\Delta \mu_{\rm \alpha} cos{\delta}$ vs.\,$\theta$ and $\Delta \mu_{\delta}$ vs.\,$\theta$ planes, respectively.
    The conclusion that the amplitude differences are due to observational uncertainties in both components at the same time is smaller than 0.002.
    Similarly to the group of N-rich Fe-rich stars, population-$a$ seems to exhibit low values of $\Delta \mu_{\rm TAN}$ relative to the N-poor Fe-rich stars \footnote{ The origin of population-a stars, which exhibit distinct metallicity than the bulk of $\omega\,$Centauri stars, is still widely debated. 
    Work based on chemical evolution models suggests that it is an extreme case of chemical enrichment \citep[e.g.][and references therein]{dantona2011}. As an alternative, recent work suggest that it could be the product of a merger \citep{calamida2020}.  Specifically, the latter hypothesis would be supported by a strong radial anisotropy of population-a stars, which is a signature of a minor-merger remnant \citep{hong2017}. Nevertheless, the fact that population-a stars are more centrally concentrated than metal-poor stars \citep[e.g.][]{bellini2009} would be a challange for the merging scenario \citep[][]{gavagnin2016}.}
     
    \item Both N-rich and N-poor stars of $\omega\,$Centauri exhibit radially anisotropic motions with some hints for differences between the level of anisotropy of the two populations in the radial interval between 1.5 and 2.5 half-light radii.  
    Between $\sim$3 and $4.5$ half-light radii N-poor stars are consistent with isotropic motions while N-rich stars have $\beta \sim -0.2$ (Figure~\ref{fig:profile subpop}).

    {Numerical studies show that tidally-filled stellar systems exhibit isotropic motions in their central regions, as a consequence of the shorter relaxation time  and the high stellar encounter rate. Moving toward the middle regions, the system starts to expand due to the relaxation process. Therefore, stars in these regions would exhibit a moderate radially anisotropic motion. Finally, since stars with radial orbits preferentially escape from the system \citep[e.g.][]{takahashi1997}, the outermost regions are characterized by isotropic motions. 
    On the other hand, tidally underfilling systems do not show isotropic pattern in the outer regions \citep{vesperini2014, tiongco2016}.
    
    Based on $N$-body simulations of multiple populations  in GCs, \citet{tiongco2019} show that the anisotropy profile of 1G\footnote{We adopt here the same naming convention used in \citet{tiongco2019}, i.e. 1G and 2G in place of 1P and 2P.} stars evolves as a tidally-filling stellar system, whereas the 2G behaves like a tidally underfilling system \citep[see also][]{tiongco2016}. Hence, their 1G and 2G stars share similar anisotropy profiles in the inner and middle regions of the clusters but exhibit different trends in the outer regions. The anisotropy profiles of N-poor Fe-poor and N-rich Fe-poor stars (bottom-middle panel of Figure~\ref{fig:profile subpop}) are qualitatively consistent with the findings by Tiongco and collaborators. Similar conclusions are drawn by \citet{bellini2015} in their investigation of the internal kinematics of stellar populations in the GC NGC\,2808.
    }

\end{itemize}

We find that the ellipses that reproduce the distribution of N-rich stars of M\,22  have higher ellipticities than those of N-poor stars, in close analogy with what is observed in $\omega\,$Centauri. 
This result is qualitatively consistent with the conclusion by \citet{lee2015}, who find that Ca-rich stars of M\,22 are more elongated than Ca-poor stars, in the hypothesis that Ca-rich stars are, on average, nitrogen enhanced with respect to the Ca-poor population.
N-poor and N-rich stars of M\,22 exhibit similar rotation patterns and radially isotropic proper motions.
The fact that M\,22 has significantly shorter relaxation times than $\omega\,$Centauri \citep[e.g.][]{baumgardt2018}, could indicate that M\,22 stars are partially mixed and have erased most of the initial dynamical differences between the distinct stellar populations. This possibility could explain why the stellar populations of M\,22 share the similar kinematics. 
However, it is worth noting that our results do not confirm the conclusion by \citet{lee2015} and \citet{lee2020} who find  metal-rich stars of M\,22 rotate faster than metal-poor stars both on the plane of the sky and along the line of sight.

The results ore based on the selection of Fe-rich and Fe-poor stars derived from the procedure I of Section~\ref{sec:cmds}. We repeated the analysis by using the sample of Fe-rich and Fe-poor stars of $\omega$\,Centauri selected by using procedures II and III of Section~\ref{sec:cmds} and confirm all the conclusions of the paper. We conclude that the results are not affected by the criteria adopted to separate stars with different metallicities.

The findings of this paper, together with results from the literature provide constraints on the formation and evolution of multiple populations in Type II GCs.
 Indeed, the present-day dynamics of stellar populations in clusters where the stars are not fully mixed provide information on the initial conditions of stellar populations in GCs.   

In this context, we emphasize that the rotation of stellar populations with different metallicities has been previously studied from radial velocities of RGB stars.
In their spectroscopic study of $\sim$400 stars in $\omega\,$Centauri, \citet{norris1997} did not find significant rotation along the line of sight among the most metal-rich stars in their sample (corresponding to $\sim$20\% of the studied stars). In contrast, the metal-poor component clearly exhibits systemic  rotation.  This result has been challenged by \citet{pancino2007} who concluded that the metal-poor, metal-intermediate, and metal-rich stars are consistent with having the same rotation patterns based on radial velocities.  

Our results on the rotation in the groups of Fe-poor and Fe-rich stars corroborate the evidence that the two main sample of stars with different metallicities share similar rotation patterns both along the line of sight and the plane of the sky.   
However, the fact that the sample of population-a stars studied in this paper exhibit lower tangential velocities relative to the bulk of $\omega\,$Centauri stars suggests that the stars of this extreme population, similarly to the other Fe-rich and N-rich stars of $\omega\,$Centauri, exhibit less pronounced rotation on the plane of the sky than the remaining cluster members, similarly to what has been suggested by \citet{norris1997} from stellar radial velocities.   

The motions on the plane of the sky of $\omega\,$Centauri stars have been recently studied by using {\it HST} relative proper motions of stars in a field located $\sim$17 arcmin south-west of the cluster center \citep{bellini2018}. The two groups of MS-I and MS-II stars studied by \citet{bellini2018} can be tentatively  associated with the populations of N-poor and N-rich stars analyzed in our paper\footnote{MS-I and MS-II stars have been identified by \citet{milone2017} along the entire MS, from the turn off towards the hydrogen-burning limit by using optical and near-infrared {\it HST} photometry. MS-I stars are consistent with having average [Fe/H]$\sim-1.7$ and primordial helium content, whereas MS-II stars are, on average, more metal-rich ([Fe/H]$\sim -1.4$) and have high helium abundance (Y$\sim$0.37--0.40). Both MSs host stellar sub-populations with different metallicities and light-element abundances, with  MS-II stars having lower oxygen and higher nitrogen content than MS-I stars.  
Due to the complexity of $\omega\,$Centauri, it is not possible to connect the stellar populations analyzed in this paper along the RGB with those identified by \citet{milone2017} along the MS. Nevertheless, based on the metallicities and the content of helium and nitrogen, we can  associate the bulk of MS-I stars with the N-poor population of this paper, whereas the majority of MS-II stars are the RGB counterparts of the N-rich population.}.      
\citet{bellini2018} shows that MS-II stars are significantly more radially anisotropic than MS-I stars, which are consistent with an isotropic velocity distribution. 
This result is consistent with our finding that at $r/R_{\rm h} \sim 3.5$ the N-rich stars have $\beta \sim -0.2$, while N-poor stars exhibit nearly isotropic motions. 
Moreover, MS-I stars exhibit excess systemic rotation in the plane of the sky  with respect to MS-II stars \citep{bellini2018}. In this paper, we find that among Fe-rich stars, the rotation curves of the N-poor population exhibit larger amplitudes than those of N-rich stars. Hence, both results from this paper and from Bellini and collaborators corroborate the conclusion that stellar populations with different nitrogen abundances exhibit distinct rotation patterns.  

A variety of scenarios predict that GCs have experienced a complex formation history and that the multiple stellar populations are a consequence of different star-formation episodes \citep[][and references therein]{renzini2015}. According to some of these scenarios, GCs host second stellar generations that formed in high-density subsystems embedded in a more-extended first generation \citep[e.g.][]{ventura2001, dercole2008, dantona2016, calura2019}.   
These scenarios are supported by the evidence that metal-rich and helium-rich stars of $\omega$\,Centauri, whose half-light relaxation time exceed the Hubble time, are more centrally concentrated than the bulk of cluster stars \citep[e.g.][]{norris1996,sollima2007,bellini2009}.

\citet{mastrobuono2013, mastrobuono2016} investigated the possibility that the formation of second-generation stars in GCs may occur in flattened and centrally-concentrated disk-like structures. They used $N$-body simulations to explore the evolution of such stellar disks embedded in first-generation stars and concluded that the signature of the initial configuration can still be observable in the present-day clusters if the relaxation time is long enough.   
The finding that N-rich stars exhibit elliptical spatial distributions with higher eccentricity than that of N-poor stars, is qualitatively consistent with the possibility that N-rich stars are the second generation of $\omega\,$Centauri and formed the a disk-like structure.
 
Based on the chemical composition of the stellar populations of $\omega\,$Centauri, \citet{marino2012} suggested that $\omega\,$Centauri has first  experienced the enrichment in iron and $\alpha$ elements (oxygen) from core-collapse supernovae. This process is followed by the formation of stellar populations from material ejected from more-massive first-generation stars, possibly in the asymptotic-giant branch phase, and processed by p-capture elements. The evidence that the groups of Fe-rich and Fe-poor stars of both $\omega\,$Centauri and M\,22 have similar spatial distributions while N-rich stars are more flattened than N-poor stars is consistent with a scenario where distinct processes are responsible for the enrichment in iron and in p-capture elements, and where the formation of N-rich stellar populations is associated with cooling flow of material in centrally-concentrated disk-like structures.

\section*{acknowledgments} 
\small
This work has received funding from the European Research Council (ERC) 
under the European Union's Horizon 2020 research innovation programme 
(Grant Agreement ERC-StG 2016, No 716082 'GALFOR', PI: Milone, \url{http://progetti.dfa.unipd.it/GALFOR}), 
and the European Union's Horizon 2020 research and innovation programme 
under the Marie Sk\l odowska-Curie (Grant Agreement No 797100, beneficiary: 
Marino). APM, ED and MT acknowledge support from MIUR through the FARE project 
R164RM93XW SEMPLICE (PI: Milone).  
APM and MT  have been supported by MIUR under PRIN program 2017Z2HSMF (PI: Bedin). AMB acknowledges support by Sonderforschungsbereich (SFB) 881 `The Milky Way System' of the German Research Foundation (DFG). HJ acknowledges support from the Australian Research Council through the Discovery Project DP150100862
 


\begin{thebibliography}{}

\bibitem[Anderson et al. (2008)]{anderson2008} Anderson, J., Sarajedini, A., Bedin, L. R., et al.\ 2008, \aj, 135, 2055

\bibitem[Anderson \& van der Marel(2010)]{anderson2010} Anderson, J., \& van der Marel, R.~P.\ 2010, \apj, 710, 1032


\bibitem[Baumgardt \& Hilker(2018)]{baumgardt2018} Baumgardt, H., \& Hilker, M. \ 2018, \mnras, 478, 1520

\bibitem[Bekki \& Freeman(2003)]{bekki2003} Bekki, K., \& Freeman, K.~C.\ 2003, \mnras, 346, L11

\bibitem[Bellazzini et al.(2008)]{bellazzini2008} Bellazzini, M., Ibata, R.~A., Chapman, S.~C., et al.\ 2008, \aj, 136, 1147

\bibitem[Bellini et al.(2009)]{bellini2009} Bellini, A., Piotto, G., Bedin, L.~R et al.\ 2009, \aas, 493, 959

\bibitem[Bellini et al.(2009)]{bellini2009b} Bellini, A., \& Bedin, L.~R et al.\ 2009, PASP, 121, 1419

\bibitem[Bellini et al.(2011)]{bellini2011} Bellini, A., Anderson, J. \& Bedin, L.~R et al.\ 2011, PASP, 123, 622

\bibitem[Bellini et al.(2015)]{bellini2015} Bellini, A., Vesperini, E. \& Piotto, G. et al.\ 2015, \apj, 810, L13

\bibitem[Bellini et al.(2018)]{bellini2018} Bellini, A., Libralato, M., Bedin, L.~R., et al.\ 2018, \apj, 853, 86 

\bibitem[Bianchini et al.(2018)]{bianchini2018} Bianchini, P., van der Marel, R.~P., del Pino, A. et al.\ 2018, \mnras, 481, 2125

\bibitem[Bianchini et al.(2019)]{bianchini2019} Bianchini, P., Ibata, R., \& Famaey, B.\ 2019, \apjl, 887, L12

\bibitem[Calamida et al.(2020)]{calamida2020} Calamida, A., Zocchi, A., Bono, G., et al.\ 2020, \apj, 891, 167

\bibitem[Calura et al.(2019)]{calura2019} Calura, F., D'Ercole, A., Vesperini, E., et al.\ 2019, \mnras, 489, 3269 

\bibitem[Carretta et al.(2009)]{carretta2009a} Carretta, E., Bragaglia, A., Gratton, R.~G., et al.\ 2009, \aap, 505, 117 

\bibitem[Cordoni et al.(2018)]{cordoni2018} Cordoni, G., Milone, A.~P., Marino, A.~F., et al.\ 2018, \apj, 869, 139 

\bibitem[Cordoni et al.(2020)]{cordoni2019} Cordoni, G., Milone, A.~P., Mastrobuono-Battisti, A., et al.\ 2020, \apj, 889, 18  

\bibitem[Da Costa et al.(2009)]{dacosta2009} Da Costa, G. S., Held, E.~V., Saviane, I. and Gullieuszik, M. 2009, \apj, 705, 1481

\bibitem[Da Costa et al.(2009)]{dacosta2011} Da Costa, G. S., \& Marino, A.\, F.\, 2011, PASA, 28, 28 

\bibitem[D'Antona et al.(2011)]{dantona2011} D'Antona, F., D'Ercole, A.,  Marino, A. F.,  Milone, A. P.;  Ventura, P. \&  Vesperini, E. 2011, \apj, 736, 7 


\bibitem[D'Antona et al.(2016)]{dantona2016} D'Antona, F., Vesperini, E., D'Ercole, A. et al.\ 2016, \mnras, 458, 2122

\bibitem[D'Ercole et al.(2010)]{dercole2010} D'Ercole, A., D'Antona, F., Ventura, P. et al.\ 2010, \mnras, 407, 854

\bibitem[D'Ercole et al.(2008)]{dercole2008} D'Ercole, A., Vesperini, E., D'Antona, F., et al.\ 2008, \mnras, 391, 825

\bibitem[Gaia Collaboration et al.(2018a)]{gaia2018a} Gaia Collaboration, Brown, A.~G.~A., Vallenari, A., et al.\ 2018, \aap, 616, A1 

\bibitem[Gavagnin et al. (2016)]{gavagnin2016} 
Gavagnin, E.,  Mapelli, M. \& Lake, G. 2016, \mnras, 461, 1276 

\bibitem[Harris(1996)]{harris1996} Harris, W.~E.\ 1996, \aj, 112, 1487 

\bibitem[Hong et al. (2017)]{hong2017} Hong, J., de Grijs, R., Askar, A., et al. 2017, \mnras, 472, 67

\bibitem[H{\'e}nault-Brunet et al.(2015)]{henault2015} H{\'e}nault-Brunet, V., Gieles, M., Agertz, O., \& Read, J.~I.\ 2015, \mnras, 450, 1164 

\bibitem[Jindal et al. (2019)]{jindal2019} Jindal, A., Webb, J.~J., \& Bovy, J.\ 2019, \mnras, 487, 3693

\bibitem[Johnson \& Pilachowski (2010)]{johnson2010} Johnson, C.~I. \& Pilachowski, C.~A, et al.\,\apj, 772, 1373

\bibitem[Johnson et al.(2015)]{johnson2015} Johnson, C.~I., Rich, R.~M., Pilachowski, C.~A, et al.\,\aj, 160, 53 

\bibitem[Landolt(1992)]{landolt1992} Landolt, A.~U.\ 1992, \aj, 104, 340 

\bibitem[Lee(2015)]{lee2015} Lee, J.-W.\ 2015, ApJS, 219, 7 

\bibitem[Lee(2020)]{lee2020} Lee, J.-W.\ 2020, \apj, 880, 6L 

\bibitem[Lee et al.(1999)]{lee1999} Lee, Y. W., Joo, J. M., Sohn, Y. J., et al.\ 1999, Nature, 402, 55

\bibitem[Lindegren et al.(2018)]{lindegren2018} Lindegren, L., Hern{\'a}ndez, J., Bombrun, A., et al.\ 2018, \aap, 616, A2 

\bibitem[Mackey et al.(2013)]{mackey2013} Mackey, A.~D., Da Costa, G.~S., Ferguson, A.~M.~N., \& Yong, D.\ 2013, \apj, 762, 65 

\bibitem[Marino et al.(2009)]{marino2009} Marino, A.~F., Milone, A.~P., Piotto, G., et al.\ 2009, \aas, 505, 1099 

\bibitem[Marino et al.(2010)]{marino2010} Marino, A.~F., Piotto, G., Gratton, R., et al.\ 2010, IAUS, 268, 183 

\bibitem[Marino et al.(2011)]{marino2011} Marino, A.~F., Sneden, C., Kraft, R.~P. et al.\ 2011, \aas, 532, 8 

\bibitem[Marino et al.(2012)]{marino2012} Marino, A.~F., Milone, A.~P., Piotto, G. et al.\ 2012, \apj, 746, 14 

\bibitem[Marino et al.(2014)]{marino2014} Marino, A.~F., Milone, A.~P., Yong, D., et al.\ 2014, \mnras, 442, 3044 

\bibitem[Marino et al.(2015)]{marino2015} Marino, A.~F., Milone, A.~P., Karakas, A.~I., et al.\ 2015, \mnras, 450, 815 

\bibitem[Marino et al.(2016)]{marino2016} Marino, A.~F., Milone, A.~P., Casagrande, L., et al.\ 2016, \mnras, 459, 610 

\bibitem[Marino et al.(2017)]{marino2017} Marino, A.~F., Milone, A.~P., Yong, D., et al.\ 2017, \apj, 843, 66 

\bibitem[Marino et al.(2019)]{marino2019} Marino, A.~F., Milone, A.~P., Renzini, A., et al.\ 2019, \mnras, 487, 3815 

\bibitem[Mastrobuono-Battisti \& Perets(2013)]{mastrobuono2013} Mastrobuono-Battisti, A., \& Perets, H.~B.\ 2013, \apj, 779, 85 
  
\bibitem[Mastrobuono-Battisti \& Perets(2016)]{mastrobuono2016} Mastrobuono-Battisti, A., \& Perets, H.~B.\ 2016, \apj, 823, 61 

\bibitem[Milone et al.(2008)]{milone2008} Milone, A.~P., Bedin, L.~R., Piotto, G., , et al.\ 2008, \apj, 673, 241 

\bibitem[Milone et al.(2012)]{milone2012} Milone, A.~P., Piotto, G., Bedin, L.~R., et al.\ 2012, \aap, 540, A16 

\bibitem[Milone et al.(2012)]{milone2012b} Milone, A.~P., Piotto, G., Bedin, L.~R., et al.\ 2012, \apj, 744, 58 

\bibitem[Milone et al.(2015)]{milone2015} Milone, A.~P., Marino, A.~F., Piotto, G., et al.\ 2015, \apj, 808, 51 

\bibitem[Milone et al.(2017)]{milone2017} Milone, A.~P., Piotto, G., Renzini, A., et al.\ 2017, \mnras, 464, 3636     

\bibitem[Milone et al.(2017)]{milone2017b} Milone, A.~P., Marino, A.~F., Bedin, L.~R., et al.\ 2017, \mnras, 469, 800     

\bibitem[Milone et al.(2018)]{milone2018} Milone, A.~P., Marino, A.~F., Mastrobuono-Battisti, A., \& Lagioia, E.~P.\ 2018, \mnras, 479, 5005 

\bibitem[Milone et al.(2020)]{milone2020} Milone, A.~P., Marino, A.~F., Da Costa, G.~S., et al.\ 2020, \mnras, 491, 515 
\bibitem[Monelli et al.(2013)]{monelli2013} Monelli, M., Milone, A.~P., Stetson, P.~B., et al.\ 2013, \mnras, 431, 2126 

\bibitem[Norris \& Da Costa (1995)]{norris1995} Norris, J.~E. \& Da Costa \apj\, 447, 680

\bibitem[Norris et al.(1996)]{norris1996} Norris, J.~E, Freeman, K.~C. \& Mighell, K.~J.\,1996 \apj\, 462, 241

\bibitem[Norris et al.(1997)]{norris1997} Norris, J.~E., Freeman, K.~C., Mayor, M., et al.\ 1997, \apjl, 487, L187

\bibitem[Pancino et al.(2000)]{pancino2000} Pancino, E., Ferraro, F.~R., Bellazini, M. et al.\ 2000, \aj, 533, L83

\bibitem[Pancino et al.(2007)]{pancino2007} Pancino, E., Galfo, A., Ferraro, F.~R., et al.\ 2007, \apjl, 661, L155

\bibitem[Piotto et al.(2012)]{piotto2012} Piotto, G., Milone, A.~P., Anderson, J., et al.\ 2012, \apj, 760, 39 

\bibitem[Hal{\i}r \& Flusser (1998)]{radim1998} Hal{\i}r, R. \& Flusser, J.\ 1998

\bibitem[Renzini et al.(2015)]{renzini2015} Renzini, A., D'Antona, F., Cassisi, S., et al.\ 2015, \mnras, 454, 4197 
  
\bibitem[Sabbi et al. (2016)]{sabbi2016} Sabbi, E., Lennon, D. J., Anderson, J. et al.\ 2016, ApJS, 222, 11

\bibitem[Sanna et al.(2020)]{sanna2020} Sanna, N., Pancino, E., Zocchi, A., et al.\ 2020, arXiv e-prints, arXiv:2003.13575

\bibitem[Silverman (1986)]{silverman1986} Silverman B. W., 1986, Monographs on Statistics and Applied Probability. Chapman and Hall, London

\bibitem[Sollima et al. (2007)]{sollima2007} Sollima, A., Ferraro, F.\,R.\, Bellazzini, M.\, et al.\ 2007 \apj, 654, 915

\bibitem[Sollima et al.(2019)]{sollima2019} Sollima, A., Baumgardt, H., Hilker, M.,\ 2019, \mnras, 485, 1460

\bibitem[Stetson(2005)]{stetson2005} Stetson, P.~B.\ 2005, \pasp, 117, 563 

\bibitem[Stetson et al.(2019)]{stetson2019} Stetson, P.~B., Pancino, E., Zocchi, A. et al.\ 2019, \mnras, 485, 3042

\bibitem[Takahashi et al.(1997)]{takahashi1997}
Takahashi, K.,  Lee, H.,  Inagaki, S.\ 1997 \mnras 
 292, 331
 
\bibitem[Tiongco et al.(2016)]{tiongco2016} Tiongco, M.~A., Vesperini, E. \& Varri, A.~L.\ 2016 \mnras, 455, 3693

\bibitem[Tiongco et al.(2019)]{tiongco2019} Tiongco, M.~A., Vesperini, E. \& Varri, A.~L.\ 2019 \mnras, 487, 5535

\bibitem[Vasiliev (2019)]{vasiliev2019} Vasiliev, E.\ 2019, \mnras, 484, 2832

 \bibitem[Ventura et al.(2001)]{ventura2001} Ventura, P., D'Antona, F., Mazzitelli, I., \& Gratton, R.\ 2001, \apjl, 550, L65 
\bibitem[van den Ven et al.(2006)]{vandenven2006} van de Ven, G., van den Bosch,R. C. E., Verolme, E. K., \&  de Zeeuw, P. T.\ 2006, \aap, 445, 513

\bibitem[Vesperini et al.(2013)]{vesperini2013} Vesperini, E., McMillan, S.~L.~W., D'Antona, F., \& D'Ercole, A.\ 2013, \mnras, 429, 1913 

\bibitem[Vesperini et al.(2014)]{vesperini2014} Vesperini, E., Varri, A. L.,  McMillan, S. L. W. \&  Zepf, S. E. \mnras, 443, 79 

\bibitem[Wand (2015)]{wand2015} Wand, M.\ 2015, KernSmooth: Functions for Kernel Smoothing Supporting Wand \& Jones (1995)

\bibitem[Yong \& Grundahl (2008)]{yong2008} Yong, D. \& Grundahl, F., \apj, 672, 29

\bibitem[Yong et al. (2014) ]{yong2014} Yong, D., Roederer, I. U., Grundahl, F., et al.\,2014, \mnras, 441, 3396

\end{thebibliography}

\bibliographystyle{aa}

\end{document}